\documentstyle[12pt,twoside]{article}
\def\ie{{\em i.e.}}
\def\eg{{\em e.g.}}

\def\bsg{B(b\to s\gamma)}
\def\beq{\begin{equation}}
\def\eeq{\end{equation}}

\catcode`\@=11 
\def\coeff#1#2{{\textstyle{#1\over #2}}}

\def\vev#1{\left\langle #1\right\rangle}
\def\lsim{\mathrel{\mathpalette\@versim<}}
\def\gsim{\mathrel{\mathpalette\@versim>}}
\def\@versim#1#2{\vcenter{\offinterlineskip
    \ialign{$\m@th#1\hfil##\hfil$\crcr#2\crcr\sim\crcr } }}
\def\etal{{\em et. al.}}
\def\JL{J. L. Lopez}
\def\DVN{D. V. Nanopoulos}
\def\AZ{A. Zichichi}

\def\XW{X. Wang}
\def\r#1{$\bf#1$}
\def\rb#1{$\bf\overline{#1}$}

\def\t1{{\tilde 1}}
\def\ov{\overline}

\def\mpt{p\hskip-5.5pt/\hskip2pt}

\def\MeV{\,{\rm MeV}}
\def\GeV{\,{\rm GeV}}
\def\TeV{\,{\rm TeV}}
\def\y{\,{\rm y}}

\def\wt{\widetilde}

\def\to{\rightarrow}
\def\pb{\,{\rm pb}}
\def\ipb{\,{\rm pb}^{-1}}
\def\NPB#1#2#3{Nucl. Phys. B {\bf#1} (19#2) #3}
\def\PLB#1#2#3{Phys. Lett. B {\bf#1} (19#2) #3}

\def\PRD#1#2#3{Phys. Rev. D {\bf#1} (19#2) #3}
\def\PRL#1#2#3{Phys. Rev. Lett. {\bf#1} (19#2) #3}
\def\PRT#1#2#3{Phys. Rep. {\bf#1} (19#2) #3}
\def\MODA#1#2#3{Mod. Phys. Lett. A {\bf#1} (19#2) #3}

\def\TAMU#1{Texas A \& M University preprint CTP-TAMU-#1}

\begin{document}

\begin{center}
\vglue 0.3cm
{\Large \bf Status of the Superworld \\
{}From Theory to Experiment
\\}
\vspace{0.2cm}
\vglue 0.5cm
{JORGE L. LOPEZ$^{(a),(b)}$, D. V. NANOPOULOS$^{(a),(b),(c)}$, and A.
ZICHICHI$^{(d)}$\\}
\vglue 0.4cm
{\em $^{(a)}$Center for Theoretical Physics, Department of Physics, Texas A\&M
University\\}
{\em College Station, TX 77843--4242, USA\\}
{\em $^{(b)}$Astroparticle Physics Group, Houston Advanced Research Center
(HARC)\\}
{\em The Mitchell Campus, The Woodlands, TX 77381, USA\\}
{\em $^{(c)}$CERN Theory Division, 1211 Geneva 23, Switzerland\\}
{\em $^{(d)}$CERN, 1211 Geneva 23, Switzerland\\}
\vglue 0.4cm
\baselineskip=12pt
\vglue 1.cm
{ABSTRACT}
\end{center}
{\rightskip=3pc
 \leftskip=3pc
\noindent
Among the most outstanding conceptual developments in particle physics we have:
the unification of all particle interactions at very-high energies (Grand
Unification), the fermion-boson symmetry (Supersymmetry), the non-point-like
structure of elementary particles (String theory), and the understanding that
all dynamical quantities (gauge couplings, masses, Yukawa couplings) run
with energy (Renormalization Group Equations). The goal is to make use of these
great developments to construct a theory which embraces all fundamental forces
of Nature, including gravity. In this review we address this problem and its
possible implications for physics in the energy range where our experimental
facilities operate. We show that what is required are not qualitative arguments
but a set of detailed calculations with definite predictions. This is why we
have chosen two specific Supergravity models: $SU(5)$ and $SU(5)\times U(1)$.
One as representative of Field Theory, the other of String Theory. This kind
of model-building at the Planck scale seems to tell us that new physics beyond
the Standard Model could be near the Fermi scale. Therefore detailed
calculations for existing facilities (Tevatron, LEP I-II, HERA, Gran Sasso,
Super Kamiokande, Dumand, Amanda, etc.) are a definite way to put string theory
-- \ie, the existence of the Superworld -- under experimental test, now.}
\begin{flushleft}
\baselineskip=12pt
{\em December 1993}
\end{flushleft}
\vfill\eject
\tableofcontents
\vfill\eject
\setcounter{page}{1}
\markright{}
\pagestyle{myheadings}
\baselineskip=14pt

\section{Introduction}
\label{Introduction}
To unify all fundamental forces of Nature is the dream of most physicists.
Important developments in this direction have taken place during the last two
decades, namely grand unification, supersymmetry, and string theory. The first
predicts that all particle interactions become of equal strength above a
very high energy scale, the second puts bosons and fermions on equal
footing, and the third abandons the point-like structure of elementary
particles in favour of  one-dimensional ``strings". A nice feature of
string theory is that it needs to be supersymmetric, thus suggesting
that the Superworld may really exist and be discovered soon, if supersymmetry
is effectively broken at low energies.

The advent of LEP has allowed to measure the three gauge couplings of the
Standard Model, $SU(3)_C$, $SU(2)_L$, $U(1)_Y$ (at $M_Z$) with unprecedented
accuracy (see \eg, Ref.~\cite{LEPC}):
\begin{center}
$\begin{array}{ccc}
\alpha^{-1}_1&=&58.83\pm0.11,\\
\alpha^{-1}_2&=&29.85\pm0.11,\\
\alpha^{-1}_3&=&8.47\pm0.57.
\end{array}$
\end{center}
Furthermore, the number of families has been determined to be \cite{LEPC}
\begin{center}
$N_f = 2.980\pm0.027.$
\end{center}
{}From these experimental data the Renormalization Group Equations (RGEs) allow
to span as many orders of magnitude as wanted, provided we know the virtual
phenomena to be accounted for in these large ranges of energy, starting at
$M_Z$.

The key features in this field of physics are therefore the quantities
above and the RGEs. From these ingredients we would like to know if it
is possible to
predict the threshold for the lightest detectable supersymmetric
particle. It is a field which we have studied since the late 70's.
We realized the importance of the new degree
of freedom (the threshold for supersymmetric particle production)
introduced in the running of the gauge couplings ($\alpha^{-1}_1,
\alpha^{-1}_2,  \alpha^{-1}_3$) for their convergence towards a unique value.
However, we also noticed the many problems to be overcome in the understanding
of where the energy level for the superparticles could be. This review
paper is organized as follows. In Section~\ref{convergence} the
problem of the convergence of the gauge couplings  $\alpha_1,
\alpha_2, \alpha_3$  is discussed. In Section~\ref{constraints} the
constraints from the unification conditions are presented in order to
allow the reader an understanding of the physical consequences hidden
in the mathematical formalism. Section~\ref{origin} discusses the popular
quantity $M_{SUSY}$ and the reasons why its use should be discontinued.
Section~\ref{need} presents the need to promote global supersymmetry to
be a local one thus giving rise to the Supergravity description of
low-energy phenomena. Sections~\ref{SU5} and ~\ref{SU5xU1} present the
two Supergravity Models $SU(5)$ and $SU(5)\times U(1)$ and in
Sections~\ref{tevatron},~\ref{LEP},~\ref{HERA},~\ref{Under}
detailed calculations for the Tevatron, LEP I-II, HERA, and the
Underground Laboratories plus Underwater Facilities are reported.
Section~\ref{indirect} is devoted to detailed calculations for
indirect experimental detection. In Section~\ref{mass} the origin of mass
problem is addressed together with a working ground to predict $m_t$.
Our conclusions are summarized in Section~\ref{conclusions}.

\section{High precision LEP data and convergence of couplings:
physics is not Euclidean geometry}
\label{convergence}
Experimentally we know that (see \eg, \cite{LP}),
\begin{eqnarray}
\alpha^{-1}_e(M_Z)&=&127.9\pm0.2,\\
\alpha_3(M_Z)&=&0.118\pm0.008,\\
\sin^2\theta_W(M_Z)&=&0.2334\pm0.0008.
\end{eqnarray}
The $U(1)_Y$ and $SU(2)_L$ gauge couplings are related to these
 by $\alpha_1={5\over3}(\alpha_e/\cos^2\theta_W)$ and
$\alpha_2=(\alpha_e/\sin^2\theta_W)$.
As mentioned above, the values of the three gauge couplings of the
Standard Model derived from Eqs.~(1),(2), and (3), are:
\begin{equation}
{\rm at}\quad Q=M_Z\qquad\left\{
\begin{array}{ccc}
\alpha^{-1}_1&=&58.83\pm0.11\\
\alpha^{-1}_2&=&29.85\pm0.11\\
\alpha^{-1}_3&=&8.47\pm0.57
\end{array}
\right.\label{initialconditions}
\end{equation}
These three gauge couplings evolve with increasing values of the scale $Q$ in
a logarithmic fashion, and may become equal at some higher scale, signaling
the possible presence of a larger gauge group. However, this need not be the
case: the three gauge couplings may meet and then depart again. Conceptually,
the presence of a unified group is essential in the discussion of unification
of couplings. In this case, the newly excited degrees of freedom will be
such that all three couplings will evolve together for scales
$Q>E_{GUT}$,  and one can then speak of a unified coupling.

The running of the gauge couplings is prescribed by a set of first-order
non-linear differential equations: the RGEs for the gauge
couplings. In general, there is one such  equation
for each dynamical variable in the theory (\ie, for each gauge
coupling, Yukawa coupling, particle and sparticle mass). These equations give
the  rate of change of each
dynamical variable as the scale $Q$ is varied. For the case of a gauge
coupling, the rate of change is proportional to (some power of) the
gauge  coupling itself, and the coefficient of proportionality is
called  the {\em beta function}. The beta functions encode the
spectrum of  the theory, and how the various gauge couplings influence
the  running of each other (a higher-order effect). Assuming that all
supersymmetric particles have a common mass $M_{SUSY}$, the RGEs
(to two-loop order) are:
\begin{equation}
{d\alpha^{-1}_i\over
dt}=-{b_i\over2\pi}-\sum^3_{j=1}{b_{ij}\alpha_j\over8\pi^2}.\label{dgidt}
\end{equation}
where $t=\ln (Q/E_{GUT})$, with $Q$ the running scale and $E_{GUT}$ the
unification mass. The one-loop ($b_i$) and two-loop ($b_{ij}$) beta functions
are given by
\begin{eqnarray}
b_i&=&\left(\coeff{33}{5},1,-3\right),
\label{coefSUSY1}
\end{eqnarray}
\begin{eqnarray}
b_{ij}&=&\left( \begin{array}{c@{\quad}c@{\quad}c}
{199\over 25} & {27\over 5} & {88\over 5} \\
{9\over 5} & 25 & 24 \\ {11\over 5} & 9 & 14
\end{array} \right).
\label{coefSUSY2}
\end{eqnarray}
These equations are valid from $Q=M_{SUSY}$ up to $Q=E_{GUT}$. For
$M_Z<Q<M_{SUSY}$
an analogous set of equations holds, but with beta functions which reflect
the non-supersymmetric nature of the theory (\ie, with all the sparticles
decoupled),
\begin{eqnarray}
b'_i&=&\left(\coeff{41}{10},-\coeff{19}{6},-7\right),
\label{coefnoSUSY1}
\end{eqnarray}
\begin{eqnarray}
b'_{ij}&=&\left( \begin{array}{c@{\quad}c@{\quad}c}
{199\over 50} & {27\over 10} & {44\over 5} \\
{9\over 10} & {35\over6} & 12 \\ {11\over 10} & {9\over2} & -26
\end{array} \right).
\label{coefnoSUSY2}
\end{eqnarray}
The non-supersymmetric equations are supplemented with the initial conditions
given in Eq.~(\ref{initialconditions}).

If the above is all the physics which is incorporated in the study of the
convergence of the gauge couplings, then it is easy to see that the couplings
will always meet at some scale $E_{GUT}$, provided that $M_{SUSY}$ is
tuned  appropriately. This is a simple consequence of euclidean
geometry, as  can be seen from Eq.~(\ref{dgidt}). Neglecting the
higher-order  terms, we see that as a function of $t$, $\alpha^{-1}_i$
are just  straight lines. In fact, the slope of these lines changes at
$Q=M_{SUSY}$, where the beta functions change. The convergence of
three  straight lines with a change in slope is then guaranteed by
euclidean  geometry, as long as the point where the slope changes is
tuned  appropriately. (This fact was pointed out by A. Peterman
and one of us in 1979 \cite{X1}.) On the other hand, the gauge couplings in
non-supersymmetric $SU(5)$ do not converge \cite{CostaAmaldi}. The precise LEP
measurements of the gauge couplings gave new life to the field
\cite{EKNI,EKNII,LL} and produced claims that the convergence of the couplings
needed a change in slope at $M_{SUSY}\sim1\TeV$ \cite{AdBF}.

\begin{figure}[p]
\vspace{6in}
\includegraphics{adbf.eps}
\vspace{1cm}
\caption{\baselineskip=12pt
In this well publicized example of unification of couplings (Fig. 2 of
Ref.~[8]), the divergence of the couplings for scales above $E_{GUT}$ ($\mu$
and $M_{GUT}$ in the figure) is clear, as is the sharp change in slope of
the lines at low energies. Note that neither threshold effects (``light'' or
``heavy'') nor the evolution of the gaugino masses (EGM) are included, and that
these are crucial in the determination of where the Superworld could start
showing evidence for its existence. Notice that the best fit for the
``geometrical'' convergence of the couplings predicts $M_{SUSY}$ at $10^3$ GeV,
and that a low value of $\alpha_3(M_Z)=0.108$ has been used in this analysis.}
\label{Amaldi_FIG}
\end{figure}

\begin{figure}[p]
\vspace{6in}
\includegraphics{figgeom.ps}
\vspace{1cm}
\caption{\baselineskip=12pt
The convergence of the gauge couplings $(\alpha_1, \alpha_2, \alpha_3)$ at
$E_{GUT}$ is followed by the unification into a unique $\alpha_{GUT}$ above
$E_{GUT}$. The RGEs include the heavy and light thresholds plus the evolution
of gaugino masses. These results are obtained using as input the world-average
value of $\alpha_3(M_Z)$ and comparing the predictions for $\sin^2\theta(M_Z)$
and $\alpha_{em}^{-1}(M_Z)$ with the experimental results. The $\chi^2$
constructed using these two physically measured  quantities allows to get the
best $E_{GUT}$, $\alpha_{GUT}(E_{GUT})$, and $\alpha_3(M_Z)$ (shown). Because
all relevant effects have been included, the RGEs can go down to $M_Z$
({\em c. f.} Fig. 1). Notice that $M_X$ corresponds to the heavy threshold. The
$\chi^2$ definition  is based on physical quantities:
$\chi^2=
{\{[\sin^2\theta(M_Z)]_{exp}-[\sin^2\theta(M_Z)]_{th}\}^2/[\sigma_s]^2}+
{\{[\alpha_{em}(M_Z)]_{exp}-[\alpha_{em}(M_Z)]_{th}\}^2/[\sigma_e]^2}$}
\label{FigureY2b}
\end{figure}

\begin{figure}[p]
\vspace{7in}
\includegraphics{fig2Layman.eps}
\vspace{1cm}
\caption{\baselineskip=12pt
This is the best proof that the convergence of the gauge
couplings can be obtained with $M_{SUSY}$ at an energy level as low as $M_Z$.
Notice that the effects of ``light" and ``heavy" thresholds have been
accounted for, as well as the Evolution of Gaugino Masses [14,17].
This is Fig. 2 of Ref.~[18]. $E_{SU}$ is the string unification scale.}
\label{FigureY2}
\end{figure}

This prediction for the likely scale of the supersymmetric spectrum (\ie,
$M_{SUSY}\sim1\TeV$ \cite{AdBF}) is in fact {\em unjustified}
\cite{ACPZI,ACPZII}. The reason is simple:  the physics at the unification
scale, which is used to predict the value of $M_{SUSY}$, has been ignored
completely. In fact, such a geometrical picture of convergence of the gauge
couplings is physically inconsistent, since
for scales $Q>E_{GUT}$ the gauge couplings will depart again, as can
be seen in Fig.~\ref{Amaldi_FIG} (of Amaldi \etal\ \cite{AdBF}). One must
consider a unified theory to be assured that the couplings will remain
unified, as shown in Fig.~\ref{FigureY2b}. This entails the study of a
new kind of effect, namely the effects on the running of the gauge
couplings produced by the degrees of freedom which are excited near the
unification  scale (\ie, the {\em heavy threshold effects})
\cite{ACPZII,EKNIII,ACZ,BH}. In fact, the whole concept of a single unification
point needs to be abandoned. The upshot of all this is that the
theoretical  uncertainties on the values of the parameters describing
the heavy GUT particles are such that the above prediction for
$M_{SUSY}$  \cite{AdBF} is washed out completely \cite{EGM,EKNIV}.
Furthermore,  the insertion of a realistic spectrum of sparticles at
low energies (as opposed to an unrealistic common $M_{SUSY}$ mass)
plus the calculation of the EGM effect (Evolution of Gaugino Masses)
blurs the issue even more \cite{EKNIII,EGM,P4,X2}. {\em Thus, it is perfectly
possible to obtain the unification of the gauge couplings, with  supersymmetric
particle  masses as low as experimentally allowed.} The most complete
analysis of a unified theory is shown in Fig.~\ref{FigureY2}. Note the
unification of the gauge couplings which continues above $E_{GUT}$.
Notice also that ``light" and ``heavy" thresholds have been duly
accounted  for, plus other important effects like the evolution of
gaugino  masses (EGM)~\cite{EGM} quoted above. This effect has in fact been
calculated at two loops \cite{X2}.

A related point is that LEP data do not uniquely demonstrate that the
gauge  couplings must unify at a scale $E_{GUT}\sim10^{16}\GeV$
\cite{ACZ}.  This is probably the simplest conclusion one could draw.
However,  this conclusion is easily altered by for example
considering all experimental and theoretical uncertainties. In  fact,
once this is done, the value of $E_{GUT}$ can reach the string
unification scale, \ie,  $E_{GUT}\sim10^{18}\GeV$, as shown in
Fig.~\ref{FigX}.

\begin{figure}[p]
\vspace{7in}
\includegraphics{figXX.eps}
\vspace{0.5cm}
\caption{\baselineskip=12pt
The dependence of $E_{GUT}$ on $\alpha_3$$(M_Z)$ and on the ratio of the two
crucial heavy-threshold masses, $M_V$$/$$M_\Sigma$. Here $m_0$ parametrizes
the squark and slepton masses, $m_{1/2}$ the gaugino masses, $m_0$ and $m_4$
the Higgs-boson masses, and $m_4$ the Higgsino mass ($m_4$ is more commonly
denoted by $\mu$). Note that the extreme value for $E_{GUT}$ is above $10^{18}$
GeV.}
\label{FigX}
\end{figure}

\begin{figure}[p]
\vspace{7in}
\includegraphics{fig3Layman.ps}
\vspace{0.5cm}
\caption{\baselineskip=12pt
The unification scale $E_{GUT}$ versus $\alpha_3(M_Z)$ for various values of
$\sin^2\theta_W(M_Z)$ within $\pm2\sigma$ of the world-average value. Also
indicated is the lower bound on $E_{GUT}$ from the lower limit on the proton
lifetime.}
\label{Figure1}
\end{figure}

\begin{figure}[p]
\vspace{7in}
\includegraphics{fig4Layman.ps}
\vspace{0.5cm}
\caption{\baselineskip=12pt
The unification scale $E_{GUT}$ versus $M_{SUSY}$ for different values of
$\alpha_3(M_Z)$ and fixed $\sin^2\theta_W(M_Z)$. Note the
anticorrelation between $M_{SUSY}$ and $E_{GUT}$. The experimental
lower bound  on $M_{SUSY}$ is shown. The lower bound on $E_{GUT}$ from
Fig. 5 is also indicated.}
\label{Figure2}
\end{figure}

\begin{figure}[p]
\vspace{7in}
\includegraphics{fig5Layman.ps}
\vspace{0.5cm}
\caption{\baselineskip=12pt
The correlation between all measured quantities, $\alpha_3(M_Z)$,
$\sin^2\theta_W(M_Z)$, $\tau_p$, the limits on the lightest detectable
supersymmetric particle (here represented by $M_{SUSY}$) and the unification
energy scale $E_{GUT}$.}
\label{Figure3}
\end{figure}

\section{Interconnections between the measured quantities due to Unification}
\label{constraints}
The convergence of the gauge couplings implies that given $\alpha_e$ and
$\alpha_3$, one is able to compute the values of $\sin^2\theta_W$, the
unification scale $E_{GUT}$, and the unified coupling $\alpha_U$. In
lowest-order
approximation (\ie, neglecting all GUT thresholds, two-loop effects, and
taking $M_{SUSY}=M_Z$) one obtains
\begin{eqnarray}
\ln{E_{GUT}\over
M_Z}&=&{\pi\over10}\left({1\over\alpha_e}-{8\over3\alpha_3}\right),\label{MU}\\
{\alpha_e\over\alpha_U}&=&{3\over20}\left(1+{4\alpha_e\over\alpha_3}\right),
\label{au}\\
\sin^2\theta_W&=&0.2+{7\alpha_e\over15\alpha_3}\quad.\label{sin2}
\end{eqnarray}
These equations provide a rough approximation to the actual values obtained
when all effects are included. Nonetheless, they embody the most important
dependences on the input parameters. In Fig.~\ref{Figure1}
we show the relation between $E_{GUT}$ and $\alpha_3$ for various values of
$\sin^2\theta_W$. One can observe that:
\begin{eqnarray}
\alpha_3\ \uparrow\quad &\Rightarrow&\quad E_{GUT}\ \uparrow
\qquad{\rm for\ fixed}\ \sin^2\theta_W\label{a3zMu}\\
\sin^2\theta_W\ \uparrow\quad &\Rightarrow&\quad E_{GUT}\ \uparrow
\qquad{\rm for\ fixed}\ \alpha_3\\
\alpha_3\ \uparrow\quad &\Rightarrow&\quad \sin^2\theta_W\ \downarrow
\qquad{\rm for\ fixed}\ E_{GUT}
\end{eqnarray}
These are the most important systematic correlations, which are not really
affected by the neglected effects. These correlations are evident in
Eqs.~(\ref{MU}--\ref{sin2}) and in Fig.~\ref{Figure1}. In this figure
we also show the lower bound on $E_{GUT}$ which follows from the proton
decay constraint. Clearly a lower bound on $\alpha_3(M_Z)$ results,
which allows the world-average value. Another interesting result is
the anticorrelation between $E_{GUT}$ and $M_{SUSY}$. This is shown in
Fig.~\ref{Figure2}, where for fixed $\sin^2\theta_W(M_Z)$ we see that
increasing $\alpha_3(M_Z)$ increases $E_{GUT}$ (as already noted in
Eq.~(\ref{a3zMu})) and  decreases  $M_{SUSY}$.  Taking for  granted
this  approach
(\ie, all supersymmetric particle masses degenerate at $M_{SUSY}$)
for comparison with the large amount of papers published following this
logic, in Fig.~\ref{Figure3} we see the narrow band left open once the
experimental limits on $\tau_p$ and $M_{SUSY}$ are imposed.
Figure~\ref{Figure3} is a guide to understand the qualitative
interconnection   between the basic experimentally measured
quantities,  $\alpha_3(M_Z)$, $\sin^2\theta_W(M_Z)$, $\tau_p$,
$M_{SUSY}$ and   the theoretically wanted  $E_{GUT}$. The
experimental lower  bounds on the proton lifetime $(\tau_p)_{exp}$ and
on  $M_{SUSY}$ produce two opposite bounds (lower and upper,
respectively) on  the unification energy
scale $E_{GUT}$.  Note that, in order to make definite predictions
on the lightest detectable supersymmetric particle, a detailed
model is needed. In particular, it is necessary to incorporate the
evolution of all masses. This has been done in Ref.~\cite{P4} and an example
of spectra is shown in Fig. \ref{FigXYZ}. Let us emphasize that the
study of the correlations between the basic
quantities,  as exemplified in Fig.~\ref{Figure3}, is interesting but should
not be mistaken as
example of prediction for the superworld. In particular, the introduction of
the quantity $M_{SUSY}$ is really misleading.
\begin{figure}[h]
\vspace{4in}
\includegraphics{figXYZ.eps}
\vspace{0.5cm}
\caption{\baselineskip=12pt
A detailed spectrum of SUSY particles showing how misleading is to
think of a unique quantity $M_{SUSY}$ in order to describe a SUSY
particle spectrum.}
\label{FigXYZ}
\end{figure}
\newpage

\section{The origin of $M_{SUSY}$ and why it should be abandoned:
masses and spectra are needed}
\label{origin}
The quantity $M_{SUSY}$ was introduced early on when one-loop RGEs for the
gauge couplings were believed to be good enough approximation and threshold
effects were altogether neglected. The running of the gauge couplings
$\alpha^{-1}_1, \alpha^{-1}_2,\alpha^{-1}_3$ was described by straight lines
(see Eqs.~\ref{dgidt}) whose slopes had to change due to the change in
the beta functions from the supersymmetric regime
Eqs.~(\ref{coefSUSY1}),(\ref{coefSUSY2}) to the non-supersymmetric regime
Eqs.~(\ref{coefnoSUSY1}),(\ref{coefnoSUSY2}). The change of slope was, for
simplicity, described by a single parameter $M_{SUSY}$, but in no case
was $M_{SUSY}$ supposed to represent a physical quantity. It is in
fact out of the question that the real spectrum of the supersymmetric
particles (an example is shown in Fig.~\ref{FigXYZ}) can be degenerate and
therefore be represented by a single mass value.

The calculations presented in Ref.~\cite{P4}, attempted to determine
the supersymmetric particle spectrum by fitting the various sparticle masses in
order to obtain the ``best possible" unification picture. These calculations
are likely the most extensive analysis of the problem with the simultaneous
evolution of seventeen dynamical variables with the radiative effects
consistently computed. It is interesting to see how the masses of the
superpartners of the various gauge  particles change as a function of the
experimentally measured quantities $\alpha_3(M_Z), \sin^2\theta_W(M_Z),
\alpha_e(M_Z)$. Examples for the \~W mass, $m_{\tilde{W}}$, and the gluino
mass, $m_{\tilde{g}}$, are shown in Figs.~\ref{FigX1} and~\ref{FigY1},
respectively. We also show in Figs.~\ref{FigT1} and~\ref{FigT2} how the squark
and slepton masses change as a function of the basic quantities quoted above
($\alpha_3(M_Z), \sin^2\theta_W(M_Z), \alpha_e(M_Z)$). Let us emphasize that
these are the only experimentally known input parameters in our system of
coupled equations~\cite{P4}. A quantity which will, hopefully, be soon
determined is the top-quark mass $m_t$. Our system of coupled equations allows
to determine the \~W mass versus $m_t$. This is shown in Fig.~\ref{FigZ}.
Notice that increasing $m_t$ corresponds to lowering the value of
$m_{\tilde{W}}$.

It is interesting to quantitatively see the effect of the three basic
quantities
($\alpha_3(M_Z), \sin^2\theta_W(M_Z),$ $\alpha_e(M_Z)$) on the variation of
$m_{\tilde{W}}$. This is summarized in Table~\ref{Table}: the dominant
effect is $\alpha_3(M_Z)$. It should be noted that these results are given
for illustrative purposes: \ie, in order to understand how radiative
effects influence the masses of the superparticles when the input
measured quantities ($\alpha_3(M_Z), \sin^2\theta_W(M_Z),
\alpha_e(M_Z)$) change within their limits of errors.

\begin{figure}[p]
\vspace{7in}
\includegraphics{figX1.ps}
\caption{\baselineskip=12pt The value of the W-ino mass,
$m_{\tilde{W}}$, vs. $\alpha_3(M_Z), \sin^2\theta(M_Z),$ and
$(m_4/m_{1/2})^2$. Each curve is the result of our iterative system of
seventeen evolution equations. The curve showing the variation of
$m_{\tilde{W}}, vs. \alpha_3(M_Z)$, or $\sin^2\theta(M_Z),$ or
$(m_4/m_{1/2})^2$ is computed keeping the other parameters fixed at
the values indicated by the arrows. For the experimentally measured
values, $\alpha_3(M_Z), \sin^2\theta(M_Z)$, the world averages have
been chosen. For the ratio of primordial masses the value
$(m_0/m_{1/2})^2 = 1$ has been taken. The top mass is kept at $m_t =$
125 GeV. Please note that our analysis is valid only above $M_Z$. The
$m_{\tilde{W}}$ values shown below this limit are for illustrative
purposes.
}
\label{FigX1}
\end{figure}

\begin{figure}[p]
\vspace{7in}
\includegraphics{figY1.ps}
\vspace{0.5cm}
\caption{\baselineskip=12pt Same as figure 9 when the gluino mass,
$m_{\tilde{g}}$, is computed.
}
\label{FigY1}
\end{figure}

\begin{figure}[p]
\vspace{7in}
\includegraphics{figT1.ps}
\vspace{0.5cm}
\caption{\baselineskip=12pt Same as figure 9 when the squark mass,
$m_{\tilde{q}}$, is computed.
}
\label{FigT1}
\end{figure}

\begin{figure}[p]
\vspace{7in}
\includegraphics{figT2.ps}
\vspace{0.5cm}
\caption{\baselineskip=12pt Same as figure 9 when the slepton masses
are computed. The left (upper curve) and right (lower curve) slepton
masses almost coincide.
}
\label{FigT2}
\end{figure}

\begin{figure}[p]
\vspace{7in}
\includegraphics{figZ.ps}
\vspace{0.5cm}
\caption{\baselineskip=12pt The value of the W-ino mass $vs. m_t$ for
given values of the other inputs, as indicated. This is the only case
($m_{\tilde{W}}$) where the dependence on $m_t$ is shown. We do not
show the result for $m_{\tilde{g}}, m_{\tilde{q}}, m_{\tilde{l}}$.
}
\label{FigZ}
\end{figure}
\begin{table}
\begin{center}
\begin{tabular}{||c||}	\hline
\\
{\Large $m_{\tilde{W}}(\alpha_3(M_Z)^{WA}-2\sigma)\over
m_{\tilde{W}}(\alpha_3(M_Z)^{WA}+2\sigma)$} {\large $\simeq 2 \times 10^4$}
\\ \\
{\em [WA values for the experimental inputs: $\sin^2\theta(M_Z)$,
$\alpha_{em}(M_Z)$]} \\ \\ \hline \\
{\Large $m_{\tilde{W}}(\sin^2\theta(M_Z)^{WA}-2\sigma)\over
m_{\tilde{W}}(\sin^2\theta(M_Z)^{WA}+2\sigma)$} {\large $= 20 \div 50$}
\\ \\
{\em [WA values for the experimental inputs: $\alpha_3(M_Z)$,
$\alpha_{em}(M_Z)$]} \\ \\ \hline \\
{\Large $m_{\tilde{W}}(1/\alpha_{em}(M_Z)^{WA}-2\sigma)\over
m_{\tilde{W}}(1/\alpha_{em}(M_Z)^{WA}+2\sigma)$} {\large $= 1.2 \div 1.3$}
\\ \\
{\em [WA values for the experimental inputs: $\sin^2\theta(M_Z)$,
$\alpha_3(M_Z)$]} \\ \\ \hline \\
Allowed range of variation for primordial SUSY breaking mass ratios:\\
\\ {\large $(m_0/m_{1/2})^2 = 10^{-2} \div 10^2$} \\ \\
{\large $(m_4/m_{1/2})^2 = 10^{-2} \div 10^2$} \\ \\ \hline
\end{tabular}
\end{center}
\caption{The variation of the W-ino mass corresponding to $\pm
2\sigma$ variation of the experimental inputs, $\alpha_3(M_Z),
\sin^2\theta(M_Z), \alpha_{em}(M_Z)$, with respect to their world
average values, and to the variation of $(m_0/m_{1/2})^2$ and
$(m_4/m_{1/2})^2$ in the indicated range. The data are derived from
our system of seventeen coupled evolution equations. The dominant
effect is clearly due to $\alpha_3(M_Z)$. Please note that our
equations are valid only above the $Z^0$ mass. Nevertheless the
results in terms of $m_{\tilde{W}}$ ratios are given, for illustrative
purposes, even when the $+2\sigma$ limit of the experimental inputs
pushes $m_{\tilde{W}}$ below $M_Z$.}
\label{Table}
\end{table}
\begin{figure}[p]
\vspace{7in}
\includegraphics{figN.eps}
\vspace{0.5cm}
\caption{\baselineskip=12pt
The predicted SUSY mass spectrum for three cases, when the measured
quantities ($\alpha_3(M_Z)$, $\sin^2\theta(M_Z)$,$\alpha_{em}(M_Z)$)
are taken at their world average values and at $\pm2\sigma$. The
ratios of the primordial parameters are kept equal to one. Note the
large range where the SUSY spectra could be on the basis of our best
experimental and theoretical knowledge. }
\label{FigN}
\end{figure}
This program has large inherent
uncertainties. As mentioned above, and first of all, because of the
fact that we cannot ignore the great uncertainties in the physics at
the GUT scale. Furthermore, even if we take the simplest approach and
assume that the physics at the GUT scale is represented by a unique
threshold, then the introduction of
experimental and theoretical uncertainties in the seventeen evolution
equations relating gauge couplings and masses,  corresponds to a variation
(within $2\sigma$) of the sparticle spectra from GeV to
PeV, as shown in Fig.\ref{FigN}. In fact, so far the only boundary
condition imposed is the ``best fit'' for the unification of the gauge
couplings. However, even if this programme would have been successful
in predicting the lightest detectable supersymmetric particle near the
Fermi scale, this would  still be far from satisfying. In fact one would
like  to know why the supersymmetric spectrum should be the way the fit would
require it to be. In  other words, the real question is: what determines the
values of the  sparticle masses? And why should these be below $\sim1\TeV$, so
that the gauge hierarchy problem  is not re-introduced?

\section{The new step forward: Supergravity}
\label{need}
In order to answer the question posed at the end of Sec.~\ref{origin},
we must abandon ``global'' supersymmetry and promote this symmetry of nature to
be ``local''. It is local supersymmetry, \ie, Supergravity which provides the
means to compute the masses of  the sparticles in a non arbitrary {\em ad hoc}
fashion. In fact, the crucial point is the breaking of local supersymmetry. In
Supergravity the breaking occurs in a  ``hidden sector" of the theory, where
``gravitational particles" (those introduced when the supersymmetry was made
local) may grow vacuum expectation values (vevs) which break supersymmetry
spontaneously in the hidden sector. These vevs are best understood as induced
dynamically by the condensation of the supersymmetric partners of the hidden
sector particles when the gauge group which describes  them becomes strongly
interacting at some large scale. The
splitting  of the particles and their partners would then be
generated, and  would be of the order of the condensation scale
($\sim10^{12-16}\GeV$). However, such huge mass splittings will be
transmitted to the ``observable" (the normal) sector of
the theory through gravitational interactions, since it is only
through these interactions that the two sectors communicate.
Gravitational interactions produce dampening in the  transmission
mechanism in such a way that the splittings in the observable sector are
usually  much more suppressed than those in the hidden sector, and
suitable  choices of hidden sectors may yield realistic low-energy
supersymmetric spectra. This picture of hidden and observable sectors becomes
completely natural in the context of superstrings, where models typically
contain both sectors and one can study explicitly the  predicted spectrum of
supersymmetric particles at low energies.

In a large number of models, the supersymmetric particle masses at the
unification scale are also ``unified". This situation is called {\em universal
soft-supersymmetry-breaking}, and the masses of all scalar partners (\eg,
squarks and sleptons) take the common value of $m_0$, the gaugino (the
partners of the gauge bosons) masses are given by $m_{1/2}$, and  there is a
third parameter ($A$) which basically parametrizes the mixing of stop-squark
mass eigenstates at low energies. The breaking of the electroweak symmetry is
obtained dynamically in the context of these models, through the so-called {\em
radiative electroweak symmetry breaking mechanism}, which involves the
top-quark mass in a fundamental way \cite{EWx,LN}.  After all these  well
motivated theoretical ingredients have been incorporated, the  models depend on
only four parameters: $m_{1/2}, m_0, A$,  and the ratio of the two Higgs vacuum
 expectations values ($\tan\beta$), plus the top-quark mass ($m_t$).

In generic Supergravity models the five-dimensional parameter space is
constrained by phenomenological requirements, such as sparticle and
Higgs-boson masses not in conflict with present experimental lower bounds,
a sufficiently long proton lifetime, a sufficiently old Universe (a
cosmological constraint on the amount of dark matter in the Universe
today),  various indirect constraints from well measured rare processes, etc.
Contemporary detailed analyses of supergravity models along these lines have
been performed shortly before the LEP era \cite{GRZ+EZ}, within the last
two years \cite{KLNPY,RR,DN,Japs,ANpd,aspects}, and also very recently
\cite{recent}. In $SU(5)\times U(1)$ Supergravity
further string-inspired theoretical constraints can be imposed which give $m_0$
and $A$ as functions of $m_{1/2}$, and thus reduce the dimension of the
parameter space  down to just two (plus the top quark mass).

It should be pointed out that a rather interesting situation  occurs
in the  so-called {\em no-scale} framework \cite{Lahanas,EKNI+II}, where all
the scales in the theory are  obtained - through radiative corrections - from
just one basic scale (\ie, the unification scale or the  Planck scale). These
models have the  unparalleled virtue of a vanishing cosmological constant at
the tree-level  {\em even after supersymmetry breaking}, and in their
unified  versions predict that, at $E_{SU}$, the universal scalar masses and
trilinear  couplings vanish (\ie, $m_0(E_{SU})=A(E_{SU})=0$) thus the universal
gaugino mass $(m_{1/2}$) is  the only seed of
supersymmetry breaking. Moreover, this unique mass can be determined in
principle by minimizing the vacuum energy at the electroweak scale. The
generic result is $m_{1/2}\sim M_Z$ \cite{Lahanas,EKNI+II}, in
agreement with  theoretical prejudices (\ie, ``naturalness" $\equiv$ the
radiative corrections of a physical quantity cannot exceed the value
of the quantity itself). Finally, it should not be forgotten that {\em
no-scale} Supergravity is the infrared solution of  superstring theory
\cite{Witten} and therefore the {\em no-scale} scenario appears very natural in
$SU(5)\times U(1)$ Supergravity.

We now discuss the two simplest Supergravity models. One as representative of
Field Theory, the other of String Theory. In fact $SU(5)$ Supergravity is not
easily derivable from Superstring theory: no one has succeeded so far. The
other is easily derivable from Superstring theory -- in fact it has already
been done \cite{revamp,LNY} -- and this is $SU(5)\times U(1)$ Supergravity.

\section{The SU(5) Supergravity Model}
\label{SU5}
The $SU(5)$ supergravity model \cite{Dickreview} needs to be specified
clearly in order to avoid the common misconception that it is simply
the so-called MSSM (Minimal Supersymmetric extension of the Standard
Model) with the low-energy gauge couplings meeting at very high
energies. Two  of its elements are particularly important: (i) it is a
supergravity model \cite{CAN} and as such the soft supersymmetry
breaking  masses which allow unification are in principle calculable
and are  assumed to be parametrized in terms of $m_{1/2},m_0,A$; and
(ii)  there exist dimension-five proton decay operators \cite{WSY},
which are  much larger than the usual dimension-six operators, and
require  either a tuning of the  supersymmetry breaking parameters or
a large  Higgs triplet mass scale, to obtain a sufficiently long
proton  lifetime \cite{ENR,EMN,ANoldpd,MATS,ANpd,HMY,LNP,LNPZ}.

The $SU(5)$ symmetry is broken down to $SU(3)\times SU(2)\times U(1)$ via a
vev of the neutral component of the adjoint \r{24} of Higgs. The low-energy
pair of Higgs doublets are contained in the \r{5},\rb{5} Higgs representations.
Of the various proposals to split the proton-decay-mediating Higgs triplets
from the light Higgs doublets, perhaps the most appealing one is the so-called
``missing partner mechanism" \cite{MPM}, whereby a \r{75} of Higgs breaks the
gauge symmetry and the \r{5},\rb{5} pentaplets are coupled to \r{50},\rb{50}
representations (${\bf50}\cdot{\bf75}\cdot{\bf5}$,
$\ov{\bf50}\cdot{\bf75}\cdot\bar{\bf5}$). The doublets remain massless, while
the triplets acquire $\sim M_U$ masses. (Note: $M_U$ and $E_{GUT}$ are used
interchangeably to represent the unification mass scale.) We should remark that
this non-minimal symmetry breaking mechanism is {\em not} the one that is
usually considered in studies of high-energy threshold effects in gauge
coupling unification, where one usually assumes that it is the \r{24} which
effects the breaking.
\subsection{Gauge and Yukawa coupling unification}
This problem can be tackled at several levels of sophistication, which entail
an increasing number of additional assumptions. The most elementary approach
consists of running the one-loop supersymmetric gauge coupling RGEs starting
with the precisely measured values of $\alpha_e,\alpha_3,\sin^2\theta_w$ at
the scale $M_Z$ and discovering that the three gauge couplings meet at the
scale $M_U\sim10^{16}\GeV$ \cite{CostaAmaldi}, neglecting that above
$M_U$ they diverge again. More interesting from
the theoretical standpoint is to assume that unification must occur,
as is the case in the $SU(5)$ supergravity model, and use this
constraint to  predict the low-energy value of $\sin^2\theta_w$ and
$\alpha_e$ in terms of $\alpha_3$~\cite{ACPZII}. The next level of
sophistication
consists  of increasing the accuracy of the RGEs to two-loop level and
parametrize the supersymmetric threshold by a single mass parameter
between  $\sim M_Z$ and $\sim{\rm few}\TeV$
\cite{EKNI,LL,AdBF,Arason,ACPZI,ACPZII}. More realistically, one
specifies the  whole light supersymmetric  spectrum in detail
\cite{EKNII,EKNIII,EKNIV,RR,HMY,EGM,P4,LP,CPW},
as well as some subtle and important effects such as the evolution of
the  gaugino masses (EGM)~\cite{EGM,EKNIV}, and the effect of the
Yukawa  couplings on the two-loop  gauge coupling RGEs \cite{BBO}. A
final step of sophistication  attempts to model the transition from
the  $SU(3)\times SU(2)\times U(1)$ theory into the $SU(5)$ theory by means of
high-energy  threshold effects which depend on the masses of the
various GUT fields
\cite{EKNIII,EKNIV,BH,HMY,ACPZII,ACZ,LP} as well as on
coefficients of possible non-renormalizable operators \cite{LP,HS}. This last
step does away completely with the concept of a single ``unification mass". In
fact, until this last step is actually accounted for somehow, one is not
dealing with a true unified theory since otherwise the gauge couplings would
diverge again past the unification scale, \ie, ``physics is not euclidean
geometry".

It is interesting to note that the original hope that precise knowledge of
the low energy gauge couplings would constrain the scale of the low-energy
supersymmetric particles, did not bear fruit \cite{BH,EGM,EKNIV}, mainly
because of the largely unknown GUT threshold effects. More precisely, the
supersymmetric particle masses can lie anywhere up to $\sim{\rm few}\TeV$
provided the parameters of the GUT theory are adjusted accordingly.

Another consequence of the $SU(5)$ symmetry is the relation $\lambda_b(M_U)=
\lambda_\tau(M_U)$ which when renormalized down to low energies gives a ratio
$m_b/m_\tau$ in fairly good agreement with experiment \cite{BEGN}.  This
problem can also be tackled with improving degree of sophistication
\cite{EKNI,Arason,EKNIII,GHS,YU,DHR,BBO,Naculich,LPII} and even postulating
some high-energy threshold effects \cite{BBO,LPII}.
In practice, the
$\lambda_b(M_U)=\lambda_\tau(M_U)$ constraint entails a relationship between
$m_t$ and $\tan\beta$, \ie, $\tan\beta=\tan\beta(m_t,m_b,\alpha_3)$, as
follows: (i) the values of $m_b$ and $m_\tau$, together with $\tan\beta$
determine the low-energy values of $\lambda_b$ and $\lambda_\tau$; (ii) the
input value of $m_t$ determines the low-energy value of $\lambda_t$; (iii)
running these three Yukawa couplings up to the unification scale one discovers
the above relation between $\tan\beta$ and $m_t$ if the Yukawa unification
constraint is satisfied. In actuality, the dependence on $m_b$ and $\alpha_3$
is quite important. We note that for arbitrary choices of $m_t$ and
$\tan\beta$, one obtains values of $m_b$ typically close to or above $5\GeV$,
whereas popular belief would like to see values below $4.5\GeV$. Strict
adherence to this prediction for $m_b$ requires that one be in a rather
constrained region of the $(m_t,\tan\beta)$ plane, where $\tan\beta\sim1$
or $\gsim40$ \cite{EKNIV,YU,BBO,LPII}, or that $m_t$ be large (above
$180\GeV$). We do not impose this stringent constraint on the parameter space,
hoping that further contributions to the quark masses (as required in $SU(5)$
GUTs to fit the lighter generations also \cite{EG}) will relax it somehow.

As noted above, the parameter space of this model can be described in terms
of five parameters: $m_t,\tan\beta,m_{1/2},m_0,A$. In Ref. \cite{LNP} we
performed an exploration of the following hypercube of the parameter space:
$\mu>0,\mu<0$, $\tan\beta=2-10\,(2)$, $m_t=100-160\,(5)$,
$\xi_0\equiv m_0/m_{1/2}=0-10\,(1)$,
$\xi_A\equiv m_A/m_{1/2}=-\xi_0,0,+\xi_0$,
and $m_{1/2}=50-300\,(6)$, where the numbers in
parenthesis represent the size of the step taken in that particular direction.
(Points outside these ranges have little (a posteriori) likelihood of being
acceptable.) Of these $92,235\times2=184,470$ points, $\approx25\%$ passed
all the standard constraints, \ie, radiative electroweak symmetry breaking
and all low-energy phenomenology as described in Ref. \cite{aspects}. The
most important constraint on this parameter space is proton decay, as discussed
below. First we discuss some aspects of the gauge coupling unification
calculation.

As a first step we used one-loop gauge coupling RGEs and a common
supersymmetric threshold at $M_Z$, to determine $M_U,\alpha_U$, and
$\sin^2\theta_w$, once $\alpha_3(M_Z)=0.113,0.120$ and
$\alpha^{-1}_e(M_Z)=127.9$ were given. In Ref. \cite{LNPZ} we refined our
study including several important features: (i) recalculation of $M_U$ using
two-loop gauge coupling RGEs including light supersymmetric thresholds, (ii)
exploration of values of $\alpha_3$ throughout its $\pm1\sigma$ allowed range,
and (iii) exploration of low values of $\tan\beta\,(<2)$ (which maximize the
proton lifetime). We used the analytical approximations to the solution of the
two-loop gauge coupling RGEs in Ref. \cite{EGM} to obtain
$M_U,\alpha_U,\sin^2\theta_w$. The supersymmetric threshold was treated in
great detail \cite{EGM} with all the sparticle masses obtained from our
procedure \cite{LNP}. Since the sparticle masses vary as one explores the
parameter space, one obtains {\it ranges} for the calculated values. In
Table~\ref{Table5} we show the one-loop value for $M_U$ ($M_U^{(0)}$), the
two-loop plus supersymmetric threshold corrected unification mass range
($M_U^{(1)}$) [as expected \cite{EGM} $M_U$ is reduced by both effects], the
ratio of the two, and the calculated range of $\sin^2\theta_w$.\footnote{ We
should note that these ranges are obtained after all constraints discussed
below have been satisfied, the proton decay being the most important one.}
Note that for $\alpha_3=0.118$ (and lower), $\sin^2\theta_w$ is outside the
experimental $\pm1\sigma$ range ($\sin^2\theta_w=0.2324\pm0.0006$ \cite{LP}),
whereas $\alpha_3=0.126$ gives quite acceptable values.

\begin{table}[t]
\hrule
\caption{
The value of the one-loop unification mass $M_U^{(0)}$, the two-loop and
supersymmetric threshold corrected unification mass range $M_U^{(1)}$, the
ratio of the two, and the range of the calculated $\sin^2\theta_w$, for the
indicated values of $\alpha_3$ (the superscript $+\,(-)$ denotes $\mu>0\,(<0)$)
and $\alpha^{-1}_e=127.9$. The $\sin^2\theta_w$ values should be compared with
the current experimental $\pm1\sigma$ range $\sin^2\theta_w=0.2324\pm0.0006$
[2]. Lower values of $\alpha_3$ drive $\sin^2\theta_w$ to values even higher
than for $\alpha_3=0.118$. All masses in units of $10^{16}\GeV$.}
\label{Table5}
\begin{center}
\begin{tabular}{|c||c|c||c|c|}\hline
&$\alpha_3=0.126^+$&$\alpha_3=0.126^-$&$\alpha_3=0.118^+$&$\alpha_3=0.118^-$
\\ \hline
$M^{(0)}_U$&$3.33$&$3.33$&$2.12$&$2.12$\\
$M^{(1)}_U$&$1.60-2.13$&$1.60-2.05$&$1.02-1.35$&$1.02-1.30$\\
$M_U^{(1)}/M_U^{(0)}$&$0.48-0.64$&$0.48-0.61$&$0.48-0.64$&$0.48-0.61$\\
$\sin^2\theta_w$&$0.2315-0.2332$&$0.2313-0.2326$&$0.2335-0.2351$&
$0.2332-0.2345$\\ \hline
\end{tabular}
\end{center}
\hrule
\end{table}

We do not specify the details of the GUT thresholds and in practice take two
of the GUT mass parameters (the masses of the $X,Y$ gauge bosons $M_V$, and
the mass of the adjoint Higgs multiplet $M_\Sigma$) to be degenerate with
$M_U$. Since below we allow $M_H<3M_U$, Table~\ref{Table5} indicates that in
our calculations $M_H<6.4\times10^{16}\GeV$. In Ref. \cite{HMY} it is argued
that a more proper upper bound is $M_H<2M_V$, but $M_V$ cannot be calculated
directly, only $(M^2_V M_\Sigma)^{1/3}<3.3\times10^{16}\GeV$ is known from
low-energy data \cite{HMY}. If we take $M_\Sigma=M_V$, this would give
$M_H<2M_V<6.6\times10^{16}\GeV$, which agrees with our present requirement.
Below we comment on the case $M_\Sigma<M_V$.
\subsection{Proton decay}
\label{su5pdecay}
In the $SU(5)$ supergravity model only the dimension-five--mediated proton
decay operators are constraining. In calculating the proton lifetime we
consider the typically dominant decay modes $p\to \bar\nu_{\mu,\tau} K^+$ and
neglect all other possible modes. Schematically the lifetime is given
by\footnote{Throughout our calculations we have used the explicit proton
decay formulas in Ref. \cite{ANoldpd}.}
\beq
\tau_p\equiv\tau(p\to\bar\nu_{\mu,\tau}K^+)
\sim\left| M_H \sin2\beta {1\over f}{1\over 1+y^{tK}}\right|^2.
\eeq
Here $M_H$ is the mass of the exchanged GUT Higgs triplet which on perturbative
grounds is assumed to be bounded above by $M_H<3M_U$ \footnote{This relation
assumes implicitly that all the components of the \r{24} superfield are nearly
degenerate in mass \cite{HMY}.} \cite{EMN,ANpd,HMY};
$\sin2\beta=2\tan\beta/(1+\tan^2\beta)$, thus $\tau_p$ ``likes" small
$\tan\beta$ (we find that only $\tan\beta\lsim6$ is allowed); $y^{tK}$
represents the calculable ratio of the third- to the second-generation
contributions to the dressing one-loop diagrams. An unkown phase appears in
this ratio (which has generally $|y^{tK}|\ll1$) and we always consider the
weakest possible case of  destructive interference. Finally $f$ represents the
sparticle-mass--dependent dressing one-loop function which decreases
asymptotically with large sparticle masses.

In Fig.~\ref{Figure11} (top row) we show a scatter plot of $(\tau_p,m_{\tilde
g})$. The various `branches' correspond to fixed values of $\xi_0$. Note that
for $\xi_0<3$, $\tau_p<\tau^{exp}_p=1\times10^{32}\y$ (at 90\% C.L.
\cite{PDG}). Also, for a given value of $\xi_0$, there is a corresponding
allowed interval in $m_{\tilde g}$. The lower end of this interval is
determined by the fact that $\tau_p\propto1/f^2$, and $f\approx
m_{\chi^+_1}/m^2_{\tilde q}\propto 1/m_{\tilde g}(c+\xi^2_0)$, in the
proton-decay--favored limit of $\mu\gg M_W$; thus
$m_{\tilde g}(c+\xi^2_0)>{\rm constant}$. The upper end of the interval follows
from the requirement $m_{\tilde q}(\propto m_{\tilde
g}\sqrt{c+\xi^2_0})<1\TeV$. Statistically speaking, the proton decay cut is
quite severe, allowing only about $\sim1/10$ of the points which passed all the
standard constraints, independently of the sign of $\mu$.

Note that if we take $M_H=M_U$ (instead of $M_H=3M_U$), then
$\tau_p\to{1\over9}\tau_p$ and all points in Fig.~\ref{Figure11} would become
excluded. To obtain a rigorous lower bound on $M_H$, we would need to explore
the lowest possible allowed values of $\tan\beta$ (in Fig.~\ref{Figure11},
$\tan\beta\ge2$). Roughly, since the dominant $\tan\beta$ dependence of
$\tau_p$ is through the explicit $\sin2\beta$ factor, the upper bound
$\tau_p\lsim8\times10^{32}\y$ for $\tan\beta=2$, would become
$\tau_p\lsim1\times10^{33}\y$ for $\tan\beta=1$. Therefore, the current
experimental lower bound on $\tau_p$ would imply $M_H\gsim M_U$. Note also that
SuperKamiokande ($\tau^{exp}_p\approx2\times10^{33}\y$) should be able to probe
the whole allowed range of $\tau_p$ values.

The actual value of $\alpha_3(M_Z)$ used in the calculations ($\alpha_3=0.120$
in Fig.~\ref{Figure11}) has a non-negligible effect on some of the final
results, mostly due to its effect on the value of $M_H=3M_U$: larger values of
$\alpha_3$ increase $M_U$ and therefore $\tau_p$, and thus open up the
parameter space, and viceversa. For example, for $\alpha_3=0.113\,(0.120)$ we
get $m_{\tilde g}\lsim550\,(800)\GeV$, $\xi_0\ge5\,(3)$, and
$\tau_p\lsim4\,(8)\times10^{32}\y$.

\begin{figure}[p]
\vspace{4.9in}
\includegraphics{proc_summer11.ps}
\vspace{-0.3in}
\caption{\baselineskip=12pt
Scatter plot of the proton lifetime
$\tau_p\equiv\tau(p\to\bar\nu_{\mu,\tau}K^+)$ versus the gluino mass for the
hypercube of the parameter space explored. The unification mass is calculated
in one-loop approximation assuming a common supersymmetric threshold at $M_Z$,
and $M_H=3M_U$ is assumed. The current experimental lower bound  is
$\tau^{exp}_p=1\times10^{32}\y$. The various `branches' correspond to fixed
values of $\xi_0$ as indicated (the labelling applies to all four windows). The
bottom row includes the cosmological constraint. The upper bound on $m_{\tilde
g}$ follows from the requirement $m_{\tilde q}<1\TeV$.}
\label{Figure11}
\end{figure}

More details on the matter instability problem are given in Sec.~\ref{pd},
where we present detailed calculations for the underground labs.

\subsection{Neutralino relic density}
\label{su5dm}
The study of the relic density of neutralinos requires the knowledge of the
total annihilation amplitude $\chi\chi\to all$. The latter depends on the model
parameters to determine all masses and couplings. Previously
\cite{LNYdmI,KLNPYdm} we have advocated the study of this problem in the
context of supergravity models with radiative electroweak symmetry breaking,
since then only a few parameters (five or less) are needed to specify the model
completely. In particular, one can explore the whole parameter space and draw
conclusions about a complete class of models. The ensuing relationships among
the various masses and couplings have been found to yield results which depart
from the conventional minimal supersymmetric standard model (MSSM) lore, where
no such relations exist. In the $SU(5)$ supergravity model we have just shown
that its five-dimensional parameter space is strongly constrained by the proton
lifetime. It was first noticed in Ref.~\cite{troubles} that the neutralino
relic density for the proton-

decay allowed points in parameter space is large, \ie, $\Omega_\chi h^2_0\gg1$,
and therefore in conflict with current
cosmological expectations: requiring that the Universe be older than the oldest
known stars implies $\Omega_0 h^2_0\le1$ \cite{KT}.

In Refs.~\cite{troubles,LNP,LNPZ} the neutralino relic density has been
computed following the methods of Refs.~\cite{LNYdmI,KLNPYdm}. In
Fig.~\ref{Figure11} (bottom row) we show the effect of the cosmological
constraint on the parameter space allowed by proton decay. Only $\sim1/6$ of
the points satisfy $\Omega_\chi h^2_0\le1$. This result is not unexpected since
proton decay is  suppressed by heavy sparticle masses, whereas $\Omega_\chi
h^2_0$ is enhanced. Therefore, a delicate balance needs to be attained to
satisfy both constraints simultaneously. Note that the subset of cosmologically
allowed points does not change the range of possible $\tau_p$ values, although
it depletes the constant-$\xi_0$ `branches'.

The effect of the cosmological constraint is perhaps more manifest when one
considers the correlation between $m_h$ and $m_{\chi^\pm_1}$ after imposing
the (\eg, {\em weaker}) proton decay constraint, but with and without imposing
the cosmological constraint. This contrast is shown in Fig.~\ref{Figure14}.

\begin{figure}[t]
\vspace{5in}
\includegraphics{proc_summer14.ps}
\vspace{-0.3in}
\caption{\baselineskip=12pt
The allowed region in parameter space which satisfies the {\em weaker} proton
decay constraint, before and after the imposition of the cosmological
constraint. Note that when the cosmological constraint is imposed (bottom
row), an interesting correlation between the two particle masses arises.}
\label{Figure14}
\end{figure}

The main conclusion is that the relic density can be small only near the
$h$- and $Z$-pole resonances, \ie, for $m_\chi\approx{1\over2}m_{h,Z}$
\cite{LNP,ANcosm}, since in this case the annihilation cross section is
enhanced. It is important to note that in this type of calculations the thermal
average of the annihilation cross section is usually computed using an
expansion around threshold (\ie, $\sqrt{s}=2m_\chi$) \cite{SWO}. In Ref.
\cite{GG} however, it has been pointed out that the resulting thermal
average can be quite inaccurate near poles and thresholds of the annihilation
cross section, which is precisely the case for the points of interest in the
$SU(5)$ model. In Ref.~\cite{ANcosm,poles} the relic density calculation has
been redone following the more accurate methods of Ref.~\cite{GG}. The result
is that the poles are broader and shallower, and thus the cosmological
constraint is weakened with respect to the standard (using the expansion)
procedure of performing the thermal average.
However, qualitatively the cosmologically allowed region of parameter space
is not changed. This result is shown in Fig.~\ref{Figure13}, where the
points in parameter space allowed by the stricter proton decay constraint and
cosmology are shown in the $(m_{\chi^\pm_1},m_h)$ plane. Note that the ``exact"
calculation of the relic density allows more points in parameter space,
therefore the plot is more populated in this case.

\begin{figure}[p]
\vspace{4.0in}
\includegraphics{proc_summer13a.ps}
\vspace{2.5in}
\includegraphics{proc_summer13b.ps}
\vspace{-1in}
\caption{\baselineskip=12pt
The points in parameter space of the $SU(5)$ supergravity model which
satisfy the stricter proton decay constraint and the cosmological constraint
with the relic density computed in approximate and accurate way. Note that
since the ``exact" calculation of the relic density allows more points in
parameter space,  the plot is more populated in this case.}
\label{Figure13}
\end{figure}

\subsection{Mass ranges and relations}
Since the proton decay constraint generally requires $|\mu|\gg M_W$ (and to a
somewhat lesser extent also $|\mu|\gg M_2$), the lightest chargino will have
mass $m_{\chi^+_1}\approx M_2\approx0.3m_{\tilde g}$, whereas the two lightest
neutralinos will have masses $m_\chi\approx M_1\approx{1\over2}M_2$ and
$m_{\chi^0_2}\approx M_2$ \cite{ANpd,ANc}. Inclusion of the cosmological
constraint does not affect significantly the range of sparticle masses in
Sec.~\ref{su5pdecay}. The value of $\alpha_3$ does not affect these mass
relations either, although the particle mass ranges do change
\beq
m_\chi<85\,(115)\GeV,\qquad m_{\chi^0_2,\chi^+_1}<165\,(225)\GeV,
\qquad{\rm for}\ \alpha_3=0.113\,(0.120).
\eeq
The reason is simple: higher values of $\alpha_3$ increase $M_U$ and therefore
$M_H\,(=3M_U)$, which in turn weakens the proton decay constraint.
We also find that the one-loop corrected lightest Higgs boson mass ($m_h$) is
bounded above by
\beq
m_h\lsim110\,(100)\GeV,
\eeq
independently of the sign of $\mu$, the value of $\alpha_3$, or the
cosmological constraint; the stronger bound holds when the stricter proton
decay constraint is enforced. In Fig.~\ref{Figure13} we have shown $m_h$ versus
$m_{\chi^+_1}$ for $\tan\beta=1.5,1.75,2$; for the maximum allowed $\tan\beta$
value ($\approx3.5$), $m_h\lsim100\GeV$. Note that for $\mu>0$,
$m_h\approx50\GeV$, and $m_{\chi^\pm_1}\gsim100\GeV$, there is a sparsely
populated area with highly fine-tuned points in parameter space
($m_t\approx100\GeV$, $\tan\beta\approx1.5$, $\xi_A\equiv A/m_{1/2}
\approx\xi_0\approx6$). This figure shows an experimentally interesting
correlation when the cosmological constraints are imposed,
\beq
m_h\gsim72\GeV \Rightarrow m_{\chi^+_1}\lsim100\GeV.\label{mV}
\eeq
The bands of points towards low values of $m_h$ represent the discrete choices
of $\tan\beta=1.5,1.75$. The voids between these bands are to be understood as
filled by points with $1.5\lsim\tan\beta\lsim1.75$. For
$m_{\chi^\pm_1}>106\,(92)\GeV$ (for $\mu>0\,(\mu<0)$), we
obtain $m_h\lsim50\,(56)\GeV$ and Higgs detection at LEP should be immediate.
The correlations among the lightest chargino and neutralino masses imply
analogous results for $(m_h,m_{\chi^0_2})$ and $(m_h,m_\chi)$,
\beq
m_h\gsim80\GeV \Rightarrow m_{\chi^0_2}\lsim90\,(110)\GeV,
\quad m_\chi\lsim48\,(60)\GeV,\label{mVI}
\eeq
for $\alpha_3=0.113\,(0.120)$.  These correlations can be understood in the
following way: since we find that $m_A\gg M_Z$, then
$m_h\approx|\cos2\beta|M_Z+({\rm rad.\ corr.})$. In the
situation we consider here, we have determined that all of the allowed points
for $m_{\tilde g}>400\;\GeV$ correspond to $\tan\beta=2$. This implies that
the tree-level contribution to $m_h$ is $\approx55\GeV$. We also find that the
cosmology cut restricts  $m_t<130(140)\;\GeV$ for $\mu>0(\mu<0)$ in this range
of $m_{\tilde g}$. Therefore, the radiative correction contribution to $m^2_h$
($\propto m^4_t$) will be modest in this range of $m_{\tilde g}$. This explains
the depletion of points for $m_h\gsim 80\;\GeV$ in Fig.~\ref{Figure13} and
leads to the mass relationships in Eqs.~(\ref{mV},\ref{mVI}).

In this model the only light particles are the lightest Higgs boson
($m_h\lsim100\GeV$), the two lightest neutralinos ($m_{\chi^0_1}\approx
{1\over2}m_{\chi^0_2}\lsim75\GeV$), and the lightest chargino ($m_{\chi^\pm_1}
\approx m_{\chi^0_2}\lsim150\GeV$). The gluino and the lightest stop can be
light ($m_{\tilde g}\approx 160-460\GeV$, $m_{\tilde t_1}\approx170-825\GeV$),
but for most of the parameter space are not within the reach of Fermilab.

In Ref. \cite{LNPWZh} it has been shown that the actual LEP lower bound on the
lightest Higgs boson mass is improved in the class of supergravity models
with radiative electroweak symmetry breaking which we consider here, one gets
$m_h\gsim60\GeV$. In Sec.~\ref{LEPI} below we discuss the details of this
procedure. For now it suffices to note that the improved bound $m_h\gsim60\GeV$
mostly restricts low values of $\tan\beta$ and therefore the $SU(5)$
supergravity model where $\tan\beta\lsim3.5$ \cite{LNPZ}. Above we obtained
upper bounds on the light particle masses in this model ($\tilde
g,h,\chi^0_{1,2},\chi^\pm_1$) for $m_h>43\GeV$. In particular, it was found
that $m_{\chi^\pm_1}\gsim100\GeV$ was only possible for $m_h\lsim50\GeV$. The
improved bound on $m_h$ immediately implies the following considerably stronger
upper bounds
\begin{eqnarray}
m_{\chi^0_1}&\lsim&52(50)\GeV, \\
m_{\chi^0_2}&\lsim&103(94)\GeV,\\
m_{\chi^\pm_1}&\lsim&104(92)\GeV,\\
m_{\tilde g}&\lsim&320\,(405)\GeV,
\end{eqnarray}
for $\mu>0$ ($\mu<0$).

A related consequence is that the mass relation $m_{\chi^0_2}>m_{\chi^0_1}+m_h$
is not satisfied for any of the remaining points in parameter space
and therefore the $\chi^0_2\to\chi^0_1 h$ decay mode is not kinematically
allowed. Points where such mode was previously allowed  led to a vanishing
trilepton signal in the reaction $p\bar p\to\chi^\pm_1\chi^0_2$ at Fermilab
(thus the name `spoiler mode') \cite{LNWZ}. The improved situation now implies
at least one event per $100\ipb$ for all remaining points in parameter space
(see Sec.~\ref{tevatron}).

\section{SU(5)xU(1) Supergravity}
\label{SU5xU1}
\subsection{General features}
The model we consider \cite{LNZI} is a generalization of that presented in
Ref. \cite{revitalized}, and contains the following $SU(5)\times U(1)$ fields:
\begin{enumerate}
\item three generations of quark and lepton fields $F_i,\bar f_i,l^c_i,\,
i=1,2,3$;
\item two pairs of Higgs \r{10},\rb{10} representations $H_i,\bar H_i,\,
i=1,2$;
\item one pair of ``electroweak" Higgs \r{5},\rb{5} representations
$h,\bar h$;
\item three singlet fields $\phi_{1,2,3}$.
\end{enumerate}
\noindent Under $SU(3)\times SU(2)$ the various $SU(5)\times U(1)$ fields
decompose as follows:
\begin{eqnarray}
F_i&=&\{Q_i,d^c_i,\nu^c_i\},\quad \bar f_i=\{L_i,u^c_i\},
\quad l^c_i=e^c_i,\\
H_i&=&\{Q_{H_i},d^c_{H_i},\nu^c_{H_i}\},\quad
\bar H_i=\{Q_{\bar H_i},d^c_{\bar H_i},\nu^c_{\bar H_i}\},\\
h&=&\{H,D\},\quad \bar h=\{\bar H,\bar D\}.
\end{eqnarray}
The most general effective\footnote{To be understood in the string context as
arising from cubic and higher order terms \cite{KLN,decisive}.}
superpotential consistent with $SU(5)\times U(1)$ symmetry is given by
\begin{eqnarray}
W&=&\lambda^{ij}_1 F_iF_jh+\lambda^{ij}_2 F_i\bar f_j \bar h
+\lambda^{ij}_3 \bar f_il^c_j h +\mu h\bar h
+\lambda^{ij}_4 H_iH_jh+\lambda^{ij}_5\bar H_i\bar H_j\bar h\nonumber\\
&+&\lambda^{ij}_{1'}H_iF_jh+\lambda^{ij}_{2'}H_i\bar f_j\bar h
+\lambda^{ijk}_6 F_i\bar H_j\phi_k+w^{ij}H_i\bar H_j+\mu^{ij}\phi_i\phi_j.
\label{W}
\end{eqnarray}
Symmetry breaking is effected by non-zero vevs $\vev{\nu^c_{H_i}}=V_i$,
$\vev{\nu^c_{\bar H_i}}=\bar V_i$, such that
$V^2_1+V^2_2=\bar V^2_1+\bar V^2_2$.
\subsubsection{Higgs doublet and triplet mass matrices}
The Higgs doublet mass matrix receives contributions from
$\mu h\bar h\to \mu H\bar H$ and $\lambda^{ij}_{2'}H_i\bar f_j\bar h$
$\to\lambda^{ij}_{2'}V_i L_j\bar H$. The resulting matrix is
\beq
{\cal M}_2=\bordermatrix{&\bar H\cr H&\mu\cr
L_1&\lambda^{i1}_{2'}V_i\cr
L_2&\lambda^{i2}_{2'}V_i\cr
L_3&\lambda^{i3}_{2'}V_i\cr}.
\eeq
To avoid fine-tunings of the $\lambda^{ij}_{2'}$ couplings we must demand
$\lambda^{ij}_{2'}\equiv0$, so that $\bar H$ remains light.

The Higgs triplet matrix receives several contributions:
$\mu h\bar h\to\mu D\bar D$;
$\lambda^{ij}_{1'}H_iF_jh$ $\to\lambda^{ij}_{1'}V_id^c_jD$;
$\lambda^{ij}_4 H_iH_jh\to\lambda^{ij}_4V_i d^c_{H_j}D$;
$\lambda^{ij}_5 \bar H_i\bar H_j\bar h
\to\lambda^{ij}_5\bar V_i d^c_{\bar H_j}\bar D$;
$w^{ij}d^c_{H_i}d^c_{\bar H_j}$. The resulting matrix is\footnote{The zero
entries in ${\cal M}_3$ result from the assumption $\vev{\phi_k}=0$ in
$\lambda_6^{ijk}F_i\bar H_j\phi_k$.}
\beq
{\cal M}_3=\bordermatrix{
&\bar D&d^c_{H_1}&d^c_{H_2}&d^c_1&d^c_2&d^c_3\cr
D&\mu&\lambda^{i1}_4V_i&\lambda^{i2}_4V_i&\lambda^{i1}_{1'}V_i
&\lambda^{i2}_{1'}V_i&\lambda^{i3}_{1'}V_i\cr
d^c_{\bar H_1}&\lambda^{i1}_5\bar V_i&w_{11}&w_{12}&0&0&0\cr
d^c_{\bar H_2}&\lambda^{i2}_5\bar V_i&w_{21}&w_{22}&0&0&0\cr}.\label{IIb}
\eeq
Clearly three linear combinations of $\{\bar D,d^c_{H_{1,2}},d^c_{1,2,3}\}$
will remain light. In fact, such a general situation will induce a mixing
in the down-type Yukawa matrix $\lambda^{ij}_1 F_iF_jh\to\lambda^{ij}_1Q_i
d^c_jH$, since the $d^c_j$ will need to be re-expressed in terms of these
mixed light eigenstates.\footnote{Note that this mixing is on top of any
structure that $\lambda^{ij}_1$ may have, and is the only source of mixing in
the typical string model-building case of a diagonal $\lambda_2$ matrix.} This
low-energy quark-mixing mechanism is an explicit realization of the general
extra-vector-abeyance (EVA) mechanism of Ref. \cite{EVA}. As a first
approximation though, in what follows we will set $\lambda^{ij}_{1'}=0$, so
that the light eigenstates are $d^c_{1,2,3}$.
\subsubsection{Neutrino see-saw matrix}
The see-saw neutrino matrix receives contributions from:
$\lambda^{ij}_2F_i\bar f_j\bar h\to m^{ij}_u\nu^c_i\nu_j$;\\
$\lambda^{ijk}_6 F_i\bar H_j\phi_k\to\lambda^{ijk}_6\bar V_j\nu^c_i\phi_k$;
$\mu^{ij}\phi_i\phi_j$. The resulting matrix is\footnote{We neglect a possible
higher-order contribution which could produce a non-vanishing $\nu^c_i\nu^c_j$
entry \cite{chorus}.}
\beq
{\cal M}_\nu=\bordermatrix{&\nu_j&\nu^c_j&\phi_j\cr
\nu_i&0&m^{ji}_u&0\cr
\nu^c_i&m^{ij}_u&0&\lambda^{ikj}_6\bar V_k\cr
\phi_i&0&\lambda^{jki}_6\bar V_k&\mu^{ij}\cr}.
\eeq
\subsubsection{Numerical scenario}
To simplify the discussion we will assume, besides\footnote{In Ref.
\cite{revitalized} the discrete symmetry $H_1\to -H_1$ was imposed so that
these couplings automatically vanish when $H_2,\bar H_2$ are not present. This
symmetry (generalized to $H_i\to -H_i$) is not needed here since it would
imply $w^{ij}\equiv0$, which is shown below to be disastrous for gauge
coupling unification.} $\lambda^{ij}_{1'}=\lambda^{ij}_{2'}\equiv0$, that
\begin{eqnarray}
\lambda^{ij}_4&=&\delta^{ij}\lambda^{(i)}_4,\quad
\lambda^{ij}_5=\delta^{ij}\lambda^{(i)}_5,\quad
\lambda^{ijk}_6=\delta^{ij}\delta^{ik}\lambda^{(i)}_6,\\
\mu^{ij}&=&\delta^{ij}\mu_i,\quad w^{ij}=\delta^{ij}w_i.
\end{eqnarray}
These choices are likely to be realized in string versions of this model
and will not alter our conclusions below. In this case the Higgs triplet mass
matrix reduces to
\beq
{\cal M}_3=\bordermatrix
{&\bar D&d^c_{H_1}&d^c_{H_2}\cr
D&\mu&\lambda^{(1)}_4V_1&\lambda^{(2)}_4V_2\cr
d^c_{\bar H_1}&\lambda^{(1)}_5\bar V_1&w_1&0\cr
d^c_{\bar H_2}&\lambda^{(2)}_5\bar V_2&0&w_2\cr}.\label{III}
\eeq
Regarding the $(3,2)$ states, the scalars get either eaten by the $X,Y$ $SU(5)$
heavy gauge bosons or become heavy Higgs bosons, whereas the fermions interact
with the $\wt X,\wt Y$ gauginos through the following mass matrix \cite{LNY}
\beq
{\cal M}_{(3,2)}=\bordermatrix
{&Q_{\bar H_1}&Q_{\bar H_2}&\wt Y\cr
Q_{H_1}&w_1&0&g_5V_1\cr
Q_{H_2}&0&w_2&g_5V_2\cr
\wt X&g_5\bar V_1&g_5\bar V_2&0\cr}.
\eeq
The lightest eigenvalues of these two matrices (denoted generally by $d^c_H$
and $Q_H$ respectively) constitute the new relatively light particles in the
spectrum, which are hereafter referred to as the ``{\em gap}" particles since
with suitable masses they bridge the gap between unification masses at
$10^{16}\GeV$ and $10^{18}\GeV$.

Guided by the phenomenological requirement on the gap particle masses, \ie,
$M_{Q_H}\gg M_{d^c_H}$ \cite{sism}, we consider the following explicit
numerical scenario
\beq
\lambda^{(2)}_4=\lambda^{(2)}_5=0,\quad
V_1,\bar V_1,V_2,\bar V_2\sim V\gg w_1\gg w_2\gg\mu,\label{V}
\eeq
which would need to be reproduced in a viable string-derived model. From Eq.
(\ref{III}) we then get $M_{d^c_{H_2}}=M_{d^c_{\bar H_2}}=w_2$, and all
other mass eigenstates $\sim V$. Furthermore, ${\cal M}_{(3,2)}$ has a
characteristic polynomial $\lambda^3-\lambda^2(w_1+w_2)-\lambda(2V^2-w_1w_2)
+(w_1+w_2)V^2=0$, which has two roots of ${\cal O}(V)$ and one root of
${\cal O}(w_1)$. The latter corresponds to $\sim(Q_{H_1}-Q_{H_2})$ and
$\sim(Q_{\bar H_1}-Q_{\bar H_2})$. In sum then, the gap particles have masses
$M_{Q_H}\sim w_1$ and $M_{d^c_H}\sim w_2$, whereas all other heavy particles
have masses $\sim V$.

The see-saw matrix reduces to
\beq
{\cal M}_\nu=\bordermatrix{&\nu_i&\nu^c_i&\phi_i\cr
\nu_i&0&m^i_u&0\cr
\nu^c_i&m^i_u&0&\lambda^{(i)}\bar V_i\cr
\phi_i&0&\lambda^{(i)}\bar V_i&\mu^i\cr},
\eeq
for each generation. The physics of this see-saw matrix has been
discussed in Ref. \cite{chorus} and more generally in Ref. \cite{ELNO},
where it was shown to lead to an interesting amount of hot dark matter
($\nu_\tau$) and an MSW-effect ($\nu_e,\nu_\mu$) compatible with all solar
neutrino data. Moreover, the out-of-equilibrium decays of the $\nu^c$
``flipped neutrino" fields in the early Universe induce a lepton number
asymmetry which is later processed into a baryon number asymmetry by
non-perturbative electroweak processes \cite{ENO,ELNO}. All these phenomena can
occur in the same region of parameter space.
\subsubsection{Proton decay}
The dimension-six operators mediating proton decay in this model are highly
suppressed due to the large mass of the $X,Y$ gauge bosons
($\sim M_U=10^{18}\GeV$). Higgsino mediated dimension-five operators exist
and are naturally suppressed in the  model of Ref. \cite{revitalized}. The
reason for this is that the Higgs triplet mixing term
$\mu h\bar h\to \mu D\bar D$ is small ($\mu\sim M_Z$), whereas the Higgs
triplet mass eigenstates obtained from Eq. (\ref{IIb}) by just keeping the
$2\times2$ submatrix in the upper left-hand corner, are always very heavy
($\sim V$). The dimension-five mediated operators are then proportional to
$\mu/V^2$ and thus the rate is suppressed by a factor of $(\mu/V)^2\ll1$
relative to the unsuppressed case found in the standard $SU(5)$ model.

In the  model presented here, the Higgs triplet mixing term is
still $\mu D\bar D$. However, the exchanged mass eigenstates are not
necessarily all very heavy. In fact, above we have demanded the existence of a
relatively light ($\sim w_1$) Higgs triplet state ($d^c_H$). In this case
the operators are proportional to $\mu\alpha_i\bar\alpha_i/{\cal M}^2_i$, where
${\cal M}_i$ is the mass of the $i$-th exchanged eigenstate and
$\alpha_i,\bar\alpha_i$ are its $D,\bar D$ admixtures. In the scenario
described above, the relatively light eigenstates ($d^c_{H_2},d^c_{\bar H_2}$)
contain no $D,\bar D$ admixtures, and the operator will again be
$\propto\mu/V^2$.

Note however that if conditions (\ref{V}) (or some analogous suitability
requirement) are not satisfied, then diagonalization of ${\cal M}_3$ in Eq.
(\ref{III}) may re-introduce a sizeable dimension-five mediated proton decay
rate, depending on the value of the $\alpha_i,\bar\alpha_i$ coefficients. To be
safe one should demand \cite{ANpd,HMY,LNPZ}
\beq
{\mu\alpha_i\bar\alpha_i\over {\cal M}^2_i}\lsim{1\over 10^{17}\GeV}.
\eeq
For the higher values of $M_{d^c_H}$ in Table \ref{Table1} (see below), this
constraint can be satisfied for not necessarily small values of
$\alpha_i,\bar\alpha_i$.
\begin{table}[p]
\hrule
\caption{The value of the gap particle masses and the unified coupling for
$\alpha_3(M_Z)=0.118\pm0.008$. We have taken $M_U=10^{18}\GeV$,
$\sin^2\theta_w=0.233$, and $\alpha^{-1}_e=127.9$.}
\label{Table1}
\begin{center}
\begin{tabular}{|c|c|c|c|}\hline
$\alpha_3(M_Z)$&$M_{d^c_H}\,(\GeV)$&$M_{Q_H}\,(\GeV)$&$\alpha(M_U)$\\ \hline
$0.110$&$4.9\times10^4\GeV$&$2.2\times10^{12}\GeV$&$0.0565$\\
$0.118$&$4.5\times10^6\GeV$&$4.1\times10^{12}\GeV$&$0.0555$\\
$0.126$&$2.3\times10^8\GeV$&$7.3\times10^{12}\GeV$&$0.0547$\\ \hline
\end{tabular}
\end{center}
\hrule
\end{table}

\begin{figure}[p]
\vspace{4in}
\includegraphics{proc_erice1.ps}
\vspace{-0.7in}
\caption{\baselineskip=12pt
The running of the gauge couplings in the $SU(5)\times U(1)$ model for
$\alpha_3(M_Z)=0.118$ (solid lines). The gap particle masses have been derived
using the gauge coupling RGEs to achieve unification at $M_U=10^{18}\GeV$. The
case with no gap particles (dotted lines) is also shown; here
$M_U\approx10^{16}\GeV$.}
\label{Figure1f}
\end{figure}

\subsubsection{Gauge coupling unification}
Since we have chosen $V\sim M_U=M_{SU}=10^{18}\GeV$, this means that the
Standard Model gauge couplings should unify at the scale $M_U$. However, their
running will be modified due to the presence of the gap particles. Note that
the underlying $SU(5)\times U(1)$ symmetry, even though not evident in this
respect, is nevertheless essential in the above discussion. The masses $M_Q$
and $M_{d^c_H}$ can then be determined, as follows \cite{sism}
\begin{eqnarray}
\ln{M_{Q_H}\over m_Z}&=&\pi\left({1\over2\alpha_e}-{1\over3\alpha_3}
-{\sin^2\theta_w-0.0029\over\alpha_e}\right)-2\ln{M_U\over m_Z}-0.63,\\
\ln{M_{d^c_H}\over m_Z}&=&\pi\left({1\over2\alpha_e}-{7\over3\alpha_3}
+{\sin^2\theta_w-0.0029\over\alpha_e}\right)
			-6\ln{M_U\over m_Z}-1.47,\label{VIb}
\end{eqnarray}
where $\alpha_e$, $\alpha_3$ and $\sin^2\theta_w$ are all measured at $M_Z$.
This is a one-loop determination (the constants account for the dominant
two-loop corrections) which neglects all low- and high-energy threshold
effects,\footnote{Here we assume a common supersymmetric threshold at $M_Z$.
In fact, the supersymmetric threshold and the $d^c_H$ mass are anticorrelated.
See Ref. \cite{sism} for a discussion.} but is quite adequate for our present
purposes. As shown in Table \ref{Table1} (and Eq. (\ref{VIb})) the $d^c_H$ mass
depends most sensitively on $\alpha_3(M_Z)=0.118\pm0.008$ \cite{Bethke},
whereas the $Q_H$ mass and the unified coupling are rather insensitive to it.
The unification of the gauge couplings is shown in Fig. \ref{Figure1f} (solid
lines) for the central value of $\alpha_3(M_Z)$. This figure also shows the
case of no gap particles (dotted lines), for which $M_U\approx10^{16}\GeV$.

\subsection{The problem of supersymmetry breaking}
\label{Scenaria}
A very important component of the model is that which triggers supersymmetry
breaking. In string models this task is performed by the hidden
sector and the universal moduli and dilaton fields. Model-dependent
calculations are required to determine the precise nature of supersymmetry
breaking in a given string model. In fact, no explicit string model exists to
date where various theoretical difficulties (\eg, suitably suppressed
cosmological constant, suitable vacuum state with perturbative gauge coupling,
etc.) have been shown to be satisfactorily overcome. Instead, it has become
apparent \cite{IL,KL,Ibanez} that a more model-independent approach to the
problem may be more profitable. In this approach one parametrizes the breaking
of supersymmetry by the largest $F$-term vacuum expectation value which
triggers supersymmetry breaking. Of all the possible fields which could be
involved (\ie, hidden sector matter fields, various moduli fields, dilaton)
the dilaton and three of the moduli fields are quite common in string
constructions and have thus received the most attention to date. In a way,
if supersymmetry breaking is triggered by these fields (\ie, $\vev{F_S}\not=0$
or $\vev{F_T}\not=0$), this would be a rather generic prediction of string
theory.

There are various possible scenarios for supersymmetry breaking which are
obtained in this model-independent way. To discriminate among these we
consider a simplified expression for the scalar masses (\eg, $m_{\tilde q}$)
$\widetilde m^2_i=m^2_{3/2}(1+n_i\cos^2\theta)$, with
$\tan\theta=\vev{F_S}/\vev{F_T}$ \cite{Ibanez}. Here $m_{3/2}$ is the gravitino
mass and the $n_i$ are the modular weights of the respective matter field.
There are two ways in which one can obtain universal scalar masses, as strongly
desired phenomenologically to avoid large flavor-changing-neutral-currents
(FCNCs) \cite{EN}: (i) setting $\theta=\pi/2$, that is $\vev{F_S}\gg\vev{F_T}$;
or (ii) in a model where all $n_i$ are the same, as occurs for $Z_2\times Z_2$
orbifolds \cite{Ibanez} and free-fermionic constructions \cite{thresholds}.

In the first scenario, supersymmetry breaking is triggered by the dilaton
$F$-term and yields universal soft-supersymmetry-breaking gaugino and scalar
masses and trilinear interactions \cite{KL,Ibanez}
\beq
m_0=\coeff{1}{\sqrt{3}}m_{1/2},\qquad A=-m_{1/2}.\label{kl}
\eeq
This supersymmetry breaking scenario has been studied recently in the context
of $SU(5)\times U(1)$ supergravity \cite{LNZII} and in the MSSM in
Ref.~\cite{BLM}.
In the second scenario, in the limit $\vev{F_T}\gg\vev{F_S}$ (\ie,
$\theta\to0$) all scalar masses at the unification scale vanish, as is the case
in no-scale supergravity models with a unified group structure \cite{LN}. In
this case we have
\beq
m_0=0,\qquad A=0. \label{nsc}
\eeq

\subsection{Phenomenology: general case}
\label{PhenoGen}
The procedure to extract the low-energy predictions of the models outlined
above is rather standard by now (see \eg, Ref. \cite{aspects}): (a) the
bottom-quark and tau-lepton masses, together with the input values of $m_t$ and
$\tan\beta$ are used to determine the respective Yukawa couplings at the
electroweak scale; (b) the gauge and Yukawa couplings are then run up to the
unification scale $M_U=10^{18}\GeV$ taking into account the extra vector-like
quark doublet ($\sim10^{12}\GeV$) and singlet ($\sim10^6\GeV$) introduced above
\cite{sism,LNZI}; (c) at the unification scale the soft-supersymmetry breaking
parameters are introduced (according to Eqs. (\ref{nsc},\ref{kl})) and the
scalar masses are then run down to the electroweak scale; (d) radiative
electroweak symmetry breaking is enforced by minimizing the one-loop effective
potential which depends on the whole mass spectrum, and the values of the Higgs
mixing term $|\mu|$ and the bilinear soft-supersymmetry breaking parameter $B$
are determined from the minimization conditions; (e) all known phenomenological
constraints on the sparticle and Higgs masses are applied (most importantly the
LEP lower bounds on the chargino and Higgs masses), including the cosmological
requirement of not-too-large neutralino relic density.
\subsubsection{Mass ranges}
We have scanned the parameter space for $m_t=130,150,170\GeV$,
$\tan\beta=2\to50$ and $m_{1/2}=50\to500\GeV$. Imposing the constraint
$m_{\tilde g,\tilde q}<1\TeV$ we find
\begin{eqnarray}
&\vev{F_M}_{m_0=0}:\qquad	&m_{1/2}<475\GeV,\quad \tan\beta\lsim32,\\
&\vev{F_D}:\qquad	&m_{1/2}<465\GeV,\quad \tan\beta\lsim46.
\end{eqnarray}
These restrictions on $m_{1/2}$ cut off the growth of most of the sparticle and
Higgs masses at $\approx1\TeV$. However, the sleptons, the lightest Higgs, the
two lightest neutralinos, and the lightest chargino are cut off at a much lower
mass, as follows\footnote{In this class of supergravity models the three
sneutrinos ($\tilde\nu$) are degenerate in mass. Also, $m_{\tilde
\mu_L}=m_{\tilde e_L}$ and $m_{\tilde\mu_R}=m_{\tilde e_R}$.}
\begin{eqnarray}
&\vev{F_M}_{m_0=0}:&\left\{
	\begin{array}{l}
	m_{\tilde e_R}<190\GeV,\quad m_{\tilde e_L}<305\GeV,
				\quad m_{\tilde\nu}<295\GeV\\
	m_{\tilde\tau_1}<185\GeV,\quad m_{\tilde\tau_2}<315\GeV\\
	m_h<125\GeV\\
	m_{\chi^0_1}<145\GeV,\quad m_{\chi^0_2}<290\GeV,
				\quad m_{\chi^\pm_1}<290\GeV
	\end{array}
		\right.\\
&\vev{F_D}:&\left\{
	\begin{array}{l}
	m_{\tilde e_R}<325\GeV,\quad m_{\tilde e_L}<400\GeV,
				\quad m_{\tilde\nu}<400\GeV\\
	m_{\tilde\tau_1}<325\GeV,\quad m_{\tilde\tau_2}<400\GeV\\
	m_h<125\GeV\\
	m_{\chi^0_1}<145\GeV,\quad m_{\chi^0_2}<285\GeV,
				\quad m_{\chi^\pm_1}<285\GeV
	\end{array}
		\right.
\end{eqnarray}
It is interesting to note that due to the various constraints on the model,
the gluino and (average) squark masses are bounded from below,
\beq
\vev{F_M}_{m_0=0}:\left\{
	\begin{array}{l}
	m_{\tilde g}\gsim245\,(260)\GeV\\
	m_{\tilde q}\gsim240\,(250)\GeV
	\end{array}
		\right.
\qquad
\vev{F_D}:\left\{
	\begin{array}{l}
	m_{\tilde g}\gsim195\,(235)\GeV\\
	m_{\tilde q}\gsim195\,(235)\GeV
	\end{array}
		\right.		\label{gmin}
\eeq
for $\mu>0(\mu<0)$. Relaxing the above conditions on $m_{1/2}$ simply allows
all sparticle masses to grow further proportional to $m_{\tilde g}$.

\begin{table}
\hrule
\caption{
The value of the $c_i$ coefficients appearing in  Eq.~(28), the ratio
$c_{\tilde g}=m_{\tilde g}/m_{1/2}$, and the average squark coefficient
$\bar c_{\tilde q}$, for $\alpha_3(M_Z)=0.118\pm0.008$. Also shown are the
$a_i,b_i$ coefficients for the central value of $\alpha_3(M_Z)$ and both
supersymmetry breaking scenaria ($M:\vev{F_M}_{m_0=0}$, $D:\vev{F_D}$).
 The results apply as well to the second-generation squark and slepton masses.}
\label{Table2}
\begin{center}
\begin{tabular}{|c|c|c|c|}\hline
$i$&$c_i\,(0.110)$&$c_i\,(0.118)$&$c_i\,(0.126)$\\ \hline
$\tilde\nu,\tilde e_L$&$0.406$&$0.409$&$0.413$\\
$\tilde e_R$&$0.153$&$0.153$&$0.153$\\
$\tilde u_L,\tilde d_L$&$3.98$&$4.41$&$4.97$\\
$\tilde u_R$&$3.68$&$4.11$&$4.66$\\
$\tilde d_R$&$3.63$&$4.06$&$4.61$\\
$c_{\tilde g}$&$1.95$&$2.12$&$2.30$\\
$\bar c_{\tilde q}$&$3.82$&$4.07$&$4.80$\\ \hline
\end{tabular}
\begin{tabular}{|c|c|c|c|c|}\hline
$i$&$a_i(M)$&$b_i(M)$&$a_i(D)$&$b_i(D)$\\ \hline
$\tilde e_L$&$0.302$&$+1.115$&$0.406$&$+0.616$\\
$\tilde e_R$&$0.185$&$+2.602$&$0.329$&$+0.818$\\
$\tilde\nu$&$0.302$&$-2.089$&$0.406$&$-1.153$\\
$\tilde u_L$&$0.991$&$-0.118$&$1.027$&$-0.110$\\
$\tilde u_R$&$0.956$&$-0.016$&$0.994$&$-0.015$\\
$\tilde d_L$&$0.991$&$+0.164$&$1.027$&$+0.152$\\
$\tilde d_R$&$0.950$&$-0.033$&$0.989$&$-0.030$\\ \hline
\end{tabular}
\end{center}
\hrule
\end{table}

\subsubsection{Mass relations}
The neutralino and chargino masses show a correlation observed before in
this class of models \cite{ANc,LNZI}, namely
\beq
m_{\chi^0_1}\approx \coeff{1}{2}m_{\chi^0_2},\qquad
m_{\chi^0_2}\approx m_{\chi^\pm_1}\approx M_2=(\alpha_2/\alpha_3)m_{\tilde g}
\approx0.28m_{\tilde g}.\label{neuchar}
\eeq
This is because throughout the parameter space $|\mu|$ is generally much larger
than $M_W$ and $|\mu|>M_2$. In practice we find $m_{\chi^0_2}\approx
m_{\chi^\pm_1}$ to be satisfied quite accurately, whereas
$m_{\chi^0_1}\approx{1\over2}m_{\chi^0_2}$ is only qualitatively satisfied,
although the agreement is better in the $\vev{F_D}$ case. In fact, these two
mass relations are much more reliable than the one that links them to
$m_{\tilde g}$. The heavier neutralino ($\chi^0_{3,4}$) and chargino
($\chi^\pm_2$) masses are determined by the value of $|\mu|$; they all approach
this limit for large enough $|\mu|$. More precisely, $m_{\chi^0_3}$ approaches
$|\mu|$ sooner than $m_{\chi^0_4}$ does. On the other hand, $m_{\chi^0_4}$
approaches $m_{\chi^\pm_2}$ rather quickly.

The first- and second-generation squark and slepton masses can be determined
analytically
\beq
\wt m_i=\left[m^2_{1/2}(c_i+\xi^2_0)-d_i{\tan^2\beta-1\over\tan^2\beta+1}
M^2_W\right]^{1/2}=a_i m_{\tilde g}\left[1+b_i\left({150\over m_{\tilde
g}}\right)^2{\tan^2\beta-1\over\tan^2\beta+1}\right]^{1/2},\label{masses}
\eeq
where $d_i=(T_{3i}-Q)\tan^2\theta_w+T_{3i}$ (\eg, $d_{\tilde u_L}={1\over2}
-{1\over6}\tan^2\theta_w$, $d_{\tilde e_R}=-\tan^2\theta_w$), and
$\xi_0=m_0/m_{1/2}=0,\coeff{1}{\sqrt{3}}$. The coefficients $c_i$ can be
calculated numerically in terms of the low-energy gauge couplings, and are
given in  Table \ref{Table2}\footnote{These are renormalized at the scale
$M_Z$. In a more accurate treatment, the $c_i$ would be renormalized at the
physical sparticle mass scale, leading to second order shifts on the sparticle
masses.} for $\alpha_3(M_Z)=0.118\pm0.008$. In the table we also give
$c_{\tilde g}=m_{\tilde g}/m_{1/2}$. Note that these values are smaller than
what is obtained in the $SU(5)$ supergravity model (where $c_{\tilde g}=2.90$
for $\alpha_3(M_Z)=0.118$) and therefore the numerical relations between the
gluino mass and the neutralino masses are different in that model.
In the table we also show the resulting values for $a_i,b_i$ for the central
value of $\alpha_3(M_Z)$. Note that the apparently larger $\tan\beta$
dependence in the $\vev{F_M}_{m_0=0}$ case (\ie, $|b_i(M)|>|b_i(D)|$) is
actually compensated by a larger minimum value of $m_{\tilde g}$ in this case
(see Eq. (\ref{gmin})).

The ``average" squark mass, $m_{\tilde q}\equiv{1\over8}(m_{\tilde
u_L}+m_{\tilde u_R}+m_{\tilde d_L}+m_{\tilde d_R}+m_{\tilde c_L}+m_{\tilde c_R}
+m_{\tilde s_L}+m_{\tilde s_R})
=(m_{\tilde g}/c_{\tilde q})\sqrt{\bar c_{\tilde q}+\xi^2_0}$, with $\bar
c_{\tilde q}$ given in Table \ref{Table2}, is determined to be
\beq
m_{\tilde q}=\left\{	\begin{array}{ll}
			(1.00,0.95,0.95) m_{\tilde g},&\quad\vev{F_M}_{m_0=0}\\
			(1.05,0.99,0.98) m_{\tilde g},&\quad\vev{F_D}
			\end{array}
		\right.
\eeq
for $\alpha_3(M_Z)=0.110,0.118,0.126$ (the dependence on $\tan\beta$ is small).
The squark splitting around the average is $\approx2\%$.

These masses are plotted in Fig. \ref{Figure2f}. The thickness and straightness
of the lines shows the small $\tan\beta$ dependence, except for $\tilde\nu$.
The results do not depend on the sign of $\mu$, except to the extent that some
points in parameter space are not allowed for both signs of $\mu$: the $\mu<0$
lines start-off at larger mass values. Note that
\beq
\vev{F_M}_{m_0=0}:\left\{
	\begin{array}{l}
	m_{\tilde e_R}\approx0.18m_{\tilde g}\\
	m_{\tilde e_L}\approx0.30m_{\tilde g}\\
	m_{\tilde e_R}/m_{\tilde e_L}\approx0.61
	\end{array}
		\right.
\qquad
\vev{F_D}:\left\{
	\begin{array}{l}
	m_{\tilde e_R}\approx0.33m_{\tilde g}\\
	m_{\tilde e_L}\approx0.41m_{\tilde g}\\
	m_{\tilde e_R}/m_{\tilde e_L}\approx0.81
	\end{array}
		\right.
\eeq

\begin{figure}[p]
\vspace{4.7in}
\includegraphics{proc_erice2a.ps}
\vspace{3.8in}
\includegraphics{proc_erice2b.ps}
\vspace{-0.7in}
\caption{\baselineskip=12pt
The first-generation squark and slepton masses as a function of
the gluino mass, for both signs of $\mu$, $m_t=150\GeV$, and both supersymmetry
breaking scenaria under consideration. The same values apply to the second
generation. The thickness of the lines and their deviation from linearity are
because of the small $\tan\beta$ dependence.}
\label{Figure2f}
\end{figure}

The third generation squark and slepton masses cannot be determined
analytically. In Fig. \ref{Figure3f} we show
$\tilde\tau_{1,2},\tilde b_{1,2},\tilde t_{1,2}$ for the choice $m_t=150\GeV$.
The variability on the $\tilde\tau_{1,2}$ and $\tilde b_{1,2}$ masses
is due to the $\tan\beta$-dependence in the off-diagonal element of the
corresponding $2\times2$ mass matrices ($\propto
m_{\tau,b}(A_{\tau,b}+\mu\tan\beta)$). The off-diagonal element in the
stop-squark mass matrix ($\propto m_t(A_t+\mu/\tan\beta)$) is
rather insensitive to $\tan\beta$ but still effects a large $\tilde t_1-\tilde
t_2$ mass splitting because of the significant $A_t$ contribution. Note
that both these effects are more pronounced for the $\vev{F_D}$ case since
there $|A_{t,b,\tau}|$ are larger than in the $\vev{F_M}_{m_0=0}$ case.
The lowest values of the $\tilde t_1$ mass go up with $m_t$ and can be as low
as
\beq
m_{\tilde t_1}\gsim\left\{	\begin{array}{ll}
		160,170,190\,(155,150,170)\GeV;&\quad\vev{F_M}_{m_0=0}\\
			88,112,150\,(92,106,150)\GeV;&\quad\vev{F_D}
			\end{array}
		\right.
\eeq
for $m_t=130,150,170\GeV$ and $\mu>0\,(\mu<0)$.

\begin{figure}[p]
\vspace{4.3in}
\includegraphics{proc_erice3a.ps}
\vspace{3.8in}
\includegraphics{proc_erice3b.ps}
\vspace{-0.5in}
\caption{\baselineskip=12pt
The $\tilde\tau_{1,2}$, $\tilde b_{1,2}$, and $\tilde t_{1,2}$ masses
versus the gluino mass for both signs of $\mu$, $m_t=150\GeV$, and both
supersymmetry breaking scenaria. The variability in the $\tilde\tau_{1,2}$,
$\tilde b_{1,2}$, and $\tilde t_{1,2}$ masses is because of the off-diagonal
elements of the corresponding mass matrices.}
\label{Figure3f}
\end{figure}

The one-loop corrected lightest CP-even ($h$) and CP-odd ($A$) Higgs boson
masses are shown in Fig. \ref{Figure4} as functions of $m_{\tilde g}$ for
$m_t=150\GeV$. Following the methods of Ref. \cite{LNPWZh} we have determined
that the LEP lower bound on $m_h$ becomes $m_h\gsim60\GeV$, as the figure
shows. The largest value of $m_h$ depends on $m_t$; we find
\beq
m_h<\left\{	\begin{array}{ll}
		106,115,125\GeV;&\quad\vev{F_M}_{m_0=0}\\
		107,117,125\GeV;&\quad\vev{F_D}
			\end{array}
		\right.
\eeq
for $m_t=130,150,170\GeV$. It is interesting to note that the one-loop
corrected values of $m_h$ for $\tan\beta=2$ are quite dependent on the sign of
$\mu$. This phenomenon can be traced back to the $\tilde t_1-\tilde t_2$ mass
splitting which enhances the dominant $\tilde t$ one-loop corrections to $m_h$
\cite{ERZ}, an effect which is usually neglected in phenomenological analyses.
The $\tilde t_{1,2}$ masses for $\tan\beta=2$ are drawn closer together than
the rest. The opposite effect occurs for $\mu<0$ and therefore the one-loop
correction is larger in this case. The sign-of-$\mu$ dependence appears in the
off-diagonal entries in the $\tilde t$ mass matrix  $\propto
m_t(A_t+\mu/\tan\beta)$, with $A_t<0$ in this case. Clearly only small
$\tan\beta$ matters, and $\mu<0$ enhances the splitting. The $A$-mass grows
fairly linearly with $m_{\tilde g}$ with a $\tan\beta$-dependent slope which
decreases for increasing $\tan\beta$, as shown in Fig. \ref{Figure4}. Note that
even though $m_A$ can be fairly light, we always get $m_A>m_h$, in agreement
with a general theorem to this effect in supergravity theories \cite{DNh}. This
result also implies that the channel $e^+e^-\to hA$ at LEPI is not
kinematically allowed in this model.

\begin{figure}[p]
\vspace{4.3in}
\includegraphics{proc_erice4a.ps}
\vspace{3.8in}
\includegraphics{proc_erice4b.ps}
\vspace{-0.3in}
\caption{\baselineskip=12pt
The one-loop corrected $h$ and $A$ Higgs masses versus the gluino
mass for both signs of $\mu$, $m_t=150\GeV$, and the two supersymmetry
breaking scenaria. Representative values of $\tan\beta$ are indicated.}
\label{Figure4}
\end{figure}

\subsubsection{Neutralino relic density}
The computation of the neutralino relic density (following the methods of
Refs. \cite{LNYdmI,KLNPYdm}) shows that $\Omega_\chi h^2_0\lsim0.25\,(0.90)$ in
the no-scale (dilaton) model. This implies that in these models the
cosmologically interesting values
$\Omega_\chi h^2_0\lsim1$ occur quite naturally. These results are in good
agreement with the observational upper bound on $\Omega_\chi h^2_0$ \cite{KT}.
Moreover, fits to the COBE data and the small and large scale structure of the
Universe suggest \cite{many} a mixture of $\approx70\%$ cold dark matter and
$\approx30\%$ hot dark matter together with $h_0\approx0.5$. The hot dark
matter component in the form of massive tau neutrinos has already been shown to
be compatible with the $SU(5)\times U(1)$ model we consider here
\cite{chorus,ELNO}, whereas the cold dark matter component implies
$\Omega_\chi h^2_0\approx0.17$ which is reachable in these models.

\subsection{Phenomenology: special cases}
\label{PhenoSp}
\subsubsection{The strict no-scale case: a striking result}
\label{strictNo-scale}
We now impose the additional constraint $B(M_U)=0$ to be added to
Eq.~(\ref{nsc}), and obtain the so-called strict no-scale case \cite{LNZI}.
Since $B(M_Z)$ is determined by the radiative electroweak symmetry breaking
conditions, this added constraint needs to be imposed in a rather indirect way.
That is, for given $m_{\tilde g}$ and $m_t$ values, we scan the possible values
of $\tan\beta$ looking for cases where $B(M_U)=0$. The most striking result is
that solutions exist {\em only} for $m_t\lsim135\GeV$ if $\mu>0$ and for
$m_t\gsim140\GeV$ if $\mu<0$. That is, the value of $m_t$ {\em determines} the
sign of $\mu$. Furthermore, for $\mu<0$ the value of $\tan\beta$ is determined
uniquely as a function of $m_t$ and $m_{\tilde g}$, whereas for $\mu>0$,
$\tan\beta$ can be double-valued for some $m_t$ range which includes
$m_t=130\GeV$ (but does not include $m_t=100\GeV$). In Fig. \ref{Figure5} (top
row) we plot the solutions found in this manner for the indicated $m_t$ values.

All the mass relationships deduced in the previous section apply here as well.
The $\tan\beta$-spread that some of them have will be much reduced though.
The most noticeable changes occur for the quantities which depend most
sensitively on $\tan\beta$. In Fig. \ref{Figure5} (bottom row) we plot the
one-loop corrected lightest Higgs boson mass versus $m_{\tilde g}$. The result
is that $m_h$ is basically determined by $m_t$; only a weak dependence on
$m_{\tilde g}$ exists. Moreover, for $m_t\lsim135\GeV\Leftrightarrow\mu>0$,
$m_h\lsim105\GeV$; whereas for $m_t\gsim140\GeV\Leftrightarrow\mu<0$,
$m_h\gsim100\GeV$. Therefore, in the strict no-scale case, once the top-quark
mass is measured, we will know the sign of $\mu$ and whether $m_h$ is above or
below $100\GeV$.

For $\mu>0$, we just showed that the strict no-scale constraint requires
$m_t\lsim135\GeV$. This implies that $\mu$ cannot grow as large as it did
previously in the general case. In fact, for $\mu>0$, $\mu_{max}\approx745\GeV$
before and $\mu_{max}\approx440\GeV$ now. This smaller value of $\mu_{max}$ has
the effect of cutting off the growth of the $\chi^0_{3,4},\chi^\pm_2$ masses
at $\approx\mu_{max}\approx440\GeV$ (c.f. $\approx750\GeV$) and of the heavy
Higgs masses at $\approx530\GeV$ (c.f. $\approx940\GeV$).

\begin{figure}[t]
\vspace{4.3in}
\includegraphics{proc_erice5.ps}
\vspace{-0.3in}
\caption{\baselineskip=12pt
The value of $\tan\beta$ versus $m_{\tilde g}$ in the strict no-scale case
(where $B(M_U)=0$) for the indicated values of $m_t$. Note that the sign of
$\mu$ is {\em determined} by $m_t$ and that $\tan\beta$ can be double-valued
for $\mu>0$. Also shown is the one-loop corrected lightest
Higgs boson mass. Note that if $\mu>0$ (for $m_t<135\GeV$) then $m_h<105\GeV$;
whereas if $\mu<0$ (for $m_t>140\GeV$) then $m_h>100\GeV$.}
\label{Figure5}
\end{figure}

\subsubsection{The special dilaton scenario case}
\label{specialDilaton}
In our analysis above, the radiative electroweak breaking conditions were used
to determine the magnitude of the Higgs mixing term $\mu$ at the electroweak
scale. This quantity is ensured to remain light as long as the supersymmetry
breaking parameters remain light. In a fundamental theory this parameter should
be calculable and its value used to determine the $Z$-boson mass. From this
point of view it is not clear that the natural value of $\mu$ should be light.
In specific models one can obtain such values by invoking non-renormalizable
interactions \cite{muproblem,decisive,Casasmu}. Another contribution to this
quantity is generically present in string supergravity models
\cite{GiMa,Casasmu,KL}. The general case with contributions from both sources
has been effectively dealt with in the previous section. If one assumes that
only supergravity-induced contributions to $\mu$ exist, then it can be shown
that the $B$-parameter at the unification scale is also determined
\cite{KL,Ibanez},
\beq
B(M_U)=2m_0=\coeff{2}{\sqrt{3}}m_{1/2},\label{klII}
\eeq
which is to be added to the set of relations in Eq. (\ref{kl}). This new
constraint effectively determines $\tan\beta$ for given $m_t$ and $m_{\tilde
g}$ values and makes this restricted version of the model highly
predictive \cite{LNZII}.

{}From the outset we note that only solutions with $\mu<0$ exist. This is not
a completely obvious result, but it can be partially understood as follows.
In tree-level approximation, $m^2_A>0\Rightarrow\mu B<0$ at the electroweak
scale. Since $B(M_U)$ is required to be positive and not small, $B(M_Z)$ will
likely be positive also, thus forcing $\mu$ to be negative. A sufficiently
small value of $B(M_U)$ and/or one-loop corrections to $m^2_A$ could alter this
result, although in practice this does not happen. A numerical iterative
procedure allows us to determine the value of $\tan\beta$ which satisfies Eq.
(\ref{klII}), from the calculated value of $B(M_Z)$. We find that
\beq
\tan\beta\approx1.57-1.63,1.37-1.45,1.38-1.40
\quad{\rm for\ }m_t=130,150,155\GeV
\eeq
is required. Since $\tan\beta$ is so small ($m^{tree}_h\approx28-41\GeV$), a
significant one-loop correction to $m_h$ is required to increase it above
its experimental lower bound of $\approx60\GeV$ \cite{LNPWZh}. This requires
the largest possible top-quark masses (and a not-too-small squark mass).
However, perturbative unification imposes an upper bound on $m_t$ for a given
$\tan\beta$ \cite{DL}, which in this case implies \cite{aspects}
\beq
m_t\lsim155\GeV,
\eeq
which limits the magnitude of $m_h$
\beq
m_h\lsim74,87,91\GeV\qquad{\rm for}\qquad m_t=130,150,155\GeV.
\eeq
Lower values of $m_t$ are experimentally disfavored.

In Table~\ref{Table3} we give the range of sparticle and Higgs masses that
are allowed in this case. Clearly, continuing top-quark searches at the
Tevatron and Higgs searches at LEPI,II should probe this restricted scenario
completely.

\begin{table}
\hrule
\caption{
The range of allowed sparticle and Higgs masses in the restricted dilaton
scenario. The top-quark mass is restricted to be $m_t<155\GeV$. All masses in
GeV.}
\label{Table3}
\begin{center}
\begin{tabular}{|c|c|c|c|}\hline
$m_t$&$130$&$150$&$155$\\ \hline
$\tilde g$&$335-1000$&$260-1000$&$640-1000$\\
$\chi^0_1$&$38-140$&$24-140$&$90-140$\\
$\chi^0_2,\chi^\pm_1$&$75-270$&$50-270$&$170-270$\\
$\tan\beta$&$1.57-1.63$&$1.37-1.45$&$1.38-1.40$\\
$h$&$61-74$&$64-87$&$84-91$\\
$\tilde l$&$110-400$&$90-400$&$210-400$\\
$\tilde q$&$335-1000$&$260-1000$&$640-1000$\\
$A,H,H^+$&$>400$&$>400$&$>970$\\ \hline
 \end{tabular}
\end{center}
\hrule
\end{table}

\section{Detailed calculations for the Tevatron}
\label{tevatron}
The sparticle and Higgs spectrum discussed so far can be directly explored
partially at present and near future collider facilities, as we now discuss for
each model considered above. Let us start with the Tevatron, where the main
topics of experimental interest are:
\begin{description}
\item (a) The search and eventual discovery of the top quark will narrow down
the parameter space of these models considerably. Moreover, in the two special
$SU(5)\times U(1)$ cases discussed in Sec.~\ref{PhenoSp} this measurement
will be very important: (i) in the strict no-scale case
(Sec.~\ref{strictNo-scale}) it will determine the sign of $\mu$ ($\mu>0$ if
$m_t\lsim135\GeV$; $\mu<0$ if $m_t\gsim140\GeV$) and whether the Higgs mass is
above or below $\approx100\GeV$, and (ii) it may rule out the restricted
dilaton scenario (Sec.~\ref{specialDilaton}) if $m_t>150\GeV$.
\item (b) The trilepton signal in $p\bar p\to \chi^0_2\chi^\pm_1X$, where
$\chi^0_2$ and $\chi^\pm_1$ both decay leptonically, is a clean test of
supersymmetry \cite{trileptons} and in particular of this class of models
\cite{LNWZ}. Examples of trilepton rates in the no-scale $SU(5)\times U(1)$ and
in the $SU(5)$ model are shown in Fig.~\ref{Figure6}. For more details see
Ref.~\cite{LNWZ}. With ${\cal L}=100\ipb$ of integrated luminosity
basically all of the parameter space of the $SU(5)$ model should be
explorable. Also, chargino masses as high as $\approx175\GeV$ in the no-scale
model could be explored, although some regions of parameter space for lighter
chargino masses would remain unexplored. We expect that somewhat weaker results
will hold for the dilaton model, since the sparticle masses are heavier in that
model, especially the sleptons which enhance the leptonic branching ratios when
they are light enough \cite{LNWZ}.
\item (c) The relation $m_{\tilde q}\approx m_{\tilde g}$ for the $\tilde
u_{L,R},\tilde d_{L,R}$ squark masses in the $SU(5)\times U(1)$ models
should allow to probe the low end of the squark and gluino allowed mass ranges,
although the outlook is more promising for the dilaton model since the allowed
range starts off at lower values of $m_{\tilde g,\tilde q}$ (see Eq.
(\ref{gmin})). An important point distinguishing the two models is that the
average squark mass is slightly below (above) the gluino mass in the no-scale
(dilaton) model, which should have an important bearing on the experimental
signatures and rates \cite{sgdetection}. In the dilaton case the $\tilde t_1$
mass can be below $100\GeV$ for sufficiently low $m_t$, and thus may be
detectable. As the lower bound on $m_t$ rises, this signal becomes less
accessible. The actual reach of the Tevatron for the above processes depends on
its ultimate integrated luminosity. The squark masses in the $SU(5)$
model ($m_{\tilde q}\gsim500\GeV$) are beyond the reach of the Tevatron.
\end{description}

\begin{figure}[p]
\vspace{4.5in}
\includegraphics{proc_summer6a.ps}
\vspace{3in}
\includegraphics{proc_erice6.ps}
\vspace{-2.2in}
\caption{\baselineskip=12pt
The number of trilepton events at the Tevatron per $100\ipb$ in the
$SU(5)$ model (``minimal $SU(5)$ in the figure) and in no-scale $SU(5)\times
U(1)$ (for $m_t=130\GeV$). Note that with $200\ipb$ and 60\% detection
efficiency it should be possible to probe basically all of the parameter space
of the $SU(5)$ model, and probe chargino masses as high as $175\GeV$ in the
no-scale model.}
\label{Figure6}
\end{figure}

Very recently the first limits from the Tevatron on trilepton searches have
been announced, as discussed in (b) above. These limits \cite{Kamon}, along
with the experimental predictions are shown in Fig.~\ref{trilep}. At present no
useful constraints on the parameter
space can be obtained from these searches. However, with the expected increase
in luminosity by a factor of four during 1994, it should be possible to probe
regions of parameter space with chargino masses as large as 100 GeV.

\begin{figure}[t]
\vspace{4.0in}
\includegraphics{Fig3-trilep.ps}
\vspace{-0.3in}
\caption{\baselineskip=12pt
The trilepton cross section at the Tevatron ($p\bar p\to
\chi^0_2\chi^\pm_1 X$; $\chi^0_2\to\chi^0_1 l^+l^-$, $\chi^\pm_1\to \chi^0_1
l^\pm\nu_l$, with $l=e,\mu$) in $SU(5)\times U(1)$ supergravity for both
no-scale and dilaton scenarios. The CDF 95\%CL upper limit is shown (solid
line), as well a possible improved limit at the end of 1994 (dashed line).}
\label{trilep}
\end{figure}

\section{Detailed calculations for LEP}
\label{LEP}
\subsection{LEP I}
\label{LEPI}
The current LEPI lower bound on the Standard Model (SM) Higgs boson mass
($m_H>62\GeV$ \cite{Hilgart}) is obtained by studying the
process $e^+e^-\to Z^*H$ with subsequent Higgs decay into two jets. The MSSM
analog of this production process leads to a cross section differing just by a
factor of $\sin^2(\alpha-\beta)$, where $\alpha$ is the SUSY Higgs mixing angle
and $\tan\beta=v_2/v_1$ is the ratio of the Higgs vacuum expectation values
\cite{HHG}. The published LEPI lower bound on the lightest SUSY Higgs boson
mass ($m_h>43\GeV$) is the result of allowing $\sin^2(\alpha-\beta)$ to vary
throughout the MSSM parameter space and by considering the $e^+e^-\to Z^*h,hA$
cross sections. It is therefore possible that in specific models (which embed
the MSSM), where $\sin^2(\alpha-\beta)$ is naturally restricted to be near
unity, the lower bound on $m_h$ could rise, and even reach the SM lower bound
if ${\rm BR}(h\to2\,{\rm jets})$ is SM-like as well. This has been shown to
be the case for the supergravity models we discuss here, and more generally
for supergravity models which enforce radiative electroweak symmetry
breaking \cite{LNPWZh}.

Non-observation of a SM Higgs signal puts the following upper bound in the
number of expected 2-jet events.
\beq
\#{\rm events}_{\rm\,SM}=\sigma(e^+e^-\to Z^*H)_{\rm SM}\times
{\rm BR}(H\to2\,{\rm jets})_{\rm SM}\times\int{\cal L}dt<3.\label{LI}
\eeq
The SM value for ${\rm BR}(H\to2\,{\rm jets})_{\rm SM}\approx
{\rm BR}(H\to b\bar b+c\bar c+gg)_{\rm SM}\approx0.92$ \cite{HHG} corresponds
to an upper bound on $\sigma(e^+e^-\to Z^*H)_{\rm SM}$. Since this is a
monotonically decreasing function of $m_H$, a lower bound on $m_H$ follows,
\ie, $m_H>62\GeV$ as noted above. We denote by $\sigma_{\rm SM}(62)$ the
corresponding value for $\sigma(e^+e^-\to Z^*H)_{\rm SM}$. For the MSSM the
following relations hold
\begin{eqnarray}
\sigma(e^+e^-\to Z^*h)_{\rm SUSY}&=&\sin^2(\alpha-\beta)\sigma(e^+e^-\to
						Z^*H)_{\rm SM},\\
{\rm BR}(h\to2\,{\rm jets})_{\rm SUSY}&=&f\cdot{\rm BR}(H\to2\,{\rm jets})_{\rm
							SM}.
\end{eqnarray}
{}From Eq.~(\ref{LI}) one can deduce the integrated luminosity achieved,\\
$\int{\cal L}dt=3/(\sigma_{\rm SM}(62){\rm BR}_{\rm SM})$. In analogy with
Eq.~(\ref{LI}), we can write
\beq
\#{\rm events}_{\rm\,SUSY}=\sigma_{\rm SUSY}(m_h)\times{\rm BR}_{\rm
SUSY}\times\int{\cal L}dt=3f\cdot\sigma_{\rm SUSY}(m_h)/\sigma_{\rm
SM}(62)<3.
\eeq
This immediately implies the following condition for {\it allowed} points in
parameter space \cite{LNPWZh,LG}
\beq
f\cdot\sin^2(\alpha-\beta)<P(62/M_Z)/P(m_h/M_Z),
\eeq
where we have used the fact that the cross sections differ simply by the
coupling factor $\sin^2(\alpha-\beta)$ and the Higgs mass dependence which
enters through a function $P$ \cite{HHG}
\begin{eqnarray}
P(y)&=&{3y(y^4-8y^2+20)\over\sqrt{4-y^2}}\cos^{-1}
\left(y(3-y^2)\over2\right)-3(y^4-6y^2+4)\ln y\nonumber\\
&&-\coeff{1}{2}(1-y^2)(2y^4-13y+47)\,.
\end{eqnarray}

The cross section $\sigma_{\rm SUSY}(m_h)$ for the $SU(5)$ model also
corresponds to the SM result since one can verify that
$\sin^2(\alpha-\beta)>0.9999$ in this case. For the flipped model there is a
small deviation ($\sin^2(\alpha-\beta)>0.95$) relative to the SM result for
some points \cite{LNPWZh}. In the calculation of ${\rm BR}(h\to2\,{\rm
jets})_{\rm SUSY}$ which enters in the ratio $f$, we have included {\it all}
contributing modes, in particular the invisible $h\to\chi^0_1\chi^0_1$ decays.
The conclusion is that these models differ little from the SM and in fact the
proper lower bound on $m_h$ is very near $60\GeV$, although it varies from
point to point in the parameter space.

In Ref.~\cite{LNPWZh} it was also shown that this phenomenon is due to a
decoupling effect of the Higgs sector as the supersymmetry scale rises, and
it is communicated to the Higgs sector through the radiative electroweak
symmetry breaking mechanism. The point to be stressed is that if the
supersymmetric Higgs sector is found to be SM-like, this could be taken as
{\em indirect} evidence for an underlying radiative electroweak breaking
mechanism, since no insight could be garnered from the MSSM itself.

Note that since the lower bound on the SM Higgs boson mass could still be
pushed up several GeV at LEPI, the strict dilaton scenario in
Sec.~\ref{specialDilaton} (which requires $m_h\approx61-91\GeV$) could be
further constrained at LEPI.

\begin{figure}[p]
\vspace{4.5in}
\includegraphics{proc_summer7a.ps}
\vspace{-2.5in}
\caption{\baselineskip=12pt
The number of ``mixed" events (1-lepton+2jets+$\mpt$) events per ${\cal
L}=100\ipb$ at LEPII versus the chargino mass in the $SU(5)$ model.}
\label{Figure7a}
\vspace{5in}
\includegraphics{proc_erice7.ps}
\vspace{-0.3in}
\caption{\baselineskip=12pt
The number of ``mixed" events (1-lepton+2jets+$\mpt$) events per ${\cal
L}=100\ipb$ at LEPII versus the chargino mass in the no-scale model (top row).
Also shown (bottom row) are the number of di-electron events per ${\cal
L}=100\ipb$  from selectron pair production versus the lightest selectron
mass.}
\label{Figure7}
\end{figure}

\subsection{LEP II}
\begin{description}
\item (a)  At LEPII the SM Higgs mass could be explored up to roughly the beam
energy minus $100\GeV$ \cite{Alcaraz}. This will allow exploration of almost
all of the Higgs parameter space in the $SU(5)$ model \cite{LNPWZ}. For
$SU(5)\times U(1)$ supergravity, only low $\tan\beta$ values could be explored,
although the strict no-scale case will probably be out of reach (see Figs.
\ref{Figure4},\ref{Figure5}). The $e^+e^-\to hA$ channel will be open for
$SU(5)\times U(1)$ for large $\tan\beta$ and low $m_{\tilde g}$. This
channel is always closed in the $SU(5)$ case (since $m_A\gsim1\TeV$).
It is important to point out that the preferred $h\to b\bar b,c\bar c,gg$
detection modes may be suppressed because of invisible Higgs decays
($h\to\chi^0_1\chi^0_1$) for $m_h\lsim80\GeV$ ($m_h\gsim80\GeV$) by as much
as 30\%/15\% (40\%/40\%) in the $SU(5)$/$SU(5)\times U(1)$ model \cite{LNPWZ}.
\item (b) Chargino masses below the kinematical limit
($m_{\chi^\pm_1}\lsim100\GeV$) should not be a problem to detect through the
``mixed" mode with one chargino decaying leptonically and the other one
hadronically \cite{LNPWZ}, \ie, $e^+e^-\to\chi^+_1\chi^-_1$, $\chi^+_1\to
\chi^0_1 q\bar q'$, $\chi^-_1\to\chi^0_1 l^-\bar\nu_l$. In
Fig.~\ref{Figure7a} and Fig.~\ref{Figure7} (top row) we show the correponding
event rates in the $SU(5)$ and no-scale $SU(5)\times U(1)$ models. Recall
that $m_{\chi^\pm_1}$ can be as high as $\approx290\GeV$ in $SU(5)\times U(1)$
supergravity, whereas $m_{\chi^\pm_1}\lsim100\GeV$ in the $SU(5)$ model.
Interestingly enough, the number of mixed events do not overlap (they are much
higher in the $SU(5)$ model) and therefore, if $m_{\chi^\pm_1}<100\GeV$
then LEPII should be able to exclude at least one of the models.
\item (c) Selectron, smuon, and stau pair production is partially accessible
for both cases of $SU(5)\times U(1)$ supergravity, no-scale and dilaton
(although more so in the no-scale case), and completely inaccessible in the
$SU(5)$ case. In Fig.~\ref{Figure7} (bottom row) we show the rates for the most
promising (dielectron) mode in $e^+e^-\to\tilde e^+_R\,\tilde e^-_R$ production
in the no-scale model.
\end{description}

\section{Detailed calculations for HERA}
\label{HERA}
 The elastic and deep-inelastic contributions to
$e^-p\to\tilde e^-_R\chi^0_1$ and $e^-p\to\tilde\nu\chi^-_1$ at HERA in the
no-scale $SU(5)\times U(1)$ model should push the LEPI lower bounds on the
lightest selectron, the lightest neutralino, and the sneutrino masses by
$\approx25\GeV$ with ${\cal L}=100\ipb$ \cite{hera}. In Fig. \ref{Figure8} we
show the elastic plus deep-inelastic contributions to the total supersymmetric
signal ($ep\to{\rm susy}\to eX+\mpt$) versus the lightest selectron mass
($m_{\tilde e_R}$) and the sneutrino mass $(m_{\tilde\nu})$ in the no-scale
model. These figures show the ``reach" of HERA in each of these variables. With
${\cal L}=1000\ipb$ HERA should be competitive with LEPII as far as the
no-scale model is concerned. In the dilaton scenario, because of the somewhat
heavier sparticle masses, the effectiveness of HERA is reduced, although
probably both channels may be accessible. HERA is not sensitive
to the $SU(5)$ model spectrum.

For elastic processes, another measurable signal is the slowed down outgoing
proton. Since the transverse momentum of the outgoing proton is very small,
the relative energy loss of the proton energy $z=(E^{in}_p-E^{out}_p)/E^{in}_p$
is given by $z=1-x_L$, where $x_L$ is the longitudinal momentum of the leading
proton. It has been pointed out \cite{DREES} that the $z$-distribution is
peaked at a value not much larger than its minimal value,
\beq
z_{min}={1\over s}(m_{\tilde e_{R,L}}+m_{\chi^0_{1,2}})^2.\label{XVII}
\eeq
Therefore, the smallest measured value in the $z$-distribution should be a good
approximation to $z_{min}$. Since the Leading Proton Spectrometer (LPS) of the
ZEUS detector at HERA can measure this distribution accurately, one may have
a new way of probing the supersymmetric spectrum, as follows. We calculate the
average $\tilde z_{min}$ weighed by the different elastic cross sections
$\sigma(\tilde e_{R,L}\chi^0_{1,2})$. The results are shown in the top row of
Fig.~\ref{LPSfig} versus the total elastic cross section. These plots show
the possible values of $\tilde z_{min}$ for a given sensitivity. For example,
if elastic cross sections could be measured down to $\approx10^{-3}\pb$, then
$\tilde z_{min}$ could be fully probed up to $\approx0.17$. Now, $\tilde
z_{min}$ can be computed from Eq.~(\ref{XVII}) and be plotted against, say
$m_{\tilde e_R}$, as shown in the bottom row of Fig.~\ref{LPSfig}. For the
example given above ($\tilde z_{min}\lsim0.17$) one could indirectly probe
$\tilde e_R$ masses as high as $\approx108\GeV$. Note that a useful constraint
on $m_{\tilde e_R}$ is possible because the correlation among the various
sparticle masses in this model makes these scatter plots be rather well
defined. This indirect experimental exploration still requires the
identification of elastic supersymmetric events with $eX+\mpt$ signature (in
order to identify protons that contribute to the relevant $z$-distribution),
but does not require a detailed reconstruction of each such event.

\begin{figure}[t]
\vspace{4.7in}
\includegraphics{proc_erice8.ps}
\vspace{-0.3in}
\caption{\baselineskip=12pt
The elastic plus deep-inelastic total supersymmetric cross section at HERA
($ep\to{\rm susy}\to eX+\mpt$) versus the lightest selectron mass ($m_{\tilde
e_R}$) and the sneutrino mass ($m_{\tilde\nu}$). The short- and long-term
limits of sensitivity are expected to be $10^{-2}\pb$ and $10^{-3}\pb$
respectively.}
\label{Figure8}
\end{figure}

\begin{figure}[t]
\vspace{4.7in}
\includegraphics{hera5.ps}
\vspace{-0.3in}
\caption{\baselineskip=12pt
The most likely value of the relative proton energy loss in elastic
processes (weighed by the various elastic cross sections) versus the total
elastic cross section for selectron-neutralino production (top row) and
$m_{\tilde e_R}$ (bottom row). The Leading Proton Spectrometer (LPS) will
allow determination of $\tilde z_{min}$, and thus an indirect measurement
of $m_{\tilde e_R}$.}
\label{LPSfig}
\end{figure}

\section{Detailed calculations for Underground Labs and Underwater facilities}
\label{Under}
\subsection{Gran Sasso and Super Kamiokande}
\label{pd}
\begin{figure}[t]
\vspace{4.3in}
\includegraphics{proc_summer12.ps}
\vspace{-1.5in}
\caption{\baselineskip=12pt
The calculated values of the proton lifetime into $p\to\bar\nu K^+$
versus the lightest chargino (or second-to-lightest neutralino) mass for both
signs of $\mu$, using the more accurate value of the unification mass (which
includes two-loop and low-energy supersymmetric threshold effects). Note that
we have taken $\alpha_3+1\sigma$ in order to maximize $\tau_p$. Note also that
future proton decay experiments should be sensitive up to
$\tau_p\approx20\times10^{32}\y$.}
\label{Figure12}
\end{figure}

The refinement on the calculation of the unification mass described in
Sec.~\ref{su5pdecay}, to include two-loop effects and light supersymmetric
thresholds, has a significant
effect on the calculated value of the proton lifetime \cite{LNPZ}, since we
take $M_H=3M_U$. With the more accurate value of $M_U$ we  simply rescale our
previously calculated
$\tau_p$ values which satisfied $\tau^{(0)}_p>\tau^{exp}_p$, and find that
$\tau^{(1)}_p=\tau^{(0)}_p[M_U^{(1)}/M_U^{(0)}]^2>\tau^{exp}_p$ for only
$\lsim25\%$ of the previously allowed points. The value of $\alpha_3$ has a
significant influence on the results since larger (smaller) values of
$\alpha_3$ increase (decrease) $M_U$, although the effect is more pronounced
for low values of $\alpha_3$. To quote the most conservative values of the
observables, in what follows we take $\alpha_3$ at its $+1\sigma$ value
($\alpha_3=0.126$). This choice of $\alpha_3$ also gives $\sin^2\theta_w$
values consistent with the $\pm1\sigma$ experimental range. Finally, in the
search  of the parameter space above, we considered only
$\tan\beta=2,4,6,8,10$ and found that $\tan\beta\lsim6$ was required. Our
present analysis indicates that this upper bound is reduced down to
$\tan\beta\lsim3.5$. Here we consider also $\tan\beta=1.5,1.75$ since low
$\tan\beta$ maximizes $\tau_p\propto \sin^22\beta$. These add new allowed
points (\ie, $\tau^{(0)}_p>\tau^{exp}_p$) to our previous set, although most of
them ($\gsim75\%$) do not survive the stricter proton decay constraint
($\tau^{(1)}_p>\tau^{exp}_p$) imposed here.

In Fig.~\ref{Figure12} we show the re-scaled values of $\tau_p$ versus the
lightest chargino mass $m_{\chi^\pm_1}$. All points satisfy $\xi_0\equiv
m_0/m_{1/2}\gsim6$ and $m_{\chi^\pm_1}\lsim150\GeV$, which are to be contrasted
with $\xi_0\gsim3$ and $m_{\chi^\pm_1}\lsim225\GeV$ derived using
the {\em weaker} proton decay constraint \cite{LNP}. The upper bound
on $m_{\chi^\pm_1}$ derives from its near proportionality to $m_{\tilde g}$,
$m_{\chi^\pm_1}\approx0.3m_{\tilde g}$ \cite{ANpd,LNP}, and the result
$m_{\tilde g}\lsim500\GeV$. The latter follows from the proton decay constraint
$\xi_0\gsim6$ and the naturalness requirement $m_{\tilde q}\approx\sqrt{m^2_0+
6m^2_{1/2}}\approx{1\over3}m_{\tilde g}\sqrt{6+\xi^2_0}<1\TeV$. With
naturalness and $H_3$ mass assumptions, we then obtain \footnote{Note that in
general, $\tau_p\propto M^2_{H_3}[m^2_{\tilde q}/
m_{\chi^\pm_1}]^2\propto M^2_{H_3}[m_{\tilde g}(6+\xi^2_0)]^2$ and thus
$\tau_p$ can be made as large as desired by increasing sufficiently either
the supersymmetric spectrum or $M_H$.}
\beq
\tau_p<3.1\,(3.4)\times10^{32}\y\quad{\rm for}\quad \mu>0\,(\mu<0).\label{taup}
\eeq
The $p\to\bar\nu K^+$ mode should then be readily observable at SuperKamiokande
and Gran Sasso since these experiments should be sensitive up to
$\tau_p\approx2\times10^{33}\y$. Note that if $M_H$ is relaxed up to its
largest possible value consistent with low-energy physics,
$M_H=2.3\times10^{17}\GeV$ \cite{HMY}, then in Eq.~(\ref{taup}) $\tau_p\to
\tau_p<4.0\,(4.8)\times10^{33}\y$, and only part of the parameter space of
the model would be experimentally accessible. However, to make this choice
of $M_H$ consistent with high-energy physics (\ie, $M_H<2M_V$) one must
have $M_V/M_\Sigma>42$.

\subsection{Neutrino Telescopes}
The basic idea is that the neutralinos $\chi$ (lightest linear combination of
the superpartners of the photon, $Z$-boson, and neutral Higgs bosons), which
are weakly interacting massive particles (WIMPs), are assumed to make up the
dark matter in the galactic halo ---an important assumption which should
not be overlooked--- and can be gravitationally captured by
the Sun or Earth \cite{PS85,Gould}, after losing a substantial amount of
energy through elastic collisions with nuclei. The neutralinos captured in
the Sun or Earth cores annihilate into all possible ordinary particles,
and the cascade decays of these particles as well as their interactions with
the solar or terrestrial media produce high-energy neutrinos as
one of several end-products. These neutrinos can then travel from the
Sun or Earth cores to the vicinity of underground detectors, and interact
with the rock underneath producing detectable upwardly-moving muons.
Such detectors are rightfully called ``neutrino telescopes", and the
possibility of indirectly detecting various WIMP candidates has been
considered in the past by several authors \cite{others}.
More recent analyses can be found
in Refs.~\cite{RS,GR89,GGR91,KAM,KEK,Bott,FH}. The calculation of the
upwardly-moving muon fluxes induced by the neutrinos from the Sun
and Earth in the still-allowed parameter space of the $SU(5)$ and $SU(5)\times
U(1)$ supergravity models has been performed in Ref.~\cite{NT}. The currently
most stringent 90\% C.L. experimental upper bounds, obtained at Kamiokande,
for neutrinos from the Sun \cite{KEKII} and Earth \cite{KEK} respectively,
{\it i.e.}
\begin{eqnarray}
\Gamma_{\rm Sun}&<&6.6\times10^{-14} {\rm cm}^{-2} {\rm s}^{-1}
=2.08\times10^{-2}{\rm m}^{-2}{\rm yr}^{-1},\label{eq:Sup}\\
\Gamma_{\rm Earth}&<&4.0\times10^{-14} {\rm cm}^{-2} {\rm s}^{-1}
=1.26\times10^{-2}{\rm m}^{-2}{\rm yr}^{-1}.\label{eq:Eup}
\end{eqnarray}
Aiming at the next generation of underground experimental facilities,
such as MACRO and other detectors at the Gran Sasso Laboratory \cite{mac},
Super-Kamiokande \cite{sk}, DUMAND, and AMANDA \cite{amd}, where
improvements in sensitivity by a factor of 2--100 are expected,
we also delineate the region of the parameter space of these models
that would become accessible with an improvement of experimental
sensitivity by modest factors of two and twelve.

\subsubsection{The capture rate}
\label{sec:cap}

In order to calculate the expected rate of neutrino production due to
neutralino annihilation, it is necessary to first evaluate the rates at
which the neutralinos are captured in the Sun and Earth.
Following the early work of Press and Spergel \cite{PS85},
the capture of WIMPs by a massive body was studied extensively
by Gould \cite{Gould}. In our calculations, we make use of Gould's formula,
and follow a procedure similar to that of Refs.~\cite{GGR91,KAM}
in calculating the capture rate.

{}From Eq.~(A10) of the second paper in Ref.~\cite{Gould}, the
capture rate of a neutralino of mass $m_\chi$ by the Sun or Earth
can be written as
\begin{equation}
C=\left({2\over 3\pi}\right)^{1\over 2}M_B
{\rho_\chi{\bar v}_\chi\over m_\chi}
\sum_i{{f_i\over m_i}\sigma_i X_i},\label{eq:nc}
\end{equation}
where $M_B$ is the mass of the Sun or Earth,
$\rho_\chi$ and ${\bar v}_\chi$ are the local neutralino density
and rms velocity in the halo respectively, $\sigma_i$ is the
elastic scattering cross-section of the neutralino with the
nucleus of element $i$ with mass $m_i$, $f_i$ is the mass fraction
of element $i$, and $X_i$ is a kinematic factor which accounts
for several important effects: (1) the motion of the Sun or
Earth relative to Galactic center; (2) the suppression due to the
mismatching of $m_\chi$ and $m_i$; (3) the loss of coherence in the
interaction due to the finite size of the nucleus
(see Ref.~\cite{Gould} for details).

In the summation in Eq.~(\ref{eq:nc}), we only include the ten most abundant
elements for the Sun or Earth respectively, and use the mass
fraction $f_i$ of these elements as listed in Table A.1
of Ref.~\cite{GGR91}. We choose
${\bar v}_\chi = 300\,{\rm km}\,{\rm sec}^{-1}$, a value within
the allowed range of the characteristic velocity of halo dark matter
particles. To take into account the effect of the actual
neutralino relic density, we follow the conservative
approach of Ref.~\cite{GGR91} for the local neutralino density $\rho_\chi$:
(a) $\rho_\chi = \rho_h = 0.3\,{\rm GeV}/{\rm cm}^3$,
if $\Omega_\chi h^2_0 > 0.05$;
while (b) $\rho_\chi = (\Omega_\chi h^2_0/0.05)\rho_h$,
if $\Omega_\chi h^2_0 \lsim 0.05$.
As for $\sigma_i$, the dominant contribution is the coherent
interaction due to the exchange of two CP-even Higgs bosons $h$ and $H$
and squarks, and we use the expressions (A10) and (A11)
of Ref.~\cite{KAM}\footnote{We have corrected a sign error for
the $H\chi\chi$ coupling in (A10) of Ref.~\cite{KAM}.}
to compute the spin-independent cross section for all the elements
included. In addition, for capture by the Sun, we also evaluate
the spin-dependent cross section due to both $Z$-boson exchange
and squark exchange for the scattering from hydrogen according
to Eq.~(A5) (EMC model case) of Ref.~\cite{KAM}. It should be noted that in all
these expressions the squarks were assumed to be degenerate. In the two
supergravity models that we consider here this need
not be the case, although for most of the parameter space this
is a fairly good approximation. Hence, we simply
use the average squark mass $m_{\tilde q}$ in this part of the
calculation.

The kinematic factor $X_i$ in Eq.~(\ref{eq:nc}), can be most
accurately evaluated once the detailed knowledge of the mass
density profile as well as the local escape velocity profile
are specified for all the elements.
In practice, this can be done by performing a numerical integration
with the physical inputs provided by standard solar model or some
sort of Earth model. Instead of performing such an involved calculation,
we approximate the integral for each element by the value of the
integrand obtained with the average effective gravitational
``potential energy'' $\phi_i$ times the integral volume. The
values of $\phi_i$ are taken from Table A.1 of Ref.~\cite{GGR91}.

\subsubsection{The detection rate}
\label{sec:det}
We next describe the procedure employed by us to calculate the
detection rate of upwardly-moving muons, resulting from the
particle production and interaction subsequent to the capture
and annihilation of neutralinos in the two supergravity models.
The annihilation process normally reaches equilibrium with the capture
process on a time scale much shorter than the age of the Sun or Earth.
We assume this is the case, so that the neutralino annihilation rate
equals half of the capture rate.
The detection rate for neutrino-induced
upwardly-moving muon events is then given by
\begin{equation}
\Gamma={C\over 8\pi R^2}\sum_{i,F}{D_iB_F\int
\left({dN\over dE_\nu}\right)_{iF}{E^2_\nu}dE_{\nu}}.\label{eq:nd}
\end{equation}
In Eq.~(\ref{eq:nd}), $D_i$ is a constant, $R$ is the distance
between the detector and the Sun or the center of the Earth,
and $(dN/dE_\nu)_{iF}$ is the differential energy spectrum of
neutrino type $i$ as it emerges at the surface of the Sun or Earth
due to the annihilation of neutralinos in the core of the Sun or Earth
into final state $F$ with a branching ratio $B_F$. It should be noted
that in Eq.~(\ref{eq:nd}) that $i$ is summed over muon neutrinos and
anti-neutrinos, and that $F$ is summed over final states that contribute
to the high-energy neutrinos. The only relevant fermion pair
final states are $\tau{\bar\tau}$, $c{\bar c}$, $b{\bar b}$, and (for
the $SU(5)\times U(1)$ model) $t{\bar t}$ when $m_\chi > m_t$. The
lighter fermions do not produce high-energy neutrinos since
they are stopped by the solar or terrestrial media before
they can decay \cite{RS}.

The branching ratio $B_F$ can be easily calculated as the relative magnitude
of the thermal-averaged product of annihilation cross section into final
state $F$ ($\sigma_F$) with the M\o ller velocity $v_M$.
Since the core temperatures of
the Sun and Earth are very low compared with the neutralino mass
($T_{\rm Sun} \sim 1.34\times 10^{-6}$ GeV;
$T_{\rm Earth} \sim 4.31\times 10^{-10}$ GeV), only the $s$-wave
contributions are relevant, hence it is enough here to use the
usual thermal average expansion up to zero-order of $T/m_\chi$
($v_M \rightarrow 0$ limit), {\it i.e.}
\begin{equation}
B_F={\langle{\sigma_F v_M}\rangle\over
\langle{\sigma_{\rm tot} v_M}\rangle}={a_F\over a_{\rm tot}}.
\label{eq:bratio}
\end{equation}
In Eq.~(\ref{eq:bratio}), {\it all} kinematically allowed final
states contribute to $a_{\rm tot}$. Besides all the fermion
pair final states, we have also included boson pair final states
$WW$, $ZZ$ and $hA$ in our calculation of $B_F$.
Due to the parameter space constraints, these channels are not
open for the $SU(5)$ model. But $WW$ and $ZZ$ channels
are generally open in the $SU(5)\times U(1)$ model, and the $hA$ channel
also opens up for large values of $\tan\beta$ in the
dilaton case. We should also remark that the annihilation channel
into lightest CP-even higgs pair $hh$ is always allowed kinematically
in some portion of the parameter space for both supergravity models
we consider, but since its $s$-wave contribution vanishes,
we do not include it in the calculation of $B_F$. However,
this channel is taken into account in the calculation of the
neutralino relic density, which does affect the capture
rate through the scaling of local density $\rho_\chi$
when $\Omega_\chi h^2_0 < 0.05$ (see Sec.~\ref{sec:cap}).
In addition, we have kept all the nonvanishing
interference terms in the evaluation of $a_{\rm tot}$ and $a_F$.

The calculation of the neutrino differential energy
spectrum is somewhat involved, since it requires a reasonably accurate
tracking of the cascade of the particles which result from neutralino
annihilation into each of the final state $F$. This involves the decay
and hadronization of the various annihilation products and their
interactions with the media of the Sun or the Earth's cores.
In addition, at high energies, neutrinos interact
with and are absorbed by  solar matter, a fact that affects the spectrum.
In Ref.~\cite{RS}, Ritz and Seckel rendered this calculation tractable
by their adaptation of the Lund Monte Carlo for this purpose. Subsequently,
analytic approximations to the Monte Carlo procedure outlined in
their paper were refined and employed by Kamionkowski to calculate
the neutrino energy spectra from neutralino annihilation for
the MSSM in Ref.~\cite{KAM}. The procedure involved is described in detail
in Ref.~\cite{NT}.

\subsubsection{Results and Discussion}
\label{sec:result}
For each point in the parameter spaces of the supergravity models we consider,
we have determined the relic abundance of neutralinos and then computed the
capture rate in the Sun and Earth (as described in Sec.~\ref{sec:cap}) and
the resulting upwardly-moving muon detection rate (as described in
Sec.~\ref{sec:det}). In Fig.~\ref{NT2}, the predicted detection rates
in the $SU(5)$ supergravity model are shown,
based on the assumption that the mass of the triplet higgsino,
which mediates dimension-five proton decay, obeys $M_{{\tilde H}_3}<3M_U$.
We have redone the calculation relaxing this assumption to
$M_{{\tilde H}_3}<10M_U$, in which case the results for the muon fluxes
remain qualitatively the same, except that the parameter space is opened
up somewhat. The dashed lines in Fig.~\ref{NT2} represent the current
Kamiokande 90\% C.L. upper limits Eqs.~(\ref{eq:Sup},\ref{eq:Eup}).
Similarly, the predictions for $SU(5)\times U(1)$ supergravity are presented in
Fig.~\ref{NT4} (\ref{NT6}) for the no-scale (dilaton) scenario, again along
with the Kamiokande upper limits (dashed lines). In Figs.~\ref{NT4},\ref{NT6},
we have taken the representative value of $m_t=150$~GeV. Similar results are
obtained for other values of $m_t$.

\begin{figure}[t]
\vspace{4.7in}
\includegraphics{NT2.ps}
\caption{\baselineskip=12pt
The upwardly-moving muon flux in underground detectors originating
from neutralino annihilation in the Sun and Earth, as a function of the
neutralino mass in the $SU(5)$ supergravity model. The dashed
lines represent the present Kamiokande 90\% C.L. experimental upper limits.}
\label{NT2}
\end{figure}
\begin{figure}[p]
\vspace{4.0in}
\includegraphics{NT4.ps}
\vspace{-0.3in}
\caption{\baselineskip=12pt
The upwardly-moving muon flux in underground detectors originating
from neutralino annihilation in the Sun and Earth, as a function of the
neutralino mass in no-scale $SU(5)\times U(1)$ supergravity. The
representative value of $m_t=150$~GeV has been used. The dashed lines
represent the present Kamiokande 90\% C.L. experimental upper limits.}
\label{NT4}
\vspace{4.1in}
\includegraphics{NT6.ps}
\vspace{-0.3in}
\caption{Same as Fig.~31 but for dilaton $SU(5)\times U(1)$ supergravity.}
\label{NT6}
\end{figure}

Several comments on these figures are in order. First, the kinematic
enhancement of the capture rate by the Earth manifests itself
in all figures as the big peaks near the Fe mass ($m_{\rm Fe}=52.0$ GeV),
as well as the smaller peaks around Si mass ($m_{\rm Si}=26.2$ GeV).
Second, there is a severe depletion of the rates near
$m_\chi={1\over 2} M_Z$ in Figs.~\ref{NT4},\ref{NT6}, which is
due to the decrease in the neutralino relic density. In the case of
Earth capture, this effect is largely compensated by the
enhancement near the Fe mass. As mentioned in Sec.~\ref{sec:cap},
in our procedure, the relic density affects the local
neutralino density $\rho_\chi$ only if $\Omega_\chi h^2_0 < 0.05$,
while in the $SU(5)$ model this almost never happens,
therefore, the effect of the $Z$-pole is not very evident in Fig.~\ref{NT2}.
Also, in Figs.~\ref{NT4},\ref{NT6}, the various
dotted curves correspond to different values of $\tan\beta$,
starting from the bottom curve with $\tan\beta = 2$, and increasing
in steps of two. These curves clearly show that the capture and
detection rates increase with increasing $\tan\beta$, since
the dominant piece of the coherent neutralino-nucleon scattering
cross section via the exchange of the lightest Higgs boson $h$ is
proportional to $(1+{\tan}^2\beta)$. The capture rate decreases
with increasing $m_\chi$, since the scattering cross section
falls off as $m^{-4}_h$ and $m_h$ increases with increasing $m_\chi$.
It is expected that the detection rate in general also decreases
for large value of $m_\chi$, since it is proportional to the capture
rate. However, the opening of new annihilation channels,
such as the $WW$, $ZZ$ and $hA$ channels in $SU(5)\times U(1)$ supergravity,
could have two compensating effects on the detection rate: (a) the
presence of a new channel to produce high-energy neutrinos
which leads to an enhancement of the detection rate; and (b) the
decrease of the branching ratios for the fermion pair channels,
which makes the neutrino yield from $\tau{\bar\tau}$,
$c{\bar c}$ and $b{\bar b}$ smaller and hence reduces the
detection rate. Therefore, these new annihilation channels
could {\it either} increase {\it or} decrease the detection
rate, depending which of these two effects wins over.
We found that, for small values of $\tan\beta$ and $\mu < 0$,
the $WW$ channel can become dominant if open,
basically because in this case the neutralino contains a rather
large neutral wino component. This explains the distortion of
the detection rate curves in the $\mu < 0$ half of Figs.~\ref{NT4}
and \ref{NT6}. The effect of the $ZZ$ channel turns out to be
negligible in $SU(5)\times U(1)$ supergravity, since neutralinos
with $m_\chi > M_Z$ have very small higgsino components. The same
argument applies to the $hA$ channel which sometimes opens up
in the dilaton case for rather large values of $\tan\beta$.

In the dilaton case, for large values of $\tan\beta$
the CP-odd Higgs boson $A$ can be rather light, and
the presence of the $A$-pole when $m_\chi \sim {1\over 2}m_A$
makes the relic density very small. $\Omega_\chi h^2_0$ as a function
of $m_\chi$, is first lower than $0.05$, it increases with $m_\chi$,
and eventually reaches values above $0.05$, when neutralinos move away
from the $A$-pole. Thus, the capture and detection rates also show
this behavior, which can be seen as the few ``anomalous'' lines in
Fig.~\ref{NT6}. For lower values of $\tan\beta$,
$\Omega_\chi h^2_0 < 0.05$, and there is no such effect.
In the $SU(5)$ model, since the allowed points include
different supersymmetry breaking scenarios and several values
of $m_t$ and $\tan\beta$, all these features are blurred.
Nonetheless, in the same range of $m_\chi$ and for same
values of $\tan\beta$ and $m_t$, we have found the results
of these two models comparable, with the rates in the
$SU(5)\times U(1)$ model slightly smaller due to the smaller
relic density.

\begin{figure}[p]
\vspace{4.0in}
\includegraphics{NT7.ps}
\vspace{-0.3in}
\caption{\baselineskip=12pt
The allowed parameter space of no-scale $SU(5)\times U(1)$
supergravity (in the $(m_{\chi^\pm_1},\tan\beta$) plane) after the
present ``neutrino telescopes" (NT) constraint has been applied. Two values
of $m_t$ (130,150 GeV) have been chosen. The crosses denote those points which
could be probed with an increase in sensitivity by a factor of two.}
\label{NT7}
\vspace{4.1in}
\includegraphics{NT9.ps}
\vspace{-0.3in}
\caption{Same as Fig.~33 but for dilaton $SU(5)\times U(1)$ supergravity.}
\label{NT9}
\end{figure}

It is clear that at present the experimental constraints
from the ``neutrino telescopes'' on the parameter space
of the two supergravity models are quite weak.
In fact, only the Kamiokande upper bound from the Earth
can be used to exclude regions of the parameter
space with $m_\chi \approx m_{\rm Fe}$ for both models,
in particular for the $SU(5)\times U(1)$ model, due to
the enhancement effect discussed above. However, it is our belief
that the results presented in this paper will be quite useful in
the future, when improved sensitivity in underground muon detection
rates become available. An improvement in experimental sensitivity
 by a factor of two should be easily possible when MACRO \cite{mac} goes
into operation, while a ten-fold improvement is envisaged when
Super-Kamiokande \cite{sk} announces its results sometime by the
end of the decade. More dramatic improvements in the sensitivity
(by a factor of 20--100) may be expected from DUMAND and
AMANDA \cite{amd}, currently under construction.
In addition, as recently argued \cite{hal},
we think that perhaps the full parameter space of a large class
of supergravity models, including the two specific ones considered here,
may only be convincingly probed by a detector with an effective
area of 1 km$^2$. It is interesting to note that, with a sensitivity
improvement by a factor of 100, a large portion of the $\mu < 0$
half parameter space of the $SU(5)$ model can be probed.
Unfortunately, the remaining portion, with fluxes below $\sim10^{-4}$,
can hardly be explored by underground experiments in the foreseeable future.
For $SU(5)\times U(1)$ supergravity, in Figs.~\ref{NT7}
and \ref{NT9} we have plotted the allowed points in the
$(m_{\chi^\pm_1},\tan\beta)$ space. These points are those obtained originally
in Refs.~\cite{LNZI,LNZII}, such that the neutrino telescopes constraint is
also satisfied; no other constraints have been imposed. It can be seen here
that the small voids of points for $m_{\chi^\pm_1}\approx100$~GeV and a variety
of values of $\tan\beta$ are excluded by the constraint from the ``neutrino
telescopes". In these figures we have marked by crosses the points in the
$(m_{\chi^\pm_1},\tan\beta)$ plane that MACRO should be able to probe (assuming
an increase in the sensitivity by a factor of two) for the no-scale and dilaton
scenarios respectively. As expected, the  constraints from future
``neutrino telescopes'' will be strictest for large values
of $\tan\beta$.

\section{Detailed calculations for indirect experimental detection}
\label{indirect}
\subsection{Flavour Changing Neutral Current (FCNC) ($b \rightarrow s \gamma$)}
There has recently been a renewed surge of interest on the
flavor-changing-neutral-current (FCNC) $b\to s\gamma$ decay, prompted by the
CLEO 95\% CL allowed range \cite{Thorndike}
\beq
{\rm BR}(b\to s\gamma)=(0.6-5.4)\times10^{-4}.
\eeq
Since the Standard Model (SM) prediction looms around $(2-5)\times10^{-4}$
depending on the top-quark mass ($m_t$), a reappraisal of beyond the SM
contributions has become topical \cite{Barger,BG,bsgamma,bsg-eps,Oshimo}.  We
use the following expression for the branching ratio $b\to s\gamma$ \cite{BG}
\beq
{{\rm BR}(b\to s\gamma)\over{\rm BR}(b\to ce\bar\nu)}={6\alpha\over\pi}
{\left[\eta^{16/23}A_\gamma
+\coeff{8}{3}(\eta^{14/23}-\eta^{16/23})A_g+C\right]^2\over
I(m_c/m_b)\left[1-\coeff{2}{3\pi}\alpha_s(m_b)f(m_c/m_b)\right]},
\eeq
where $\eta=\alpha_s(M_Z)/\alpha_s(m_b)$, $I$ is the phase-space factor
$I(x)=1-8x^2+8x^6-x^8-24x^4\ln x$, and $f(m_c/m_b)=2.41$ the QCD
correction factor for the semileptonic decay. The $A_\gamma,A_g$ are the
coefficients of the effective $bs\gamma$ and $bsg$ penguin operators
evaluated at the scale $M_Z$. Their simplified expressions are given in
the Appendix of Ref.~\cite{BG}, where the gluino and neutralino contributions
have been justifiably neglected \cite{Bertolini} and the squarks are considered
degenerate in mass, except for the $\tilde t_{1,2}$ which are significantly
split by $m_t$.

For the $SU(5)$ supergravity model we find \cite{bsgamma}
\beq
2.3\,(2.6)\times10^{-4}<{\rm BR}(b\to
s\gamma)_{minimal}<3.6\,(3.3)\times10^{-4},
\eeq
for $\mu>0\,(\mu<0)$, which are all within the CLEO limits. One can show
that ${\rm BR}(b\to s\gamma)$ would need to be measured with better than $20\%$
accuracy to start disentangling the $SU(5)$ supergravity model from the
SM. In Fig.~\ref{Figure15} we present the analogous results for the
$SU(5)\times U(1)$ supergravity cases: no-scale (top row) and strict no-scale
(bottom row). The results are strikingly different than in the prior case.  One
observes that part of the parameter space is actually excluded by the new CLEO
bound, for a range of sparticle masses. Perhaps the most surprising feature of
the results is the strong suppression of ${\rm BR}(b\to s\gamma)$ which occurs
for a good portion of the parameter space for $\mu>0$. It has proven to be
non-trivial to find a simple explanation for the observed cancellation. We
discuss in Ref.~\cite{bsg-eps} the implications of this indirect constraint on
the prospects for direct experimental detection of these models.

\begin{figure}[t]
\vspace{4.7in}
\includegraphics{proc_summer15.ps}
\vspace{-0.3in}
\caption{\baselineskip=12pt
The calculated values of ${\rm BR}(b\to s\gamma)$ versus the chargino mass in
no-scale (top row) and strict no-scale (bottom row, for the indicated
values of $m_t$) $SU(5)\times U(1)$ supergravity. Note the fraction of
parameter space excluded by the CLEO allowed range (in between the dashed
lines).}
\label{Figure15}
\end{figure}

\subsection{Anomalous magnetic moment of the muon}
The long standing experimental values of $a_\mu$ for each sign of the muon
electric charge \cite{oldg} can be averaged to yield \cite{kinoII}
\beq
a^{exp}_\mu=1\ 165\ 923(8.5)\times10^{-9}.
\eeq
The uncertainty on the last digit is indicated in parenthesis. On the other
hand, the various standard model contributions to $a_\mu$ have been estimated
to be as follows \cite{kinoII}
\begin{eqnarray}
{\rm QED:}&&1\ 165\ 846\ 984(17)(28)\times10^{-12}\\
{\rm had.1:}&&7\ 068(59)(164)\times10^{-11}\\
{\rm had.2:}&&-90(5)\times10^{-11}\\
{\rm had.3:}&&49(5)\times10^{-11}\\
{\rm Total\ hadronic:}&&7\ 027(175)\times10^{-11}\\
{\rm Electroweak:}&&195(10)\times10^{-11}
\end{eqnarray}
Here had.1 is the hadronic vacuum-polarization contribution to the second-order
muon vertex, had.2 comes from higher-order hadronic terms,
and had.3 from the hadronic light-by-light scattering. The total standard model
prediction is then \cite{kinoII}
\beq
a^{SM}_\mu=116\ 591\ 9.20(1.76)\times10^{-9}.
\eeq
Subtracting the experimental result gives \cite{kinoII}
\beq
a^{SM}_\mu-a^{exp}_\mu=-3.8(8.7)\times10^{-9},\label{diff}
\eeq
which is perfectly consistent with zero. The uncertainty in the theoretical
prediction is dominated by the uncertainty in the lowest order hadronic
contribution (had.1), which ongoing experiments at Novosibirsk hope to reduce
by a factor of two in the near future. This is an important preliminary step
to testing the electroweak contribution, which is of the same order. The
uncertainty in the experimental determination of $a_\mu$ is expected to be
reduced significantly (down to $0.4\times10^{-9}$) by the new E821 Brookhaven
experiment \cite{newg}, which is scheduled to start taking data in late 1994.
Any beyond-the-standard-model contribution to $a_\mu$ (with presumably
negligible uncertainty) will simply be added to the central value in
Eq.~(\ref{diff}). Therefore, we can obtain an allowed interval for any
supersymmetric contribution, such that $a^{susy}_\mu+a^{SM}_\mu-a^{exp}_\mu$ is
consistent with zero at some given confidence level,
\begin{eqnarray}
-4.9\times10^{-9}&<a^{susy}_\mu&<12.5\times10^{-9},\qquad{\rm at\ 1\sigma};\\
-10.5\times10^{-9}&<a^{susy}_\mu&<18.1\times10^{-9},\qquad{\rm at\ 90\%CL};\\
-13.2\times10^{-9}&<a^{susy}_\mu&<20.8\times10^{-9},\qquad{\rm at\
95\%CL}.\label{bounds}
\end{eqnarray}

The supersymmetric contributions to $a_\mu$ have been computed to various
degrees of completeness and in the context of several models, including
the minimal supersymmetric standard model (MSSM)
\cite{Fayet,GM,KKS,YACN,Vendramin}, an $E_6$ string-inspired model \cite{E6g},
and a non-minimal MSSM with an additional singlet \cite{NMSSMg,abel}. Because
of the large number of parameters appearing in the typical formula for
$a^{susy}_\mu$, various contributions have often been neglected and numerical
results are basically out of date. More importantly, a contribution which is
roughly proportional to the ratio of Higgs vacuum expectation values
($\tan\beta$), even though known for a while \cite{KKS,YACN,Vendramin,abel},
has to date remained greatly unappreciated. This has been the case because in
the past only small values of $\tan\beta$ were usually considered and the
enhancement of $a^{susy}_\mu$, which is the focus of this section, was not
evident. In fact, such enhancement can easily make $a^{susy}_\mu$ run in
conflict with the bounds given in Eq.~(\ref{bounds}), even after the LEP lower
bounds on the sparticle masses are imposed. In Ref.~\cite{g-2} a reappraisal
of this calculation has been done in the context of $SU(5)\times U(1)$
supergravity.

There are two sources of one-loop supersymmetric contributions to $a_\mu$: (i)
with neutralinos and smuons in the loop; and (ii) with charginos and sneutrinos
in the loop. The general
formula for the lowest order supersymmetric contribution to $a_\mu$ has been
given in Refs.~\cite{KKS,YACN,Vendramin,abel}. Here we use the expression in
Ref.~\cite{abel},
\begin{eqnarray}
a^{susy}_\mu=-\frac{g^2_2}{8\pi^2}
\Biggl\{\sum_{\chi^0_i,\tilde \mu_j} \frac{m_\mu}{m_{\chi^0_i}}
	&\biggl[&\!\!\!\!(-1)^{j+1}\sin(2\theta)B_1(\eta_{ij})\tan\theta_W N_{i1}
		[\tan\theta_W N_{i1}+N_{i2}]\nonumber\\
&+&\frac{m_\mu}{2M_W\cos\beta}
	B_1(\eta_{ij})N_{i3}[3\tan\theta_W N_{i1}+N_{i2}]\nonumber\\
&+&\left(\frac{m_{\mu}}{m_{\chi^0_i}}\right)^{2}A_1(\eta_{ij})
\Bigl\{\coeff{1}{4}[\tan\theta_W N_{i1}+N_{i2}]^{2}+
[\tan\theta_W N_{i1}]^{2}\Bigr\}\biggr]\nonumber\\
 &&\hskip-1.3in -\sum_{\chi_j^\pm}\left[\frac{m_{\mu}m_{\chi_j^\pm}}
{m_{\tilde \nu}^2}
\frac{m_{\mu}}{\sqrt{2}M_W\cos\beta} B_2(\kappa_j)V_{j1}U_{j2}+
\left(\frac{m_{\mu}}
{m_{\tilde\nu}}\right)^{2}\frac{A_1(\kappa_j)}{2}V^2_{j1}\right]\Biggr\}.
\label{formula}
\end{eqnarray}
where $N_{ij}$ are elements of the matrix which diagonalizes the neutralino
mass matrix, and $U_{ij},V_{ij}$ are the corresponding ones for the chargino
mass matrix, in the notation of Ref.~\cite{HK}. Also,
\beq
\eta_{ij}=\left[1
-\left(\frac{m_{\tilde\mu_j}}{m_{\chi^0_i}}\right)^2\right]^{-1},\qquad
\kappa_j=\left[1
-\left(\frac{m_{\chi^\pm_j}}{m_{\tilde\nu}}\right)^2\right]^{-1},
\eeq
and
\begin{eqnarray}
B_1(x)&=&x^2-\coeff{1}{2}x+x^2(x-1)\ln\left(\frac{x-1}{x}\right),\\
A_1(x)&=&x^3-\coeff{1}{2}x^2-\coeff{1}{6}x+x^3(x-1)\ln\left(\frac{x-1}{x}\right),\\
B_2(x)&=&-x^2-\coeff{1}{2}x-x^3\ln\left(\frac{x-1}{x}\right).
\end{eqnarray}

As has been pointed out, the mixing angle of the smuon eigenstates is small
(although it can be enhanced for large $\tan\beta$) and it makes the
neutralino-smuon contribution suppressed. Moreover, the various
neutralino-smuon contributions (the first three lines
in Eq.~(\ref{formula})) tend to largely cancel among themselves
\cite{Vendramin}.\footnote{The original Fayet formula \cite{Fayet} is obtained
from the third neutralino-smuon contribution in the limit of a massless
photino and no smuon mixing.} This means that the chargino-sneutrino
contributions (on the fourth line in Eq.~(\ref{formula})) will likely be the
dominant ones. In fact, as we stress in this paper, the first
chargino-sneutrino contribution (the ``gauge-Yukawa" contribution) is enhanced
relative to the second one (the ``pure gauge" contribution) for large values
of $\tan\beta$. This can be easily seen as follows.

Picturing the chargino-sneutrino one-loop diagram, with the photon being
emitted off the chargino line, there are two ways in which the helicity of the
muon can be flipped, as is necessary to obtain a non-vanishing $a_\mu$:
\begin{description}
\item (i) It can be flipped by an explicit muon mass insertion on one of the
external muon lines, in which case the coupling at the vertices is between a
left-handed muon, a sneutrino, and the wino component of the chargino and has
magnitude $g_2$. It then follows that $a_\mu$ will be proportional to
$g^2_2(m_\mu/\tilde m)^2 |V_{j1}|^2$, where $\tilde m$ is a supersymmetric mass
in the loop and the $V_{j1}$ factor picks out the wino component of the $j$-th
chargino. This is the origin of the ``pure gauge" contribution to
$a^{susy}_\mu$.
\item (ii) Another possibility is to use the muon Yukawa coupling on one of
the vertices, which flips the helicity and couples to the Higgsino component
of the chargino. One also introduces a chargino mass insertion to switch to the
wino component and couple with strength $g_2$ at the other vertex. The
contribution is now proportional to $g_2\lambda_\mu(m_\mu m_{\chi^\pm_j}/\tilde
m^2)V_{j1}U_{j2}$, where $U_{j2}$ picks out the Higgsino component of the
$j$-th chargino. The muon Yukawa coupling is given by $\lambda_\mu=g_2
m_\mu/(\sqrt{2}M_W\cos\beta)$. This is the origin of the gauge-Yukawa
contribution to $a^{susy}_\mu$.
\end{description}
The ratio of the ``pure gauge" to the ``gauge-Yukawa" contributions is
roughly then
\beq
g^2_2\,(m_\mu/\tilde m)/(g_2\lambda_\mu)\sim g_2/\sqrt{1+\tan^2\beta},
\eeq
for $\tilde m\sim100\GeV$. Thus, for small $\tan\beta$ both contributions
are comparable, but for large $\tan\beta$ the ``gauge-Yukawa" contribution
is greatly enhanced.\footnote{A similar enhancement in the second
neutralino-smuon contribution is suppressed by small Higgsino admixtures
(\ie, $|N_{13}|,|N_{23}|\ll1$).} This phenomenon was first noticed in
Ref.~\cite{KKS}. It is interesting to note that an analogous $\tan\beta$
enhancement also occurs in the $b\to s\gamma$ amplitude \cite{bsgamma},
although its effect is somewhat obscured by possible strong cancellations
against the QCD correction factor.

\begin{figure}[p]
\vspace{4.0in}
\includegraphics{g-2_1a.ps}
\vspace{4.1in}
\includegraphics{g-2_1b.ps}
\vspace{-0.3in}
\caption{\baselineskip=12pt
The supersymmetric contribution to the muon anomalous magnetic
moment in  no-scale and dilaton $SU(5)\times U(1)$ supergravity, plotted
against the gluino mass for the indicated values of $m_t$
and $\tan\beta$ (which increase in steps of two). The dashed lines represent
the 95\%CL experimentally allowed range.}
\label{g-2}
\end{figure}

The results of the calculation in the no-scale and dilaton cases \cite{g-2} are
plotted in Fig.~\ref{g-2} respectively, against the gluino mass, for the
indicated values of $m_t$. As anticipated, the values of $\tan\beta$ increase
as the corresponding curves move away from the zero axis. Note that
$a^{susy}_\mu$ drops off faster than naively expected (\ie, $\propto1/m_{\tilde
g}$) since the $U_{12}$ mixing element decreases as the limit of pure wino and
Higgsino is approached for large $m_{\tilde g}$. Note also that $a^{susy}_\mu$
can have either sign, in fact, it has the same sign as the Higgs mixing
parameter $\mu$.\footnote{For comparison with earlier work, our sign convention
for $\mu$ is opposite to that in Ref.~\cite{HK}.} The incorrect perception that
$a^{susy}_\mu$ is generally negative appears to be based on several model
analyses where either $\mu$ was chosen to be negative or only some of the
neutralino-smuon pieces were kept (which are mostly negative). Interestingly,
the largest allowed values of $\tan\beta$ do not exceed the $a_\mu$ constraint
since consistency of the models (\ie, the radiative breaking constraint)
requires larger gluino masses as $\tan\beta$ gets larger.

It is hard to tell what will happen when the E821 experiment reaches its
designed accuracy limit. However, one point should be quite clear, the
supersymmetric contributions to $a_\mu$ can be so much larger than the present
hadronic uncertainty ($\approx\pm1.76\times10^{-9}$) that the latter is
basically irrelevant for purposes of testing a large fraction of the allowed
parameter space of the models. This is not true for the electroweak
contribution and will also not hold for small values of $\tan\beta$. Should the
actual measurement agree very well with the standard model contribution, then
either $\tan\beta\sim1$ or the sparticle spectrum would need to be in the TeV
range. This situation is certainly a window of opportunity for sparticle
detection before LEPII starts operating. Moreover, a significant portion of the
explorable parameter space (those points with $m_{\chi^\pm_1}\gsim100\GeV$
and equivalently $m_{\tilde g}\gsim350\GeV$) is in fact beyond the reach of
LEPII.

\subsection{$\bf{\epsilon_{1,2,3,b}}$}
A complete study of one-loop electroweak radiative corrections in the
supergravity models we consider here has been made in
Refs.~\cite{ewcorr,bsg-eps,eps1-epsb}. These calculations include some recently
discovered important $q^2$-dependent effects, which occur when light charginos
($m_{\chi^\pm}\lsim 60-70\GeV$) are present \cite{BFC}, and lead to strong
correlations between the chargino and the top-quark mass. Specifically, one
finds that at present the $90\%$ CL upper limit on the top-quark mass is
$m_t\lsim175\GeV$ in no-scale $SU(5)\times U(1)$ supergravity. These bounds can
be strengthened for increasing chargino masses in the $50-100\GeV$ interval.
For example, in$SU(5)\times U(1)$ supergravity, for
$m_{\chi^\pm_1}\gsim60\,(70)\GeV$, one finds $m_t\lsim165\,(160)\GeV$. As
expected, the heavy sector of both models decouples quite rapidly with
increasing sparticle masses, and at present, only $\epsilon_1$ leads to
constraints on the parameter spaces of these models. For future reference, it
is important to note that global SM fits to all of the low-energy and
electroweak data \cite{PDG} constrain $m_t<194,178,165\GeV$ for
$m_{H_{SM}}=1000,250,50\GeV$ at the $90\%$ CL respectively.

One can show that an expansion of the vacuum polarization tensors to order
$q^2$, results in three independent physical parameters. In the first
scheme introduced to study these effects \cite{PT}, namely the $(S,T,U)$
scheme, a SM reference value for $m_t,m_{H_{SM}}$ is used, and the deviation
from this reference is calculated and is considered to be ``new" physics. This
scheme is only valid to lowest order in $q^2$, and is therefore not applicable
to a theory with new, light $(\sim M_Z)$ particles. In the supergravity
models we consider here, each point in parameter space is actually a distinct
model, and a SM reference point is not meaningful. For these reasons, in
Ref.~\cite{ewcorr} the scheme of Refs. \cite{AB,BFC} was chosen, where the
contributions are {\it absolute} and valid to higher order in $q^2$. This
is the so-called $\epsilon_{1,2,3}$ scheme. More recently, a new observable
($\epsilon_b$), which parametrizes the one-loop vertex corrections to the $Z\to
b\bar b$ vertex, has been added to this set \cite{ABC,ABCII}. Regardless of the
scheme used, all of the global fits to the three physical parameters are {\it
entirely consistent} with the SM at $90\%$ CL.

With the assumption that the dominant ``new" contributions arise from the
process-independent (\ie, ``oblique") vacuum polarization amplitudes, one
can combine several observables in suitable ways such that they are most
sensitive to new effects. It is important to note that not all observables
can be included in the experimental fits which determine the $\epsilon_{1,2,3}$
paremeters, if only the oblique contributions are kept \cite{ewcorr}. The
most important non-oblique corrections are encoded in $\epsilon_b$

It is well known in the MSSM that the largest contributions to $\epsilon_1$
(\ie, $\delta\rho$ if $q^2$-dependent effects are neglected) are expected to
arise from the $\tilde t$-$\tilde b$ sector, and in the limiting case of a very
light stop, the contribution is comparable to that of the $t$-$b$ sector
\cite{DH}. The remaining squark, slepton, chargino, neutralino, and Higgs
sectors all typically contribute considerably less. For increasing sparticle
masses, the heavy sector of the theory decouples, and only SM effects  with a
{\it light} Higgs survive. However, for very light chargino, a $Z$-wavefunction
renormalization threshold effect can introduce a substantial $q^2$-dependence
in the calculation, thus modifying significantly the standard $\delta\rho$
results.  For completeness, in Ref.~\cite{ewcorr} the complete vacuum
polarization contributions from the Higgs sector, the supersymmetric
chargino-neutralino and sfermion sectors, and also the corresponding
contributions in the SM were included.

In Fig.~\ref{Figure16} we show the calculated values of $\epsilon_1$ versus the
lightest chargino mass ($m_{\chi^\pm_1}$) for the sampled points in the $SU(5)$
model and in no-scale $SU(5)\times U(1)$) supergravity. In the no-scale
$SU(5)\times U(1)$ case three representative values of $m_t$ were used,
$m_t=100,130,160\GeV$,
whereas in the $SU(5)$ case several other values for $m_t$ in the range
$90\GeV\le m_t\le 160\GeV$ were sampled. In both models, but most clearly in
the no-scale model one can see how quickly the sparticle spectrum decouples as
$m_{\chi^\pm_1}$ increases, and the value of $\epsilon_1$ asymptotes to the SM
value appropriate to each value of $m_t$ and for a {\it light} ($\sim 100
\GeV$) Higgs mass. The threshold effect of $\chi^\pm_1$ is manifest as
$m_{\chi^\pm_1}\rightarrow {1\over2} M_Z$ and is especially visible for $\mu<0$
in both models. The calculation of this effect is not expected to be very
accurate as $m_{\chi^\pm_1}\rightarrow {1\over 2} M_Z$. However, according to
Ref.~\cite{BFC}, for $m_{\chi^\pm_1}>50\GeV$, this estimate agrees to better
than $10\%$ with the result obtained in a more accurate way.

\begin{figure}[t]
\vspace{4.7in}
\includegraphics{proc_summer16.ps}
\vspace{-0.3in}
\caption{\baselineskip=12pt
The total contribution to $\epsilon_1$ as a function
of the lightest chargino mass $m_{\chi^\pm_1}$ for the $SU(5)$ model
(upper row) and the no-scale $SU(5)\times U(1)$ model (bottom row). Points
between the two horizontal solid lines are allowed at $90\%$ CL. The three
distinct curves (from lowest to highest) in the no-scale case correspond to
$m_t=100,130,160 \GeV$.}
\label{Figure16}
\end{figure}

Recent values for $\epsilon_{1,2,3}$ obtained from a global fit to the
LEP (\ie, $\Gamma_l,A^{l,b}_{FB},A^\tau_{pol})$ and $M_W/M_Z$ measurements
are \cite{BB},
\beq
\epsilon_1=(-0.9\pm
3.7)10^{-3},\quad\epsilon_2=(9.9\pm8.0)10^{-3},\quad\epsilon_3=(-0.9\pm
4.1)10^{-3}.
\eeq
For $\epsilon_1$ it is clear that virtually all the sampled points in the
$SU(5)$ supergravity model are within the $\pm 1.64 \sigma$
($90\%$ CL) bounds (denoted by the two horizontal solid lines in the figures).
Since several values for $90\le m_t\le 160\GeV$ were sampled, the trends for
fixed $m_t$ are not very clear from the figure. Nonetheless, the points
just outside the $1.64 \sigma$ line correspond to $m_t=160\GeV$, which
are therefore excluded at the $90\%$ CL. In the no-scale model, the upper bound
on $m_t$ depends sensitively on the chargino mass. For example, for
$m_t=160\GeV$, only light chargino masses would be acceptable at $90\%$ CL. In
fact,  in Ref.~\cite{ewcorr} the region $130\GeV\le m_t\le 190\GeV$ was
scanned in increments of $5\GeV$ and obtained the maximum values
for $m_{\chi^\pm_1}$ allowed by the experimental value for $\epsilon_1$ at
$90\%$ CL. There is a strong correlation between $m_t$ and $m_{\chi^\pm_1}$: as
$m_t$ rises, the upper limit to $m_{\chi^\pm_1}$ falls, and vice versa. In
particular, for $m_t\le150\GeV$ all values of $m_{\chi^\pm_1}$ are allowed,
while one could have $m_t$ as large as $160\,(175)\GeV$ for $\mu>0\,(\mu<0)$ if
the chargino mass were light enough.

\begin{figure}[p]
\vspace{5in}
\includegraphics{Easpects5a.ps}
\includegraphics{Easpects5b.ps}
\vspace{-0.1in}
\caption{\baselineskip=12pt
The correlated predictions for the $\epsilon_1$ and
$\epsilon_b$ parameters in units of $10^{-3}$ in the no-scale and dilaton
$SU(5)\times U(1)$ supergravity scenarios. The ellipses represent the
$1\sigma$, 90\%CL, and 95\%CL contours obtained from all LEP data. The values
of $m_t$ are as indicated.}
\label{ellipses}
\end{figure}

The calculated values of $\epsilon_b$ are shown in Fig.~\ref{ellipses} in
conjuction with the values of $\epsilon_1$, since these two observables are
the only ones which constrain supersymmetric models at present. The 1-$\sigma$
experimental ellipse is taken from Ref.~\cite{Altarelli}. Clearly smaller
values of $m_t$ fit the data better. The regions of the plot where the points
accumulate correspond to the Standard Model limit, with a light Higgs boson.

\section{The problem of mass and $m_t$}
\label{mass}
\subsection{Generalities}
The origin of mass is one of the most profound problems in Physics. Modern
Field Theories try to answer this problem by invoking spontaneous broken gauge
symmetry through the vacuum expectation value (vev) of an elementary (or
composite) scalar field, the Higgs field. In general the masses of all
particles (scalars, fermions, gauge bosons) are proportional to this (or these)
vev(s).  The proportionality coefficients are: the quartic couplings
($\lambda$) for the scalars, the Yukawa couplings ($y$) for the fermions, and
the gauge couplings ($g$) for the gauge bosons. Therefore we have (in
principle)
\begin{eqnarray}
	m_s &=& \lambda\vev{\rm vev},\\
	m_f &=& y \vev{\rm vev},\\
	m_g &=& g \vev{\rm vev}.
\end{eqnarray}
The above general picture looks convincingly simple, but its implementation in
realistic models is cumbersome.  There are several reasons that prevent us from
a complete and satisfactory solution, at present, of the mass problem.  The
quark and lepton mass spectrum, neglecting neutrinos, spans a range of at least
five orders of magnitude $m_e=0.5\MeV$ to $m_t\gsim 130\GeV$.
If we take as ``normal" the electroweak gauge boson masses, ${\cal O}(80-90)
\GeV$,  then a seemingly ``heavy" top quark ${\cal O}(150\GeV)$ looks perfectly
normal, while all other quark and lepton masses look peculiarly small.
Clearly, a natural theory cannot support fundamental Yukawa couplings extending
over five orders of magnitude.  The hope has always been that several of these
Yukawa couplings are {\em naturally zero} at the classical level, and then
quantum (radiative) corrections create a Yukawa coupling that reproduces
reality. A modern version of this programm has arisen in string theory and will
be discussed shortly.  For the moment, let us emphasize that, despite the
pessimism expressed above, certain features of the fermion mass spectrum has
been already ``explored" and have led to some spectacular predictions.  Namely,
in GUTs or SUSY-GUTs the difference between quark and lepton masses is
attributed to the strong interactions (QCD) that make the quarks much heavier
than the leptons (of the sa

me generation).

The successful prediction of the ratio $m_b/m_\tau\approx 2.9$  \cite{BEGN} has
also led to the highly correlated prediction of $N_f = 3$, much before any
primordial nucleosynthesis prediction. This was spectacularly confirmed at LEP
$N_f = 2.980 \pm 0.027$ \cite{LEPC}. Another rather generic feature of
supergravity GUTs, is their ability to trigger radiative spontaneous breaking
of the EW-Symmetry \cite{EWx,LN}, thus explaining naturally why
$m_W/m_{Pl}\approx 10^{-16}$. However, the triggering of this radiative
breaking is only possible when the theory contains a Yukawa coupling of the
order of ``$g$", i.e.
$y={\cal O}(g)$, which is naturally identified with the top-quark Yukawa
coupling.  In other words, in supergravity GUTs it is not only that ``heavy"
t-quark is natural but it is needed, if we want to have a dynamical
understanding of the EW-spontaneous breaking. Finally, string theory -- more
precisely its infrared limit -- which encompasses naturally supergravity GUTs,
is characterised by two features of relevance to us here (see \eg,
Refs.~\cite{revamp,decisive}):
\begin{enumerate}
\item Most of the Yukawa couplings are naturally zero at the tree level and
``pick up" non-zero values progressively at higher orders (through effective
``non-renormalizable" terms), consistent with the problem of fermion masses
observed in Nature.
\item Non-zero Yukawa couplings at tree level are automatically of ${\cal
O}(g)$.
\end{enumerate}

Once more a ``heavy" top quark is a natural possibility and for the first time
in string theory we may have a dynamical explanation of the origin of its large
Yukawa coupling, \ie, ${\cal O}(g)$.

\subsection{A working ground to predict $m_t$}
A ``working ground" for further discussing and testing the above ideas is
provided by $SU(5)\times U(1)$ supergravity, viewed as the infrared limit of
string theory.  The singularly unique properties of $SU(5)\times U(1)$ theory
has been repeatedly emphasized above. Here we will concentrate on its features
related to the specific purpose of predicting the best value for $m_t$.   As a
``descendant" of string theory, $SU(5)\times U(1)$ supergravity is
characterized by two basic features:
\begin{description}
\item a) A large top-Yukawa coupling: ${\cal O}(g)$.
\item b) The no-scale structure.
\end{description}
Notice that b) in conjunction with a), not only triggers EW-radiative breaking
but, in principle, may also determine dynamically the magnitude of the SUSY
breaking scale.  As discussed above, the ``allowable" ``free parameters" are
reduced to a ``minimal" number, compared with any other model available at
present.  This unique feature of $SU(5)\times U(1)$ supergravity allows to put
its ``predictions" under severe experimental scrutiny.
One of these ``predictions" is the best value of $m_t$.  This is obtained
exploring the dependence on $m_t$ of the ``allowable" parameters space in
$SU(5)\times U(1)$ supergravity, after all presently known constraints
(theoretical, phenomenological, cosmological) have been duly taken into
account, as described above. Interestingly enough, we find that the presently
available constraints, once considered all together simultaneously (not one at
a time) allow to put an upper bound on $m_t$ \cite{Easpects}. This result is
illustrated in Fig.~\ref{counts}. Note that for $m_t\gsim190\GeV$ no points in
parameter space are allowed anyway because the top-quark Yukawa coupling would
reach the Landau pole below the string scale \cite{aspects}.

    This bottom-up approach indicates an $m_t$-range which is entirely
consistent \cite{t-paper} with the top-bottom approach sketched above. This
result is best appreciated in a plot of the top-quark Yukawa coupling at the
string scale versus the top-quark mass, for various values of $\tan\beta$, as
shown in Fig.~\ref{t-paperfigure}. The dashed lines indicate typical string
model-building predictions for the Yukawa coupling.
\newpage

\begin{figure}[t]
\vspace{4.7in}
\includegraphics{counts.ps}
\vspace{-0.3in}
\caption{The number of allowed points in parameter space of no-scale and
dilaton $SU(5)\times U(1)$ supergravity as a function of $m_t$ when the
basic theoretical and experimental LEP constraints have been imposed
(``theory+LEP"), and when all known direct and indirect experimental
constraints have been additionally imposed (``ALL").}
\label{counts}
\end{figure}

\begin{figure}[t]
\vspace{4.7in}
\includegraphics{yuks.ps}
\vspace{-0.5in}
\caption{\baselineskip=12pt
The top-quark Yukawa coupling at the string scale in $SU(5)\times U(1)$
supergravity versus the top-quark mass for fixed values of $\tan\beta$ (larger
values of $\tan\beta$ overlap with the $\tan\beta=10$ curve). The dashed lines
indicate typical string-like predictions for the Yukawa coupling.}
\label{t-paperfigure}
\end{figure}

\section{Conclusions}
\label{conclusions}
We have reviewed the remarkable amount of work done during these last
few years in order to disentangle one of the most fascinating problems
we are confronted with: the existence of the Superworld. We know that
new physics must exist beyond the Standard Model and the Superworld is
the best candidate.  On the other hand the hope to simplify the enormous
variety of phenomena and have them describable in a theory with a very small
number of parameters -- possibly only one --  can only be provided by String
Theory. This is why we have chosen the two simplest possible supergravity
models: $SU(5)$, a representative of field theory; and $SU(5)\times U(1)$,
a representative of string theory. We have emphasized the relevance of
performing detailed calculations in order to put the predictions under
experimental tests with existing facilities. We have also pointed out that all
experimental data have to be used simultaneously in order to constrain the
parameter space of a theory. To appreciate the value of this procedure we show
one example in Fig.~\ref{PS} where all possible experimental constraints are
applied to the parameter space of $SU(5)\times U(1)$ supergravity in the
$(m_{\chi^\pm_1},\tan\beta)$ plane for $m_t=150\GeV$ in both no-scale and
dilaton scenarios.

The main properties of the two archetypical supergravity models discussed in
this review are shown in Tables~\ref{Table7} and \ref{Table4}  for the
$SU(5)$ supergravity model and $SU(5)\times U(1)$ supergravity, respectively.

\addcontentsline{toc}{section}{Acknowledgements}
\section*{Acknowledgements}
We would like to thank our various collaborators on the projects which are
described here: Franco Anselmo, Luisa Cifarelli, Raj Gandhi, Andre Petermann,
Gye Park, Heath Pois, Xu Wang, and Kajia Yuan. This work has been supported in
part by DOE grant DE-FG05-91-ER-40633.

\begin{table}[t]
\hrule
\caption{Major features of the $SU(5)$ supergravity model and its
spectrum (All masses in GeV).}
\label{Table7}
\begin{center}
\begin{tabular}{|l|}\hline
\hfil$\bf SU(5)$\hfil\\ \hline
$\bullet$ Not easily string-derivable, no known examples\\
$\bullet$ Symmetry breaking to Standard Model due to vevs of \r{24}\\
\quad and independent of supersymmetry breaking\\
$\bullet$ No simple doublet-triplet splitting mechanism\\
$\bullet$ Proton decay: $d=5$ operators large, strong constraints needed\\
$\bullet$ Baryon asymmetry ?\\ \hline
\end{tabular}
\end{center}
\begin{center}
\begin{tabular}{|l|l|}\hline
\hfil Spectrum\hfil\\ 	\hline
$\bullet$ Parameters 5: $m_{1/2},m_0,A,\tan\beta,m_t$\\
$\bullet$ Universal soft-supersymmetry breaking automatic\\
$\bullet$ $m_0/m_{1/2}>3$, $\tan\beta\lsim3.5$\\
$\bullet$ Dark matter: $\Omega_\chi h^2_0\gg1$, 1/6 of points excluded\\
$\bullet$ $m_{\tilde g}<400\GeV$,
$m_{\tilde q}>m_{\tilde l}>2m_{\tilde g}\gsim500\GeV$\\
$\bullet$ $m_{\tilde t_1}>45\GeV$\\
$\bullet$ $60\GeV<m_h<100\GeV$\\
$\bullet$ $2m_{\chi^0_1}\approx m_{\chi^0_2}\approx m_{\chi^\pm_1}\approx
0.28 m_{\tilde g}\lsim100$\\
$\bullet$ $m_{\chi^0_3}\sim m_{\chi^0_4}\sim m_{\chi^\pm_2}\sim\vert\mu\vert$
\\
$\bullet$ Chargino and Higgs easily accessible soon\\
\hline
\end{tabular}
\end{center}
\hrule
\end{table}

\begin{table}[p]
\hrule
\caption{Major features of $SU(5)\times U(1)$ supergravity and a comparison of
the two supersymmetry breaking scenaria considered. (All masses in GeV).}
\label{Table4}
\begin{center}
\begin{tabular}{|l|}\hline
\hfil$\bf SU(5)\times U(1)$\hfil\\ \hline
$\bullet$ Easily string-derivable, several known examples\\
$\bullet$ Symmetry breaking to Standard Model due to vevs of \r{10},\rb{10}\\
\quad and tied to onset of supersymmetry breaking\\
$\bullet$ Natural doublet-triplet splitting mechanism\\
$\bullet$ Proton decay: $d=5$ operators very small\\
$\bullet$ Baryon asymmetry through lepton number asymmetry\\
\quad (induced by the decay of heavy neutrinos) as processed by\\
\quad non-perturbative electroweak interactions\\
\hline
\end{tabular}
\end{center}
\begin{center}
\begin{tabular}{|l|l|}\hline
\hfil$\vev{F_M}_{m_0=0}$ (no-scale)\hfil&\hfil$\vev{F_D}$ (dilaton)\hfil\\
									\hline
$\bullet$ Parameters 3: $m_{1/2},\tan\beta,m_t$&
$\bullet$ Parameters 3: $m_{1/2},\tan\beta,m_t$\\
$\bullet$ Universal soft-supersymmetry&$\bullet$ Universal soft-supersymmetry\\
\quad breaking automatic&\quad breaking automatic\\
$\bullet$ $m_0=0$, $A=0$&$\bullet$ $m_0=\coeff{1}{\sqrt{3}}m_{1/2}$,
$A=-m_{1/2}$\\
$\bullet$ Dark matter: $\Omega_\chi h^2_0<0.25$
&$\bullet$ Dark matter: $\Omega_\chi h^2_0<0.90$\\
$\bullet$ $m_{1/2}<475\GeV$, $\tan\beta<32$
&$\bullet$ $m_{1/2}<465\GeV$, $\tan\beta<46$\\
$\bullet$ $m_{\tilde g}>245\GeV$, $m_{\tilde q}>240\GeV$
&$\bullet$ $m_{\tilde g}>195\GeV$, $m_{\tilde q}>195\GeV$\\
$\bullet$ $m_{\tilde q}\approx0.97m_{\tilde g}$
&$\bullet$ $m_{\tilde q}\approx1.01m_{\tilde g}$\\
$\bullet$ $m_{\tilde t_1}>155\GeV$
&$\bullet$ $m_{\tilde t_1}>90\GeV$\\
$\bullet$ $m_{\tilde e_R}\approx0.18m_{\tilde g}$,
$m_{\tilde e_L}\approx0.30m_{\tilde g}$
&$\bullet$ $m_{\tilde e_R}\approx0.33m_{\tilde g}$,
$m_{\tilde e_L}\approx0.41m_{\tilde g}$\\
\quad $m_{\tilde e_R}/m_{\tilde e_L}\approx0.61$
&\quad $m_{\tilde e_R}/m_{\tilde e_L}\approx0.81$\\
$\bullet$ $60\GeV<m_h<125\GeV$&$\bullet$ $60\GeV<m_h<125\GeV$\\
$\bullet$ $2m_{\chi^0_1}\approx m_{\chi^0_2}\approx m_{\chi^\pm_1}\approx
0.28 m_{\tilde g}\lsim290$
&$\bullet$ $2m_{\chi^0_1}\approx m_{\chi^0_2}\approx m_{\chi^\pm_1}\approx
0.28 m_{\tilde g}\lsim285$\\
$\bullet$ $m_{\chi^0_3}\sim m_{\chi^0_4}\sim m_{\chi^\pm_2}\sim\vert\mu\vert$
&$\bullet$ $m_{\chi^0_3}\sim m_{\chi^0_4}\sim m_{\chi^\pm_2}\sim\vert\mu\vert$
\\
$\bullet$ Spectrum easily accessible soon
&$\bullet$ Spectrum accessible soon\\ \hline
$\bullet$ Strict no-scale: $B(M_U)=0$
&$\bullet$ Special dilaton: $B(M_U)=2m_0$\\
\quad $\tan\beta=\tan\beta(m_t,m_{\tilde g})$
&\quad $\tan\beta=\tan\beta(m_t,m_{\tilde g})$\\
\quad $m_t\lsim135\GeV\Rightarrow\mu>0,m_h\lsim100\GeV$
&\quad $\tan\beta\approx1.4-1.6$, $m_t<155\GeV$\\
\quad $m_t\gsim140\GeV\Rightarrow\mu<0,m_h\gsim100\GeV$
&\quad $m_h\approx61-91\GeV$\\
\hline
\end{tabular}
\end{center}
\hrule
\end{table}

\begin{figure}[p]
\vspace{4.7in}
\includegraphics{faessler1.ps}
\vspace{4.0in}
\includegraphics{faessler2.ps}
\vspace{-1.5in}
\caption{\baselineskip=12pt
The parameter space of the no-scale and dilaton scenarios in $SU(5)\times
U(1)$ supergravity in the $(m_{\chi^\pm_1},\tan\beta)$ plane for
$m_t=150\GeV$. The periods indicate points that passed
all constraints, the pluses fail the $\bsg$ constraint, the crosses fail the
$(g-2)_\mu$ constraint, and the diamonds fail the neutrino telescopes (NT)
constraint. The dashed line indicates the direct reach of LEPII for
chargino masses. Note that when various symbols overlap a more complex symbol
is obtained.}
\label{PS}
\end{figure}


\clearpage


\newpage
\begin{thebibliography}{999}
\addcontentsline{toc}{section}{References}
\bibitem{LEPC} The LEP Collaborations (ALEPH, DELPHI, L3, OPAL) and the LEP
Electroweak Working Group, in {\em Proceedings of the International
Europhysics Conference on High Energy Physics}, Marseille, France, July 22--28,
1993, ed. by J. Carr and M. Perrottet (Editions Frontieres, Gif-sur-Yvette,
1993) CERN/PPE/93-157 (August 1993).
\bibitem{LP}P. Langacker and N. Polonsky, \PRD{47}{93}{4028}.
\bibitem{X1} A. Zichichi, EPS Conference, Geneva 1979, Proceedings; and
Nuovo Cimento Rivista {\bf2} (1979) 1. This is the first instance in which it
was pointed out that supersymmetry may play an important role in the
convergence of the gauge couplings.
\bibitem{CostaAmaldi}U. Amaldi, \etal, \PRD{36}{87}{1385}; G. Costa, \etal,
\NPB{297}{88}{244}.
\bibitem{EKNI}J. Ellis, S. Kelley, and \DVN, \PLB{249}{90}{441}.
\bibitem{EKNII}J. Ellis, S. Kelley, and \DVN, \PLB{260}{91}{131}.
\bibitem{LL}P. Langacker and M. Luo, \PRD{44}{91}{817}; L. Clavelli,
\PRD{45}{92}{3276}.
\bibitem{AdBF}U. Amaldi, W. de Boer, and H. F\"urstenau, \PLB{260}{91}{447}.
\bibitem{ACPZI}F. Anselmo, L. Cifarelli, A. Peterman, and A. Zichichi,
Nuovo Cim. {\bf104A} (1991) 1817.
\bibitem{ACPZII}F. Anselmo, L. Cifarelli, A. Peterman, and A. Zichichi,
Nuovo Cim. {\bf105A} (1992) 1025.
\bibitem{EKNIII}J. Ellis, S. Kelley and D. V.  Nanopoulos, \NPB{373}{92}{55}.
\bibitem{ACZ}F. Anselmo, L. Cifarelli, and A. Zichichi, Nuovo Cim. {\bf105A}
(1992) 1335.
\bibitem{BH}R. Barbieri and L. Hall, \PRL{68}{92}{752}; K. Hagiwara and Y.
Yamada, \PRL{70}{93}{709}; A. Faraggi, B. Grinstein, and S. Meshkov,
\PRD{47}{93}{5018}.
\bibitem{EGM}F. Anselmo, L. Cifarelli, A. Peterman, and A. Zichichi,
Nuovo Cim. {\bf105A} (1992) 581.
\bibitem{EKNIV}J. Ellis, S. Kelley and D. V.  Nanopoulos, \PLB{287}{92}{95}.
\bibitem{P4}F. Anselmo, L. Cifarelli, A. Peterman, and A. Zichichi,
Nuovo Cim. {\bf105A} (1992) 1179.
\bibitem{X2}F. Anselmo, L. Cifarelli, A. Peterman, and A. Zichichi,
Nuovo Cim. {\bf105A} (1992) 1201.
\bibitem{Talk} A. Zichichi, ``Understanding where the supersymmetry threshold
should be", CERN-PPE/92-149, 7 September 1992.
\bibitem{EWx}L. Ib\'a\~nez and G. Ross, \PLB{110}{82}{215}; K. Inoue, \etal,
Prog. Theor. Phys. 68 (1982) 927; L. Ib\'a\~nez, \NPB{218}{83}{514} and
\PLB{118}{82}{73}; H. P. Nilles, \NPB{217}{83}{366}; J. Ellis, \DVN, and
K. Tamvakis, \PLB{121}{83}{123}; J. Ellis, J. Hagelin, \DVN, and K. Tamvakis,
\PLB{125}{83}{275}; L. Alvarez-Gaum\'e, J. Polchinski, and M. Wise,
\NPB{221}{83}{495}; L. Iba\~n\'ez and C. L\'opez, \PLB{126}{83}{54} and
\NPB{233}{84}{545}; C. Kounnas, A. Lahanas, \DVN, and M. Quir\'os,
\PLB{132}{83}{95} and C. Kounnas, A. Lahanas, \DVN, and M. Quir\'os,
\NPB{236}{84}{438}.
\bibitem{LN}For a review see A. B. Lahanas and D. V. Nanopoulos,
\PRT{145}{87}{1}.
\bibitem{GRZ+EZ} G. Gamberini, G. Ridolfi, and F. Zwirner, \NPB{331}{90}{331};
J. Ellis and F. Zwirner, \NPB{338}{90}{317}.
\bibitem{KLNPY}S. Kelley, \JL, \DVN, H. Pois, and K. Yuan, \PLB{273}{91}{423}.
\bibitem{RR}G. Ross and R. Roberts, \NPB{377}{92}{571}.
\bibitem{DN} M. Drees and M.M. Nojiri, \NPB{369}{92}{54}.
\bibitem{Japs} K. Inoue, M. Kawasaki, M. Yamaguchi, and T. Yanagida,
\PRD{45}{92}{328}.
\bibitem{ANpd}R. Arnowitt and P. Nath, \PRL{69}{92}{725}; P. Nath and
R. Arnowitt, \PLB{287}{92}{89}.
\bibitem{aspects}S. Kelley, \JL, \DVN, H. Pois, and K. Yuan, \NPB{398}{93}{3}.
\bibitem{recent} R. Roberts and L. Roszkowski, \PLB{309}{93}{329};
M. Olechowski and S. Pokorski, \NPB{404}{93}{590}; B. de Carlos and J. Casas,
\PLB{309}{93}{320}; M. Carena, L. Clavelli, D. Matalliotakis, H. Nilles,
and C. Wagner, \PLB{317}{93}{346}; G. Leontaris, \PLB{317}{93}{569};
S. Martin and P. Ramond, NUB-3067-93TH (June 1993); D. Casta\~no, E.
Piard,  and P. Ramond, UFIFT-HP-93-18 (August 1993); W. de Boer, R.
Ehret,  and D. Kazakov, IEKP-KA/93-13 (August 1993); A. Faraggi and B.
Grinstein, SSCL-Preprint-496 (August 1993); M. Bastero-Gil, V. Manias, and J.
Perez-Mercader, LAEFF-93/012 (September 1993); M. Carena, M. Olechowski, S.
Pokorski, and C. Wagner, CERN-TH.7060/93 (October 1993); V. Barger, M. Berger,
and P. Ohmann, MAD/PH/801 (November 1993); A. Lahanas, K. Tamvakis, and N.
Tracas, CERN-TH.7089/93 (November 1993); G. Kane, C. Kolda, L. Roszkowski, and
J. Wells, UM-TH-93-24 (December 1993). For an elementary introduction
and a review of this field see \JL, \DVN, and \AZ, CERN-TH.7077/93
(October 1993), to be published in Rivista del Nuovo Cimento.
\bibitem{Lahanas} J. Ellis, A. Lahanas, \DVN, and K. Tamvakis,
\PLB{134}{84}{429}.
\bibitem{EKNI+II} J. Ellis, C. Kounnas, and \DVN, \NPB{241}{84}{406},
\NPB{247}{84}{373}.
\bibitem{Witten} E. Witten, \PLB{155}{85}{151}.
\bibitem{revamp}I. Antoniadis, J. Ellis, J. Hagelin, and \DVN,
\PLB{231}{89}{65}.
\bibitem{LNY}\JL, \DVN, and K. Yuan, \NPB{399}{93}{654}.
\bibitem{Dickreview}For reviews see \eg, R. Arnowitt and P. Nath, {\it Applied
N=1 Supergravity} (World Scientific, Singapore 1983); H. P. Nilles,
\PRT{110}{84}{1}.
\bibitem{CAN}A. Chamseddine, R. Arnowitt, and P. Nath, \PRL{49}{82}{970}.
\bibitem{WSY}S. Weinberg, \PRD{26}{82}{287}; N. Sakai and T. Yanagida,
\NPB{197}{82}{533}.
\bibitem{ENR}J. Ellis, \DVN, and S. Rudaz, \NPB{202}{82}{43};
B. Campbell, J. Ellis, and \DVN, \PLB{141}{84}{229}.
\bibitem{EMN}K. Enqvist, A. Masiero, and \DVN, \PLB{156}{85}{209}.
\bibitem{ANoldpd}P. Nath, A. Chamseddine, and R. Arnowitt, \PRD{32}{85}{2348};
P. Nath and R. Arnowitt, \PRD{38}{88}{1479}.
\bibitem{MATS}M. Matsumoto, J. Arafune, H. Tanaka, and K. Shiraishi,
\PRD{46}{92}{3966}.
\bibitem{HMY}J. Hisano, H. Murayama, and T. Yanagida, \PRL{69}{92}{1014} and
\NPB{402}{93}{46}.
\bibitem{LNP}\JL, \DVN, and H. Pois, \PRD{47}{93}{2468}.
\bibitem{LNPZ}\JL, \DVN, H. Pois, and A. Zichichi, \PLB{299}{93}{262}.
\bibitem{MPM} A. Masiero, \DVN, K. Tamvakis, and T. Yanagida,
\PLB{115}{82}{380}; B. Grinstein, \NPB{206}{82}{387}.
\bibitem{Arason}H. Arason, \etal, \PRL{67}{91}{2933}, \PRD{46}{92}{3945},
\PRD{47}{92}{232}.
\bibitem{CPW}M. Carena, S. Pokorski, and C. Wagner, \NPB{406}{93}{59}.
\bibitem{BBO}V. Barger, M. Berger, and P. Ohman, \PRD{47}{93}{1093}.
\bibitem{HS}L. Hall and U. Sarid, \PRL{70}{93}{2673}.
\bibitem{BEGN} A. J. Buras, J. Ellis, M. K. Gaillard, D. V. Nanopoulos,
\NPB{135}{78}{66}; D. V. Nanopoulos and D. A. Ross, \NPB{157}{79}{273},
\PLB{108}{82}{351}, and \PLB{118}{82}{99}.
\bibitem{GHS} A. Giveon, L. Hall, and U. Sarid, \PLB{271}{91}{138}.
\bibitem{YU}S. Kelley, \JL, and \DVN, \PLB{274}{92}{387}.
\bibitem{DHR}S. Dimopoulos, L. Hall, and S. Raby, \PRL{68}{92}{1984},
\PRD{45}{92}{4192}; G. Anderson, S. Dimopoulos, L. Hall, and S. Raby,
\PRD{47}{92}{R3702}.
\bibitem{Naculich}S. Naculich, JHU-TIPAC-930002 (January 1993).
\bibitem{LPII}P. Langacker and N. Polonsky, UPR-0556T (May 1993).
\bibitem{EG}J. Ellis and M. Gaillard, \PLB{88}{79}{315};
\DVN\ and M. Srednicki, \PLB{124}{83}{37}.
\bibitem{PDG}Particle Data Group, \PRD{45}{92}{S1}.
\bibitem{LNYdmI}\JL, \DVN, K. Yuan, \NPB{370}{92}{445}.
\bibitem{KLNPYdm}S. Kelley, \JL, \DVN, H. Pois, and K. Yuan,
\PRD{47}{93}{2461}.
\bibitem{troubles}\JL, \DVN, and A. Zichichi, \PLB{291}{92}{255}.
\bibitem{KT}See \eg, E. Kolb and M. Turner, {\em The Early Universe}
(Addison-Wesley, 1990).
\bibitem{ANcosm}R. Arnowitt and P. Nath, \PLB{299}{93}{58} and {\bf307} (1993)
403(E); P. Nath and R. Arnowitt, \PRL{70}{93}{3696}.
\bibitem{SWO}M. Srednicki, R. Watkins, and K. Olive, \NPB{310}{88}{693}.
\bibitem{GG}P. Gondolo and G. Gelmini, \NPB{360}{91}{145}; K. Griest and D.
Seckel, \PRD{43}{91}{3191}.
\bibitem{poles}\JL, \DVN, and K. Yuan, \PRD{48}{93}{2766}.
\bibitem{ANc}P. Nath and R. Arnowitt, \PLB{289}{92}{368}.
\bibitem{LNPWZh}\JL, \DVN, H. Pois, X. Wang, and A. Zichichi,
\PLB{306}{93}{73}.
\bibitem{LNWZ}\JL, \DVN, X. Wang, and A. Zichichi, \PRD{48}{93}{2062}.
\bibitem{LNZI}\JL, \DVN, and A. Zichichi, \PRD{49}{94}{343}.
\bibitem{revitalized}I. Antoniadis, J. Ellis, J. Hagelin, and \DVN,
\PLB{194}{87}{231}.
\bibitem{KLN}S. Kalara, J. Lopez, and \DVN, \PLB{245}{90}{421},
\NPB{353}{91}{650}.
\bibitem{decisive}J. L. Lopez and \DVN, \PLB{251}{90}{73}, \PLB{256}{91}{150},
and \PLB{268}{91}{359}.
\bibitem{EVA}S. Kelley, \JL, and \DVN, \PLB{261}{91}{424}.
\bibitem{chorus}J. Ellis, \JL, and \DVN, \PLB{292}{92}{189}.
\bibitem{sism}S. Kelley, \JL, and \DVN, \PLB{278}{92}{140}.
\bibitem{ELNO}J. Ellis, \JL, \DVN, and K. Olive, \PLB{308}{93}{70}.
\bibitem{ENO}J. Ellis, \DVN, and K. Olive, \PLB{300}{93}{121}.
\bibitem{Bethke}S. Bethke, in Proceedings of the XXVI International Conference
on High Energy Physics, Dallas, August 1992, ed. by J. R. Sanford (AIP
Conference Proceedings No. 272), p. 81.
\bibitem{IL}L. Ib\'a\~nez and D. L\"ust, \NPB{382}{92}{305}.
\bibitem{KL}V. Kaplunovsky and J. Louis, \PLB{306}{93}{269}.
\bibitem{Ibanez} A. Brignole, L. Ib\'a\~nez, and C. Mu\~noz, FTUAM-26/93
(August 1993).
\bibitem{EN}J. Ellis and \DVN, \PLB{110}{82}{44}.
\bibitem{thresholds} S. Kalara, J.L. Lopez, and D.V. Nanopoulos,
Phys. Lett. B {\bf269} (1991) 84; \JL, \DVN, and K. Yuan, in preparation.
\bibitem{LNZII}\JL, \DVN, and A. Zichichi, \PLB{319}{93}{451}.
\bibitem{BLM} R. Barbieri, J. Louis, and M. Moretti, \PLB{312}{93}{451} and
\PLB{316}{93}{632}(E).
\bibitem{ERZ}J. Ellis, G. Ridolfi, and F. Zwirner, \PLB{262}{91}{477}.
\bibitem{DNh}M. Drees and M. Nojiri, \PRD{45}{92}{2482}.
\bibitem{many}See \eg, R. Schaefer and Q. Shafi, Nature {\bf359} (1992) 199;
 A. N. Taylor and M. Rowan-Robinson, Nature {\bf359} (1992) 393.
\bibitem{muproblem}J. E. Kim and H. P. Nilles, \PLB{138}{84}{150} and
\PLB{263}{91}{79}; E. J. Chun, J. E. Kim, and H. P Nilles, \NPB{370}{92}{105}.
\bibitem{Casasmu}J. Casas and C. Mu\~noz, \PLB{306}{93}{288}.
\bibitem{GiMa}G. Giudice and A. Masiero, \PLB{206}{88}{480}.
\bibitem{DL}L. Durand and \JL, \PLB{217}{89}{463}, \PRD{40}{89}{207}.
\bibitem{trileptons}J. Ellis, J. Hagelin, \DVN, and M. Srednicki,
\PLB{127}{83}{233}; P. Nath and R. Arnowitt, \MODA{2}{87}{331}; R. Barbieri,
F. Caravaglios, M. Frigeni, and M. Mangano, \NPB{367}{91}{28};
H. Baer and X. Tata, \PRD{47}{93}{2739}.
\bibitem{sgdetection} See \eg, J. White, in {\em Recent Advances in the
Superworld}, Proceedings of the HARC Workshop, edited by \JL\ and \DVN
(World Scientific, Singapore 1994); R. M. Barnett, J. Gunion, and H. Haber,
\PLB{315}{93}{349}; H. Baer, C. Kao, and X. Tata, \PRD{48}{93}{R2978}.
\bibitem{Kamon} Talks given by J. T. White (D0 Collaboration) and Y. Kato (CDF
Collaboration) at the 9th Topical Workshop on Proton-Antiproton Collider
Physics, Tsukuba, Japan, October 1993; T. Kamon, private communication.
\bibitem{Hilgart} J. Schwindling, in {\em Proceedings of the International
Europhysics Conference on High Energy Physics}, Marseille, France, July 22--28,
1993, ed. by J. Carr and M. Perrottet (Editions Frontieres, Gif-sur-Yvette,
1993).
\bibitem{HHG}See \eg, {\it The Higgs Hunter's Guide}, J. Gunion, H. Haber, G.
Kane, and S. Dawson (Addisson-Wesley, Redwood City, 1990).
\bibitem{LG}\JL, \DVN, and X. Wang, \PLB{313}{93}{241}.
\bibitem{Alcaraz} See \eg, J. Alcaraz, M. Felcini, M. Pieri, and B. Zhou,
CERN-PPE/93-28.
\bibitem{LNPWZ}\JL, \DVN, H. Pois, X. Wang, and A. Zichichi,
\PRD{48}{93}{4062}.
\bibitem{hera} \JL, \DVN, \XW, and \AZ, \PRD{48}{93}{4029}.
\bibitem{DREES}M. Drees, and D. Zeppenfeld, \PRD{39}{89}{2536}.
\bibitem{PS85}W.H. Press and D.N. Spergel, Astrophys. J. {\bf 296}
(1985) 679.
\bibitem{Gould}A. Gould,  Astrophys. J. {\bf 321} (1987) 560, 571;
{\bf 328} (1988) 919; {\bf 388} (1992) 338.
\bibitem{others} J. Silk, K. Olive, and M. Srednicki, Phys. Rev. Lett.
{\bf 55} (1985) 257; T. Gaisser, G. Steigman, and S. Tilav, Phys. Rev.
D {\bf34} (1986) 2206; J. Hagelin, K. Ng, and K.  Olive,
Phys. Lett. B {\bf180} (1987) 375; M. Srednicki, K. Olive, and J. Silk,
Nucl. Phys. B {\bf 279} (1987) 804; K. Ng, K. Olive, and M. Srednicki,
Phys. Lett. B {\bf188} (1987) 138;
K. Olive and M. Srednicki, Phys. Lett. B {\bf205} (1988) 553; L. Krauss,
M. Srednicki, and F. Wilczek, Phys. Rev. D {\bf33} (1986) 2079; K. Freese,
Phys. Lett. {\bf 167B} (1986) 295.
\bibitem{RS}S. Ritz and D. Seckel, Nucl. Phys. B {\bf 304} (1988) 877.
\bibitem{GR89}G.F. Giudice and E. Roulet, Nucl. Phys. B {\bf 316} (1989)
429.
\bibitem{GGR91}G. Gelmini, P. Gondolo, and E. Roulet, Nucl. Phys.
B {\bf 351} (1991) 623.
\bibitem{KAM}M. Kamionkowski, Phys. Rev D {\bf 44} (1991) 3021.
\bibitem{KEK}M. Mori {\it et al.} (Kamiokande Collaboration),
Phys. Lett. B {\bf 270} (1991) 89.
\bibitem{Bott}A. Bottino, V. de Alfaro, N. Fornengo, G. Mignola,
and S. Scopel, Phys. Lett. B {\bf 265} (1991) 57;
Astroparticle Phys. {\bf 1} (1992) 61.
\bibitem{FH}F. Halzen, M. Kamionkowski, and T. Stelzer, Phys. Rev D
{\bf 45} (1992) 4439.
\bibitem{NT}R. Gandhi, \JL, \DVN, K. Yuan, and A. Zichichi, CERN-TH.6999/93.
\bibitem{KEKII}M. Mori {\it et al.} (Kamiokande Collaboration),
Phys. Lett. B {\bf 289} (1992) 463.
\bibitem{mac}B. Barish in the Proceedings of the Supernova Watch Workshop
(Santa Monica, California, 1990) (unpublished).
\bibitem{sk}Y. Suzuki in Proceedings of the 3rd International Workshop
on Neutrino Telescopes, Venice, March 1992, ed. M. Baldo-Ceolin, Venice
(1992).
\bibitem{amd}J. Learned in Proceedings of the European Cosmic Ray
Symposium, Geneva, July 1992, eds. P. Grieder and B. Pattison,
Nucl. Phys. B (1993), to appear.
\bibitem{hal} F. Halzen, Talk presented at the Johns Hopkins Workshop
on particles and the Universe, Budapest, July 1993, MAD-PH-785 (1993).
\bibitem{Thorndike} E. Thorndike, Bull. Am. Phys. Soc. {\bf38}, 922 (1993);
R. Ammar, \etal, CLEO Collaboration, \PRL{71}{93}{674}.
\bibitem{Barger}V. Barger, M. Berger, and R. J. N. Phillips,
\PRL{70}{93}{1368}; J. Hewett, \PRL{70}{93}{1045}.
\bibitem{BG}R. Barbieri and G. Giudice, \PLB{309}{93}{86}.
\bibitem{bsgamma}\JL, \DVN, and G. Park, \PRD{48}{93}{R974}.
\bibitem{bsg-eps}\JL, \DVN, G. Park, and A. Zichichi, \PRD{49}{94}{355}.
\bibitem{Oshimo} N. Oshimo, \NPB{404}{93}{20}; Y. Okada, \PLB{315}{93}{119}; R.
Garisto and J. N. Ng, \PLB{315}{93}{372}; F. Borzumati, DESY 93-090 (August
1993); M. Diaz, VAND-TH-93-13 (October 1993).
\bibitem{Bertolini}S. Bertolini, F. Borzumati, A. Masiero, and G. Ridolfi,
\NPB{353}{91}{591}.
\bibitem{oldg}J. Bailey \etal, \NPB{150}{79}{1}.
\bibitem{kinoII} For a recent review see, T. Kinoshita, Z. Phys. C{\bf56}
(1992) S80, and in {\em Frontiers of High Energy Spin Physics}, Proceedings
of the 10th International Symposium on High Energy Spin Physics, edited by
T. Hasegawa, N. Horikawa, A. Masaike, and S. Sawada (Universal Academy Press,
1993).
\bibitem{newg}M. May, in AIP Conf. Proc. USA Vol. 176 (AIP, New York, 1988)
p. 1168; B. L. Roberts, Z. Phys. {\bf C56} (1992) S101.
\bibitem{Fayet}P. Fayet, in Unification of the fundamental particle
interactions, eds. S. Ferrara, J. Ellis, and P. van Nieuwenhuizen (Plenum, New
York, 1980) p. 587.
\bibitem{GM}J. Grifols and A. Mendez, \PRD{26}{82}{1809}; J. Ellis, J. Hagelin
and \DVN, \PLB{116}{82}{283}; R. Barbieri and L. Maiani, \PLB{117}{82}{203};
J. C. Romao, A. Barroso, M. C. Bento and G. C. Branco, \NPB{250}{85}{295}.
\bibitem{KKS}D. A. Kosower, L. M. Kraus and N. Sakai, \PLB{133}{83}{305}.
\bibitem{YACN} T. C. Yuan, R. Arnowitt, A. H. Chamseddine, and P. Nath, Z.
Phys. C{\bf26} (1984) 407.
\bibitem{Vendramin}I. Vendramin, Nuovo Cimento {\bf 101A} (1989) 731.
\bibitem{E6g}J. A. Grifols, J. Sola and A. Mendez, \PRL{57}{86}{2348};
D. A. Morris, \PRD{37}{88}{2012}.
\bibitem{NMSSMg} M. Frank and C. S. Kalman, \PRD{38}{88}{1469};
R. M. Francis, M. Frank and C. S. Kalman, \PRD{43}{91}{2369};
H. K\"onig, Z. Phys. {\bf C52} (1991) 159 and Mod. Phys. Lett. {\bf A7} (1992)
279.
\bibitem{abel}S. A. Abel, W. N. Cottingham, and I. B. Whittingham,
\PLB{259}{91}{307}.
\bibitem{g-2}\JL, \DVN, and X. Wang, \PRD{49}{94}{366}.
\bibitem{HK}H. Haber and G. Kane, \PRT{117}{85}{75}.
\bibitem{ewcorr}\JL, \DVN, G. Park, H. Pois, and K. Yuan, \PRD{48}{93}{3297}.
\bibitem{eps1-epsb} \JL, \DVN, G.~T.~Park, and A. Zichichi, \TAMU{68/93}.
\bibitem{BFC}R. Barbieri, M. Frigeni, and F. Caravaglios, \PLB{279}{92}{169}.
\bibitem{PT}M. Peskin and T. Takeuchi, \PRL{65}{90}{964};
W. Marciano and J. Rosner, \PRL{65}{90}{2963};
D. Kennedy and P. Langacker, \PRL{65}{90}{2967}.
\bibitem{AB}G. Altarelli and R. Barbieri, \PLB{253}{90}{161}; G. Altarelli, R.
Barbieri, and S. Jadach, \NPB{369}{92}{3}.
\bibitem{ABC}G. Altarelli, R. Barbieri, and F. Caravaglios, \NPB{405}{93}{3}.
\bibitem{ABCII}G. Altarelli, R. Barbieri, and F. Caravaglios,
\PLB{314}{93}{357}.
\bibitem{DH}M. Dress and K. Hagiwara, \PRD{42}{90}{1709}.
\bibitem{BB}R. Barbieri, CERN Report No. CERN-TH.6659/92 (unpublished).
\bibitem{Altarelli} G. Altarelli, in {\em Proceedings of the International
Europhysics Conference on High Energy Physics}, Marseille, France, July 22--28,
1993, ed. by J. Carr and M. Perrottet (Editions Frontieres, Gif-sur-Yvette,
1993) CERN-TH.7045/93 (October 1993).
\bibitem{Easpects} \JL, \DVN, G. Park, X. Wang, and A. Zichichi, \TAMU{74/93}.
\bibitem{t-paper} \JL, \DVN, and A. Zichichi, \TAMU{78/93}.
\end{thebibliography}
\end{document}